\documentclass[aps,11pt,prd,groupedaddress,nofootinbib,notitlepage,eqsecnum,preprintnumbers]{revtex4-2}
\usepackage[utf8]{inputenc}
\usepackage{hyperref}
\usepackage{xcolor}
\usepackage{graphicx}
\usepackage{amsmath,amssymb}
\usepackage{bm}
\usepackage{comment}
\usepackage[shortlabels]{enumitem}
\usepackage{overpic}
\usepackage{ulem}
\begin{document}
\title{Induced gravitational waves from slow-roll inflation after an enhancing phase}

\author{\textsc{Shyam Balaji$^{a,b}$}}
    \email{{sbalaji}@{lpthe.jussieu.fr}}
\author{\textsc{Guillem Dom\`enech$^{c}$}}
    \email{{domenech}@{pd.infn.it}}
\author{\textsc{Joseph Silk$^{b,d,e}$}}
    \email{{silk}@{iap.fr}}

\affiliation{$^a$Laboratoire de Physique Th\'{e}orique et Hautes Energies (LPTHE), \\
    UMR 7589 CNRS \& Sorbonne Universit\'{e}, 4 Place Jussieu, F-75252, Paris, France}
\affiliation{$^b$Institut d’Astrophysique de Paris, UMR 7095 CNRS \& Sorbonne Universit\'{e}, 98 bis boulevard Arago, F-75014 Paris, France}
\affiliation{$^c$INFN Sezione di Padova, via Marzolo 8, I-35131 Padova, Italy}
\affiliation{$^d$Department of Physics and Astronomy, The Johns Hopkins University, 3400 N. Charles  Street, Baltimore, MD 21218, U.S.A.}
\affiliation{$^e$Beecroft Institute for Particle Astrophysics and Cosmology, University of Oxford, Keble Road, Oxford OX1 3RH, U.K.}

\begin{abstract}
The primordial spectrum of fluctuations may present a large peak as a result of enhancing features during inflation. This may include, but is not limited to, bumps in the inflaton's potential, phases of ultra-slow-roll or turns in multi-field space. However, in many models, inflation does not end immediately after the enhancing feature and it is likely to continue with a second phase of slow-roll. We show that the resulting induced gravitational waves may probe the primordial spectrum from the second inflationary phase, even if its amplitude is too small to directly induce detectable gravitational waves. This is because, if there are sharp peaks in the primordial spectrum, the total gravitational wave spectrum is not simply the sum of gravitational waves induced by a peaked and 
scale-invariant primordial spectra separately, but cross terms from interaction between these modes also become important. We also find that such cross terms always have a characteristic slope. We discuss the parameter space that may be probed by future gravitational wave detectors in the presence of these signals. 
\end{abstract}

\maketitle

\section{Introduction \label{sec:intro}}

Gravitational waves (GWs) provide a promising window to test the physics of the early universe. In the range of future GW detectors, we may discover GW signals from phase transitions, cosmic strings and primordial fluctuations, among many other physical scenarios (see \textit{e.g.} Refs.~\cite{Guzzetti:2016mkm,Caprini:2018mtu} for a review). In this work, we focus on the GWs induced by primordial fluctuations in the early universe \cite{Tomita,Matarrese:1992rp,Matarrese:1993zf,Ananda:2006af,Baumann:2007zm,Saito:2008jc,Saito:2009jt} (see \textit{e.g.} Refs.~\cite{Yuan:2021qgz,Domenech:2021ztg} for recent reviews). Strongly supported by observations of large scale fluctuations in the Cosmic Microwave Background (CMB) \cite{Akrami:2018odb}, these primordial perturbations emerge from quantum fluctuations during a period of cosmic inflation.\footnote{But see Ref.~\cite{Brandenberger:2016vhg} for a review on the status of bouncing cosmologies.} While the primordial spectrum of fluctuations is very well measured on CMB scales, it is mostly unknown on much smaller scales. Encouragingly, such small scales may induce observable GWs \cite{Yuan:2021qgz,Domenech:2021ztg}. Thus, the so-called induced GWs behave as a messenger from the physics of the yet unprobed inflation epoch \cite{Garcia-Bellido:2016dkw,Gong:2017qlj,Ando:2018nge,Byrnes:2018txb,Gao:2019kto,Xu:2019bdp,Liu:2020oqe,Cai:2019amo,Ozsoy:2019lyy,Ozsoy:2020kat,Ragavendra:2020sop,Fumagalli:2020nvq,Braglia:2020eai,Atal:2021jyo,Braglia:2020taf,Fumagalli:2021cel,Bastero-Gil:2021fac,Fumagalli:2021mpc,Fumagalli:2021dtd,Saikawa:2018rcs,Pi:2021dft,Fujita:2022ait,Lozanov:2022yoy,Inomata:2022ydj,Balaji:2022rsy,Domenech:2021and,Addazi:2022ukh,Balaji:2020yrx,Balaji:2018qyo} and subsequent period of reheating \cite{Assadullahi:2009nf,Inomata:2019zqy,Inomata:2019ivs,Inomata:2020lmk,Papanikolaou:2020qtd,Domenech:2020ssp,Domenech:2021wkk,Dalianis:2020gup,Hajkarim:2019nbx,Bhattacharya:2019bvk,Domenech:2019quo,Domenech:2020kqm,Dalianis:2020cla,Abe:2020sqb,Witkowski:2021raz}, especially if such physics leads to an enhancement of the primordial spectrum of curvature fluctuations on the smallest scales \cite{Saito:2008jc,Saito:2009jt,Assadullahi:2009jc,Bugaev:2009zh,Bugaev:2009kq,Bugaev:2010bb,Inomata:2018epa}.\footnote{Note that GWs induced during matter domination are significantly enhanced \cite{Assadullahi:2009nf,Alabidi:2013lya}. In this case, even a primordial spectrum with low amplitude might yield a detectable induced GW signal \cite{Inomata:2019ivs,Inomata:2019zqy,Dalianis:2020gup}.}

An enhancement of the primordial spectrum of curvature fluctuations might also lead to Primordial Black Hole (PBH) formation in the early universe \cite{Zeldovich:1967lct,Hawking:1971ei,Carr:1974nx,Meszaros:1974tb,Carr:1975qj,Khlopov:1985jw,Niemeyer:1999ak}. The PBH scenario is very attractive as it provides possible explanations for recent observations, such as some of the binary black hole merger events reported by LIGO \cite{Inomata:2016rbd,Nakama:2016gzw,Ando:2017veq,Clesse:2017bsw,Kohri:2018qtx,Garcia-Bellido:2020pwq,Franciolini:2021tla}. We refer the reader to Refs.~\cite{Khlopov:2008qy,Sasaki:2018dmp,Carr:2020gox,Carr:2020xqk,Green:2020jor,Escriva:2021aeh} for recent reviews. The resulting interest of the cosmology community in PBHs and induced GWs, has led to a thorough study of several mechanisms to enhance the primordial spectrum of fluctuations during inflation. Some examples are \cite{Kawasaki:1997ju,Frampton:2010sw,Kawasaki:2012wr,Inomata:2017okj,Pi:2017gih,Cai:2018tuh,Cai:2019jah,Chen:2019zza,Ashoorioon:2019xqc,Chen:2020uhe,Garcia-Bellido:1996mdl,Yokoyama:1998pt,Kohri:2012yw,Clesse:2015wea,Cheng:2016qzb,Espinosa:2017sgp,Inomata:2017okj,Kannike:2017bxn,Garcia-Bellido:2017mdw,Cheng:2018yyr,Ando:2018nge,Espinosa:2018eve,Inomata:2018cht,Braglia:2020eai,Atal:2018neu,Ng:2021hll,Byrnes:2018txb,Carrilho:2019oqg,Palma:2020ejf,Fumagalli:2020adf,Inomata:2021tpx,Cole:2022xqc,Zhou:2020kkf,Ragavendra:2020sop}, which include phases of ultra slow-roll, bumps in the inflaton's potential, sudden turns in the inflationary trajectory and resonances during inflation, among others.

Most of the recent attention has focused on the GWs induced by the enhanced peak in the primordial spectrum, the obvious reason for such a scenario being that the peak produces the largest amount of induced GWs.\footnote{The peak is even more relevant for PBH formation, as the PBH fraction depends exponentially on the amplitude of the primordial spectrum.} However, unless the feature is very broad or very close to the end of inflation, inflation must continue after the enhancing feature and it is likely to do so with a second phase of slow-roll preceding the end of inflation. Although the value for the slow-roll parameter of the second phase, and therefore the amplitude of the generated fluctuations, is in general terms a free parameter, there are some potential examples in the literature. Here we point out a few recent examples. In single field inflation, we see that step-like features in the inflaton potential \cite{Leach:2000yw,Inomata:2021uqj,Inomata:2021tpx}, punctuated inflation \cite{Ragavendra:2020sop} and constant rate inflation \cite{Ng:2021hll} yield a peak and a plateau in the primordial spectrum. In two-field inflationary models, we have curvatons \cite{Chen:2019zza}, axions \cite{Ando:2018nge,Ozsoy:2020ccy}, scalarons in Starobinsky inflation \cite{Pi:2017gih} and turns in field space \cite{Braglia:2020eai,Fumagalli:2020adf} can also lead to such structure. Note that in more general cases the plateau might instead be a slightly red-tilted spectrum. Nevertheless, we expect that the peak-plateau-like structure in the primordial spectrum is typical of inflationary models with a relatively sharp enhancing feature in the primordial power spectrum.

In this work, we explore the detectability of the GWs induced by the primordial spectrum from the second slow-roll phase. We do so by considering an arbitrary amplitude of the second slow-roll power spectrum. This is a reasonable assumption as the slow-roll parameter of the second phase is arbitrary. We find that, since induced GWs are a second-order effect, there could be a considerable range of frequencies where the main source of GWs are the GWs induced by the interaction between the peak and plateau modes in the primordial spectrum. This means that, in some cases and especially for sharp peaks, we may see a signal from the second stage of inflation without seeing the corresponding plateau in the induced GW spectrum. We also show that this effect is more pronounced if the propagation speed of fluctuations in the early universe is close to unity and that such an interaction signal has a very characteristic slope.

This paper is organized as follows. In \S~\ref{sec:inducedGWs} we introduce the formalism for induced GWs and classify their respective contributions from the different inflationary stages. In \S~\ref{sec:template} we derive GW spectrum templates for sharp peaks in the primordial spectrum followed by a plateau. Then, in \S~\ref{sec:forecasts} we forecast the parameter space that may be probed by future GW detectors. Finally we conclude with further discussions in \S~\ref{sec:conclusions}. We provide details of the calculations in the appendices. Throughout this work, we use reduced Planck units where $8\pi G=c=1$.

\section{Induced gravitational waves from two stages of inflation\label{sec:inducedGWs}}

In this section, we consider an inflation scenario consisting of two phases joined by a special feature. The first phase is the standard slow-roll inflation that yields the right amplitude of primordial fluctuations to explain the CMB anisotropies, say ${\cal A}_{\rm CMB}$. Then, there is a feature during inflation that enhances primordial fluctuations and yields a peak in the primordial spectrum of curvature fluctuations with amplitude ${\cal A}_{\rm peak}$. Subsequent to this, inflation continues with a second phase of slow-roll. For simplicity, we take the primordial spectrum from the second phase to be a 
scale-invariant spectrum with arbitrary amplitude, say ${\cal A}_{\rm flat}$. The arbitrariness of ${\cal A}_{\rm flat}$ stems from the fact that the first slow-roll parameter $\epsilon$ during the second phase is basically a free parameter. The smaller the value of $\epsilon$, the larger the value of ${\cal A}_{\rm flat}$. From now on, we assume that ${\cal A}_{\rm CMB}\ll{\cal A}_{\rm flat}$ and neglect the contribution from ${\cal A}_{\rm CMB}$. We justify a posteriori that this has no effect on our results.

With the above motivation in mind, let us consider the following approximate primordial spectrum of curvature fluctuations
\begin{align}\label{eq:PR}
{\cal P}_{\cal R}(k)={\cal A}_{\rm peak}{\cal P}_{{\cal R},{\rm peak}}(k/k_p)+{\cal A}_{\rm flat}{\cal P}_{{\cal R},{\rm flat}}(k/k_p)\,,
\end{align}
where ${\cal P}_{{\cal R},{\rm peak}}(k/k_p)$ is a peaked function at $k=k_p$ and ${\cal P}_{{\cal R},{\rm flat}}$ is either unity or a function that has a step at $k=k_p$ and is non-vanishing for $k>k_p$. We expect the template \eqref{eq:PR} to be a good approximation to most models where the feature during inflation has sharp transitions from and to the first and second slow-roll phases. For example, see Refs.~\cite{Pi:2017gih,Ando:2018nge,Atal:2018neu,Chen:2019zza,Braglia:2020eai,Ragavendra:2020sop,Fumagalli:2020adf,Ng:2021hll}. In cases where transitions between these phases are gradual, the distinction between the peak and the plateau used in Eq.~\eqref{eq:PR} might not be as accurate but is still useful for order-of-magnitude estimates.

Our assumption that the primordial spectrum \eqref{eq:PR} can be split into two different contributions results in a clear separation of the total induced GW spectrum: the GWs induced by ${\cal P}_{{\cal R},{\rm peak}}$ and ${\cal P}_{{\cal R},{\rm flat}}$ but also by the interaction between them, which we call cross terms. For very sharp peaks in ${\cal P}_{{\cal R},{\rm peak}}$, the physical picture is the following. Since it will be the most relevant case, let us focus on tensor modes that enter the horizon before the scalar mode $k_p$, corresponding to the position of the peak in ${\cal P}_{{\cal R},{\rm peak}}$. That is to say, we consider $k>k_p$. When $k\tau<1$, tensor modes have a constant second order source and grow as $\tau^2$. Then, most of the GWs are induced at horizon crossing, except for possible resonances. For a sharply peaked scalar spectrum, most GWs are induced when $k_p$ enters the horizon, as the source quickly decays afterwards. For a scale invariant scalar spectrum, GW production effectively stops when the tensor mode $k$ enters the horizon. But, when both a peaked and a flat spectra are present, tensor modes with $k>k_p$ acquire growth due to the peak, which scales as $(k_p\tau)^2$, until horizon crossing at $\tau=1/k$. This implies that the cross term in the induced GW spectrum for $k>k_p$ behaves as a power-law with spectral index equal to $-4$. Interestingly, it turns out that when $c_s^2\sim1$ the source term oscillates with frequency $c_sk\sim k$. Then, there is a resonance which persists since the tensor mode $k$ enters the horizon until the scalar mode $k_p$ also enters. Such resonance instead yields $\Omega_{\rm GW,cross}\sim k^{-2}$. We shall explicitly derive these two results and present a heuristic explanation of the source and the resonance in Appendix~\ref{app:derivation}.

%While most of the GWs are induced when the scalar mode $k_p$, corresponding to the position of the peak, enters the horizon, it is immediately followed by other scalar modes from ${\cal P}_{{\cal R},{\rm flat}}$. Since induced GWs are a second order effect, GWs induced by scalar modes with $k>k_p$ will pick up some of the larger amplitude of ${\cal P}_{{\cal R},{\rm peak}}$. For $k\gtrsim k_p$, the amplitude of the scalar mode with $k_p$ has not decayed significantly and the amplitude of the cross term is larger than for $k\gg k_p$. Thus, we expect that the cross term connects the GWs induced separately by ${\cal P}_{{\cal R},{\rm peak}}$ and ${\cal P}_{{\cal R},{\rm flat}}$ with a smooth decreasing function of $k$. Let us proceed to compute the total GW spectrum in more detail.

For simplicity, we consider that the universe after inflation is dominated by radiation but with an arbitrary speed of sound $c_s$. Then, the resulting induced GW spectrum during a radiation-like domination phase evaluated today is given by \cite{Inomata:2016rbd,Domenech:2021ztg}
\begin{align}\label{eq:omega0}
\Omega_{\rm GW,0}h^2= \Omega_{r,0}h^2 \left(\frac{g_*(T_{\rm c})}{g_{*,0}}\right)\left(\frac{g_{*s}(T_{\rm c})}{g_{*s,0}}\right)^{-4/3}\Omega_{\rm GW,c}\,,
\end{align}
where $\Omega_{r,0}$ is the density fraction of radiation today and $g_{*}(T)$ and $g_{*s}(T)$ are the effective number of degrees of freedom in the energy density and entropy at a temperature $T$ respectively. From Planck satellite data, we have $\Omega_{r,0}h^2\approx4.18\times 10^{-5}$ \cite{Aghanim:2018eyx}. Fitting functions for $g_{*}(T)$ and $g_{*s}(T)$ can be found in Ref.~\cite{Saikawa:2018rcs}. In particular, we have $g_{*,0} = 3.36$ and $g_{*s,0} = 3.91$. One also finds for $T>100 {\rm GeV}$ that, assuming only Standard Model particle content, $g_{*}(T) = g_{*s}(T) =106.75$. Following the notation of \cite{Inomata:2016rbd}, the subscript ``c'' in Eq.~\eqref{eq:omega0}indicates evaluation at a time when the spectral density is constant, that is when tensor modes are contained sufficiently within the cosmological horizon. Then, $\Omega_{\rm GW,c}$ is the spectral density of induced GWs evaluated deep inside the radiation domination era, which is given by \cite{Kohri:2018awv,Espinosa:2018eve,Domenech:2019quo,Domenech:2021ztg} 
\begin{align}\label{eq:inducedGWs}
\Omega_{\rm GW,c}=\int_0^\infty dv\int_{|1-v|}^{1+v}du\,{\cal T}(u,v,c_s){{\cal P}_{\cal R}(ku)}{{\cal P}_{\cal R}(kv)}\,,
\end{align}
where $v$ and $u$ are dimensionless variables related to the scalar internal momentum $q$ by $v=q/k$ and $u=|k-q|/k$ and we have defined the transfer function, also referred to as a kernel, ${\cal T}(u,v,c_s)$ as 
\begin{align}\label{eq:kernel}
{\cal T}(u,v,c_s)\equiv\frac{y^2}{3c_s^4}&\left(\frac{4v^2-(1-u^2+v^2)^2}{4u^2v^2}\right)^2\nonumber\\&\times
\left\{\frac{\pi^2}{4}y^2\Theta[c_s(u+v)-1]
+\left(1-\frac{1}{2}y \ln\left|\frac{1+y}{1-y}\right|\right)^2\right\}\,,
\end{align}
with
\begin{align}\label{eq:y}
y=\frac{u^2+v^2-c_s^{-2}}{2uv}\,.
\end{align}
The transfer function \eqref{eq:kernel} has been derived in Refs.~\cite{Domenech:2019quo,Domenech:2020kqm,Domenech:2021ztg} assuming that the early universe is dominated by a radiation-like fluid, i.e. the equation of state of the fluid is $w=1/3$, but with an arbitrary sound speed of scalar fluctuations $c_s$. For an adiabatic perfect fluid we have $c_s^2=w=1/3$. For a canonical scalar field in a suitable exponential potential \cite{Lucchin:1984yf} we have $w=1/3$ and $c_s^2=1$. We consider $c_s^2$ as a free parameter because little is known about the composition of the early universe at the times when such induced GWs are produced. Most importantly though is that, as will be shown later, the high-frequency part of the GW spectrum is sensitive to the value of $c_s^2$.

It should be noted that \eqref{eq:inducedGWs} is only valid for Gaussian initial conditions. Large local non-Gaussianity might introduce some changes in the shape of the resulting induced GW spectrum. For instance, for sharp peaks in the primordial spectrum, GWs induced by primordial local non-Gaussianity appear noticeably in the range $2k_p<k<4k_p$ \cite{Cai:2018dig,Unal:2018yaa,Adshead:2021hnm}, beyond the cut-off of the Gaussian contribution. But for a broken power-law primordial spectrum, local non-Gaussianity does not significantly change the shape of the induced GW spectrum \cite{Atal:2021jyo,Adshead:2021hnm}. Nevertheless, it is important to mention that the contribution of local non-Gaussianity to the induced GW spectrum often becomes important when $F_{\rm NL}^2{\cal A}_{\cal R}\gtrsim {\cal O}(1)$, with $F_{\rm NL}$ denoting the local non-Gaussianity parameter, which is beyond the perturbative regime \cite{Atal:2021jyo}. Furthermore, we expect that in concrete multi-field models local non-Gaussianity might only be important around the scale of the peak $k_p$ but not for the fluctuations generated during the second slow-roll phase. Also, in some single field models, local non-Gaussianity can be linked to the slope of the primordial spectrum after the peak \cite{Atal:2019erb,Atal:2021jyo}. Thus, having a plateau-like structure implies small local non-Gaussianity in single field models. While it would be interesting to study the general impact of local non-Gaussianity in the cross term, it is out of the scope of this paper and we leave such an analysis for future work.

For concreteness, we shall substitute the primordial curvature power spectrum \eqref{eq:PR} into \eqref{eq:inducedGWs} and separate the subsequent contributions to the total GW spectral density as
\begin{align}\label{eq:split}
\Omega_{\rm GW,c}={\cal A}_{\rm peak}^2\Omega_{\rm GW,peak}+2{\cal A}_{\rm peak}{\cal A}_{\rm flat}\Omega_{\rm GW,\rm cross}+{\cal A}_{\rm flat}^2\Omega_{\rm GW,flat}\,.
\end{align}
We note that this separation is only possible because we started with a primordial spectrum which is a sum of two different contributions \eqref{eq:PR}. Nevertheless, it will prove to be convenient later.  In Eq.~\eqref{eq:split}, $\Omega_{\rm GW,peak}$ and $\Omega_{\rm GW,flat}$ are given by \eqref{eq:inducedGWs} replacing ${\cal P}_{\cal R}$ for ${\cal P}_{\cal R,\rm peak}$ and ${\cal P}_{{\cal R},{\rm flat}}$ respectively. We also define
\begin{align}\label{eq:inducedGWscross}
\Omega_{\rm GW,cross}=\int_0^\infty dv \,{{\cal P}_{\cal R,\rm peak}(v/v_p)}\,\int_{|1-v|}^{1+v}du\,{\cal P}_{{\cal R},{\rm flat}}(u/v_p)\,{\cal T}(u,v,c_s)\,,
\end{align}
where $v_p\equiv k_p/k$ and we used the fact that the integral is symmetric with respect to the exchange of variables $u\leftrightarrow v$ to write a single cross term with an additional factor $2$ in Eq.~\eqref{eq:split}. The contributions $\Omega_{\rm GW,peak}$ and $\Omega_{\rm GW,flat}$ have been studied in the literature for $c_s^2=w=1/3$, for example, in Refs.~\cite{Kohri:2018awv,Pi:2020otn,Atal:2021jyo}. The new results of this work are the generalisation to $c_s^2=1$ and the inclusion of $\Omega_{\rm GW,cross}$.

Before we study two useful examples, let us present a series of general analytical approximations for the low frequency/infrared (IR), that is for $k\ll k_p$, and the high frequency/ultraviolet (UV) regimes, that is for $k\gg k_p$, in the case where ${\cal P}_{\cal R,\rm peak}$ is sharply peaked.

\subsection{Low frequency (infrared) approximation}

In the limit when $k\ll k_p$, or equivalently $v_p\gg1$, we find that, since ${\cal P}_{\cal R,\rm peak}$ is sharply peaked, only those momenta $v$ very close to $v_p$ contribute to the $v$ integral. Thus, we may assume that for all practical purposes $v\gg1$. This also implies that since $u$ is bounded by $|1-v|<u<1+v$, that $u\sim v\gg 1$. In this regime, the transfer function \eqref{eq:kernel} is approximated by
\begin{align}\label{eq:kernelIR}
{\cal T}(u\sim v \gg 1,c_s)\approx \frac{1}{3c_s^4}v^{-4}\ln^2v\,.
\end{align}
In fact, by the mean value theorem for integrals, we may also approximate the $u$ integral by evaluating the integrand at $u=v$. This may result in at most an $\mathcal{O}(1)$ error in the numerical coefficient \cite{Atal:2021jyo}. However, if the peak is sharp enough, this already gives a good order-of-magnitude estimate. With this approximation, we have that Eqs.~\eqref{eq:inducedGWs} and \eqref{eq:inducedGWscross} are approximately given by
\begin{align}\label{eq:inducedGWspeakIR}
\Omega^{\rm IR}_{\rm GW,peak/cross}\approx\frac{1}{3c_s^4}\left(\frac{k}{k_p}\right)^3&\ln^2\left(\frac{k}{k_p}\right)\nonumber\\&
\times\left\{
\begin{aligned}
{{\cal P}_{\cal R,\rm peak}(k=k_p)}\\
{{\cal P}_{\cal R,\rm flat}(k=k_p)}
\end{aligned}
\right\}
\int_0^\infty dV \,{{\cal P}_{\cal R,\rm peak}(V)}\,V^{-4}\,,
\end{align}
where we have defined $V\equiv {v}/{v_p}$. We then conclude that both $\Omega_{\rm GW,peak}$ and $\Omega_{\rm GW,cross}$ decay as $k^3\ln^2k$ in the IR. This is the well-known universal IR behaviour for localised sources \cite{Cai:2019cdl}. Note that the logarithmic correction is typical of GWs induced during radiation domination \cite{Yuan:2019wwo,Domenech:2020kqm}. Therefore, as long as ${\cal A}_{\rm peak}\gg {\cal A}_{\rm flat}$, the cross term never dominates the IR tail of the GW spectrum. As we shall see, the high frequency regime becomes more interesting and qualitatively different for the cross term.

\subsection{High frequency (ultraviolet) approximation}

We now study the limit when $k\gg k_p$, which corresponds $v_p\ll1$. As explained in Ref.~\cite{Atal:2021jyo}, we find that because ${\cal P}_{\cal R,\rm peak}$ is peaked, only for those momenta $v$ or $u$ close to $v_p$ which contribute to the integral. This means that we may focus only on those regions with $v\ll1$ and $u\sim 1$, and by symmetry also $u\ll1$ and $v\sim 1$. We shall therefore restrict ourselves only  to the region $v\ll1$ and $u\sim 1$. For $\Omega_{\rm GW,peak}$ the symmetry in $u\leftrightarrow v$ introduces an additional factor $2$. For $\Omega_{\rm GW,cross}$ we already used that symmetry, and the factor $2$ is therefore already included in Eq.~\eqref{eq:split}. 

In contrast to the IR regime, the behaviour of the transfer function in the UV regime is different for $c_s^2=1/3$ and $c_s^2=1$. This is mainly because of a competition in the integration plane between the resonance at $u+v=c_s^{-2}$ and the boundaries at $u=1+v$ and at $u=|1-v|$ where the integrand vanishes.\footnote{More precisely, the factor between brackets in the first line of \eqref{eq:kernel}, which comes from the projection of the scalar mode on the tensor polarisation plane, vanishes when the scalar mode $\mathbf{q}$ is parallel to the tensor mode $\mathbf{k}$, which occurs when $u=1+v$ or $u=|1-v|$.} In particular, when $c_s^2=1$ the resonance is completely ``killed'' at the boundary $u=1-v$. The fact that the kernel has two different behaviours in the UV is also clear from the Taylor expansion of the variable $y$ for $v\ll1$ and $u\sim1$, which yields
\begin{align}
y\approx \frac{1-c_s^{-2}}{2v} + \frac{v}{2}\,.
\end{align}
When $c_s^2=1$ the first term exactly cancels and $y\ll1$ for $v\ll1$. However, for $c_s^{2}<1$, we see that $|y|\gg1$ for $v\ll1$.  For the moment, let us consider the two cases at hand, $c_s^2=1/3$ and $c_s^2=1$. We provide the details on how to compute the integral over $u$ of the transfer function ${\cal T}(u,v,c_s)$ in Appendix~\ref{app:generalcs} for general values of $c_s$. There, we also show that the limit $c_s^2\to1$ is continuous. Expanding \eqref{eq:inducedGWs} and \eqref{eq:inducedGWscross} for small $v$, we find on one hand for $c_s^2=1/3$ that
\begin{align}\label{eq:inducedGWspeakUVcs1/3}
\Omega^{\rm UV}_{\rm GW,peak/cross}&(c_s^2<1)\nonumber\\&\approx \frac{8}{27}\frac{1}{(1-c_s^2)^2}\left(\frac{k}{k_p}\right)^{-4}\left\{
\begin{aligned}
2{{\cal P}_{\cal R,\rm peak}(k/k_p)}\\
{{\cal P}_{\cal R,\rm flat}(k/k_p)}
\end{aligned}
\right\}
\int_0^\infty dV \,{{\cal P}_{\cal R,\rm peak}(V)}V^3\,,
\end{align}
while for $c_s^2=1$ we instead obtain
\begin{align}\label{eq:inducedGWspeakUVcs1}
\Omega^{\rm UV}_{\rm GW,peak/cross}&(c_s^2=1)\nonumber\\&\approx 2\frac{35+24\pi^2}{8505}\left(\frac{k}{k_p}\right)^{-2}\left\{
\begin{aligned}
2{{\cal P}_{\cal R,\rm peak}(k/k_p)}\\
{{\cal P}_{\cal R,\rm flat}(k/k_p)}
\end{aligned}
\right\}
\int_0^\infty dV \,{{\cal P}_{\cal R,\rm peak}(V)}V\,.
\end{align}
From Eqs.~\eqref{eq:inducedGWspeakUVcs1/3} and \eqref{eq:inducedGWspeakUVcs1} we see that while for $c_s^2=1/3$ the UV tail decays as $k^{-4}{\cal P}_{\cal R,\rm peak/flat}(k)$, it decays much slower for $c_s^2=1$ as $k^{-2}{\cal P}_{\cal R,\rm peak/flat}(k)$. This also means that the case of $c_s^2=1$ is much more interesting for future prospects of detecting the GW signal from the cross term. It is worth emphasizing that the $k^{-2}$ slope is not unique to $c_s^2=1$. In fact, as shown in Appendix~\ref{app:generalcs}, we find that a substantial $k^{-2}$ slope appears right after the peak in the cross term even for $c^2_s\gtrsim0.6$. Also note that the power of $k$ in Eqs.~\eqref{eq:inducedGWspeakUVcs1/3} and \eqref{eq:inducedGWspeakUVcs1} depends on the equation of state of the universe $w$ ($w=1/3$ for radiation) and on the spectral index of the plateau respectively. Using the formulas provided in Ref.~\cite{Atal:2021jyo}, \eqref{eq:inducedGWspeakUVcs1/3} and \eqref{eq:inducedGWspeakUVcs1}, we expect that in the general case there is an additional factor $k^{-b-n}$, where $b=(1-3w)/(1+3w)$. However, so as not to obscure the main discussion of the paper, we leave the study for general $w$ for future work. One of the main results of our paper is that for $w=1/3$, the 
cross term decays with a characteristic slope of $k^{-4}$ for $c_s^2\lesssim 0.6$ and of $k^{-2}$ for $c_s^2>0.6$. We now proceed to show rigorous examples with the approximations developed here.

\section{Template spectra\label{sec:template}}

In most cases, the primordial spectrum ${\cal P}_{\cal R,\rm peak}$ resulting from a feature during inflation is broadly classified by either a broken power-law or a log-normal function \cite{Kawasaki:1997ju,Frampton:2010sw,Kawasaki:2012wr,Inomata:2017okj,Pi:2017gih,Cai:2018tuh,Cai:2019jah,Chen:2019zza,Ashoorioon:2019xqc,Chen:2020uhe,Byrnes:2018txb,Cole:2022xqc}. In general terms, a broken power-law is typical of single field inflation models \cite{Byrnes:2018txb,Cole:2022xqc} while a log-normal appears in multi-field models \cite{Pi:2017gih,Chen:2019zza,Braglia:2020eai,Palma:2020ejf,Fumagalli:2020adf}. In this section, we focus on the case of a log-normal as it yields the most interesting phenomenology involving the cross term \eqref{eq:inducedGWscross}. We show in Appendix~\ref{app:brokenpowerlaw} that in the case where ${\cal P}_{\cal R,\rm peak}$ is a broken power-law, the cross term does not contribute significantly unless the broken power-law is very sharp, which shares similarities with a sharp log-normal. We proceed to study two examples which will allow us to infer properties for more general cases. 

\subsection{Toy model: Dirac delta plus flat primordial spectra \label{sec:dirac}}

We start with an exact analytical model, which is a Dirac delta plus a scale invariant spectrum, namely we take
\begin{align}\label{eq:Pdelta}
{\cal P}_{\cal R,\rm peak}=\delta\left(\ln\left(\frac{k}{k_p}\right)\right)\quad{\rm and}\quad {\cal P}_{\cal R,\rm flat}=1.
\end{align}
In this case, we have from \eqref{eq:inducedGWs} that
\begin{align}\label{eq:peakOmega}
\Omega_{\rm GW,peak}&=v_p^2{\cal T}(u=v_p,v=v_p,c_s)\Theta(2v_p-1)\,,
\end{align}
where the Heaviside function $\Theta(x)$ comes from momentum conservation, or mathematically from the integration boundaries of $u$.
We also find numerically that
\begin{align}\label{eq:flatOmega}
\Omega_{\rm GW,flat}&\approx\left\{
\begin{aligned}
0.82 & \quad (c_s^2=1/3)\\
0.14 & \quad (c_s^2=1)\,
\end{aligned}
\right.\,.
\end{align}
Lastly, from \eqref{eq:inducedGWscross} we obtain
\begin{align}\label{eq:deltacross}
\Omega_{\rm GW,cross}=v_p\int_{|1-v_p|}^{1+v_p}du{\cal T}(u,v=v_p,c_s)\,.
\end{align}

While the Dirac delta example \eqref{eq:Pdelta} is the simplest case, Eqs.~\eqref{eq:flatOmega} and \eqref{eq:deltacross} are not particularly useful in the current form. Nevertheless, we may gain insight by studying the IR and UV approximations given by Eqs.~\eqref{eq:inducedGWspeakIR}, \eqref{eq:inducedGWspeakUVcs1/3} and \eqref{eq:inducedGWspeakUVcs1}. First, using \eqref{eq:inducedGWspeakIR} we find that the IR tail goes as
\begin{align}\label{eq:IRcrossdelta}
\Omega^{\rm IR}_{\rm GW,peak}\approx\frac{1}{3c_s^4}\left(\frac{k}{k_p}\right)^2\ln^2\left(\frac{k}{k_p}\right)
\quad{\rm and}\quad
\Omega^{\rm IR}_{\rm GW,cross}\approx\frac{1}{3c_s^4}\left(\frac{k}{k_p}\right)^3\ln^2\left(\frac{k}{k_p}\right)\,,
\end{align}
which confirms our expectation that the cross term never dominates in the IR regime of the total GW spectrum \eqref{eq:split} if ${\cal A}_{\rm peak}>{\cal A}_{\rm flat}$ as it is suppressed by a factor $k/k_p$. We now turn to the UV tail. Inserting \eqref{eq:Pdelta} into \eqref{eq:inducedGWspeakUVcs1/3} and \eqref{eq:inducedGWspeakUVcs1} we obtain
\begin{align}\label{eq:UVtaildelta}
\Omega^{\rm UV}_{\rm GW,cross}\approx 
\left\{
\begin{aligned}
&\frac{2}{3}\left(\frac{k}{k_p}\right)^{-4}&(c_s^2=1/3)\\
&2\frac{35+24\pi^2}{8505}\left(\frac{k}{k_p}\right)^{-2}&(c_s^2=1)
\end{aligned}
\right.\,.
\end{align}
In Eq.~\eqref{eq:UVtaildelta}, we only show the cross term as $\Omega_{\rm GW,peak}$ in Eq.~\eqref{eq:peakOmega} has a well-known cut-off at $k=2k_p$. Therefore, for the Dirac delta case, the cross term always dominates for $k>2k_p$. 

In light of the analytical approximations, we can also estimate the range of $k$ where the cross term dominates the total GW spectrum in Eq.~\eqref{eq:split}. We start by noting that the first break point of the total GW spectrum occurs at $k_{b1}=2k_p$. From the UV tail of the cross term \eqref{eq:UVtaildelta} and the amplitude of $\Omega_{\rm GW, flat}$ \eqref{eq:flatOmega}, we find that the second break point of the total GW spectrum \eqref{eq:split} occurs at
\begin{align}\label{eq:break2}
\frac{k_{b2}}{k_p}\approx \left\{
\begin{aligned}
&1.13A_{\rm rel}^{-1/4}&(c_s^2=1/3)\\
&0.96A_{\rm rel}^{-1/2}&(c_s^2=1)
\end{aligned}
\right.\,,
\end{align}
where we have introduced the relative amplitude given by
\begin{align}\label{eq:arel}
A_{\rm rel}\equiv\frac{{\cal A}_{\rm flat}}{{\cal A}_{\rm peak}}\,.
\end{align}
We then conclude that the visible width of the cross term is given by
\begin{align}\label{eq:widthdelta}
\frac{\Delta k_{\rm cross}}{{k_{b1}}}=\frac{k_{b2}-k_{b1}}{k_{b1}}\approx\left\{
\begin{aligned}
&0.57A_{\rm rel}^{-1/4}-1&(c_s^2=1/3)\\
&0.48A_{\rm rel}^{-1/2}-1&(c_s^2=1)
\end{aligned}
\right.\,.
\end{align}
Eq.~\eqref{eq:widthdelta} also shows that for $A_{\rm rel}\gg 1$, the logarithmic width of the cross terms for $c_s^2=1$ is twice that of the $c_s^2=1/3$ case. These results for the cross term, namely Eqs.~\eqref{eq:IRcrossdelta}, \eqref{eq:deltacross} and \eqref{eq:widthdelta}, also hold approximately for sharp enough peaks such as a narrow log-normal, which we study below.

\begin{figure}
\includegraphics[width=0.45\columnwidth]{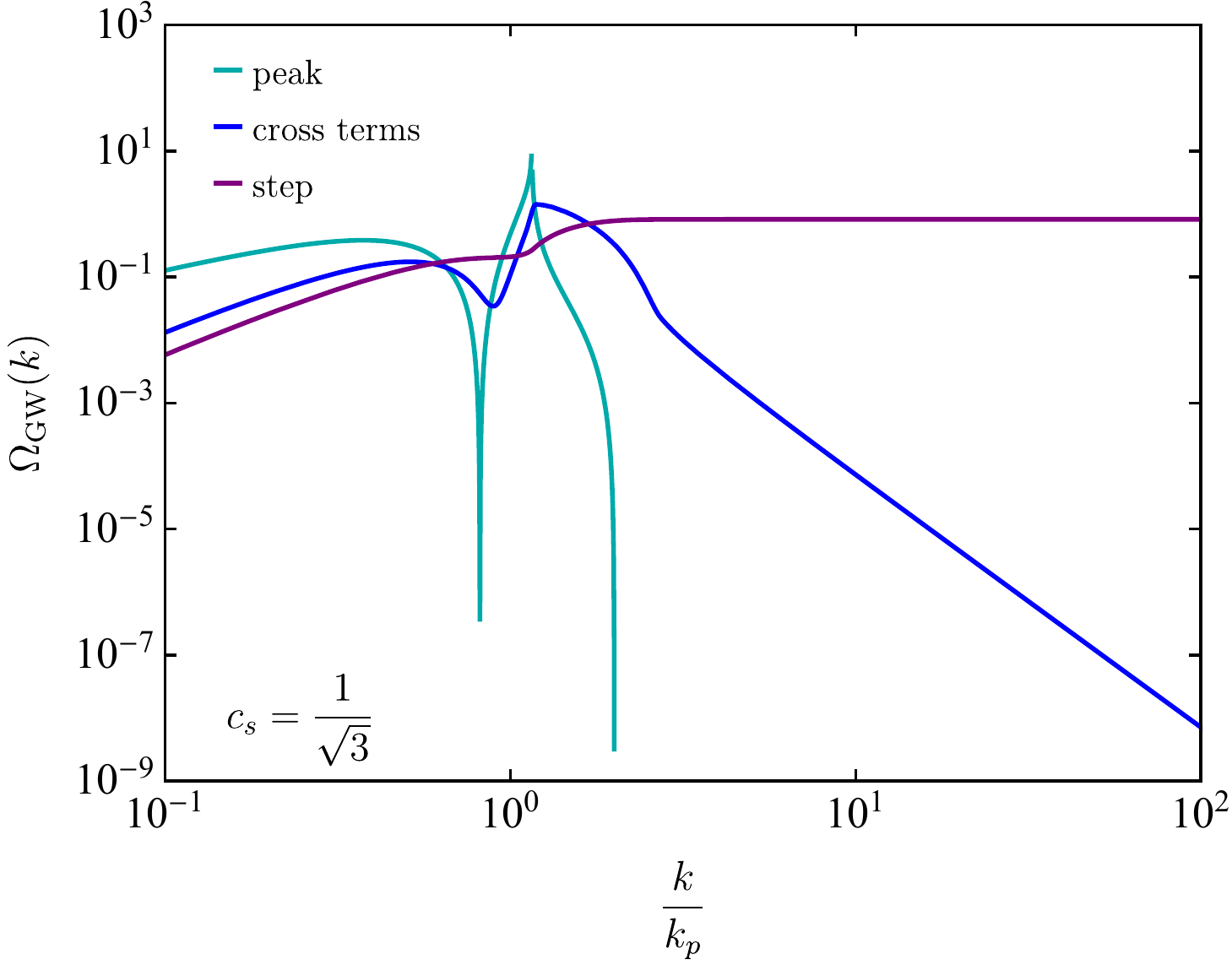}
\includegraphics[width=0.45\columnwidth]{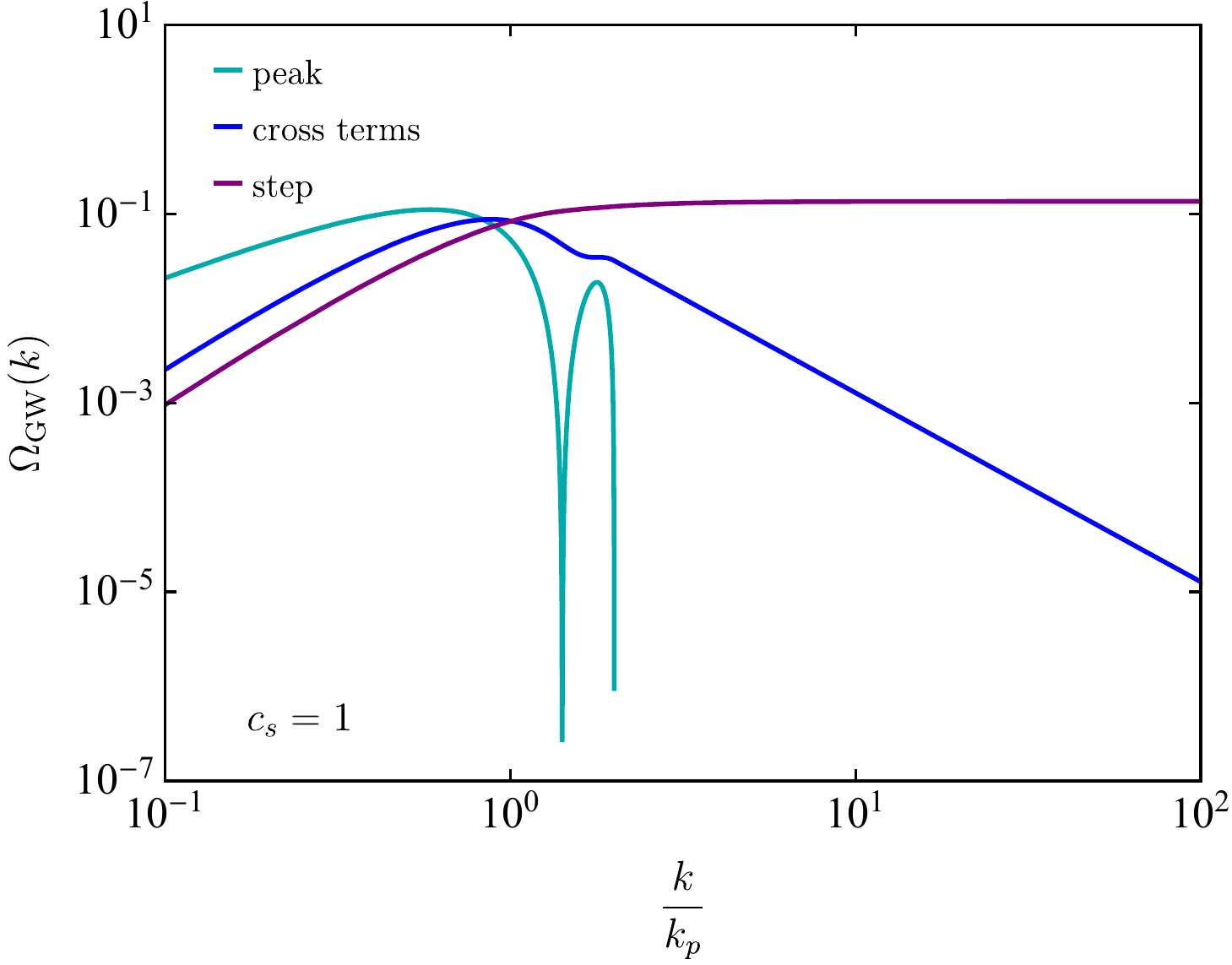} 
\includegraphics[width=0.45\columnwidth]{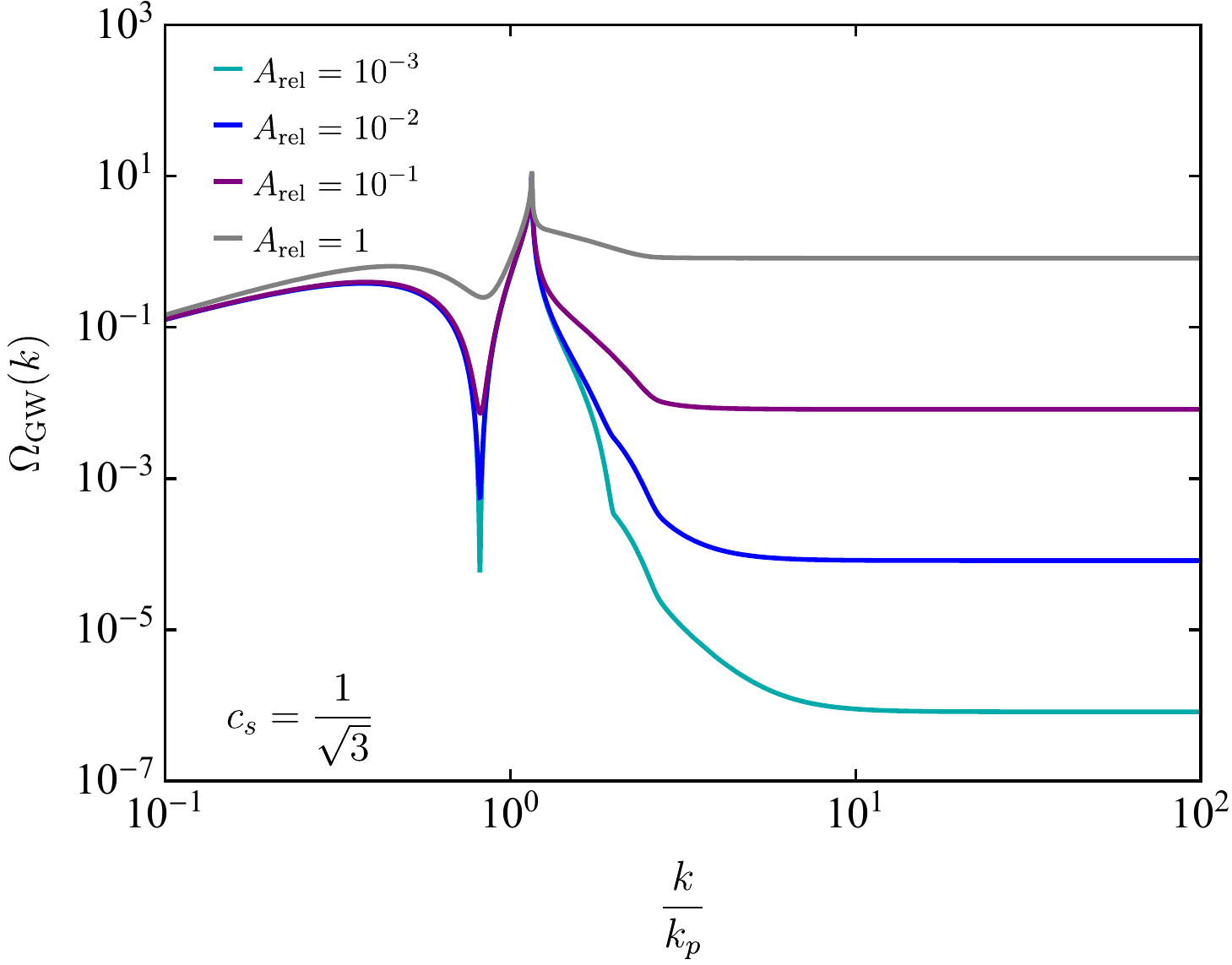}
\includegraphics[width=0.45\columnwidth]{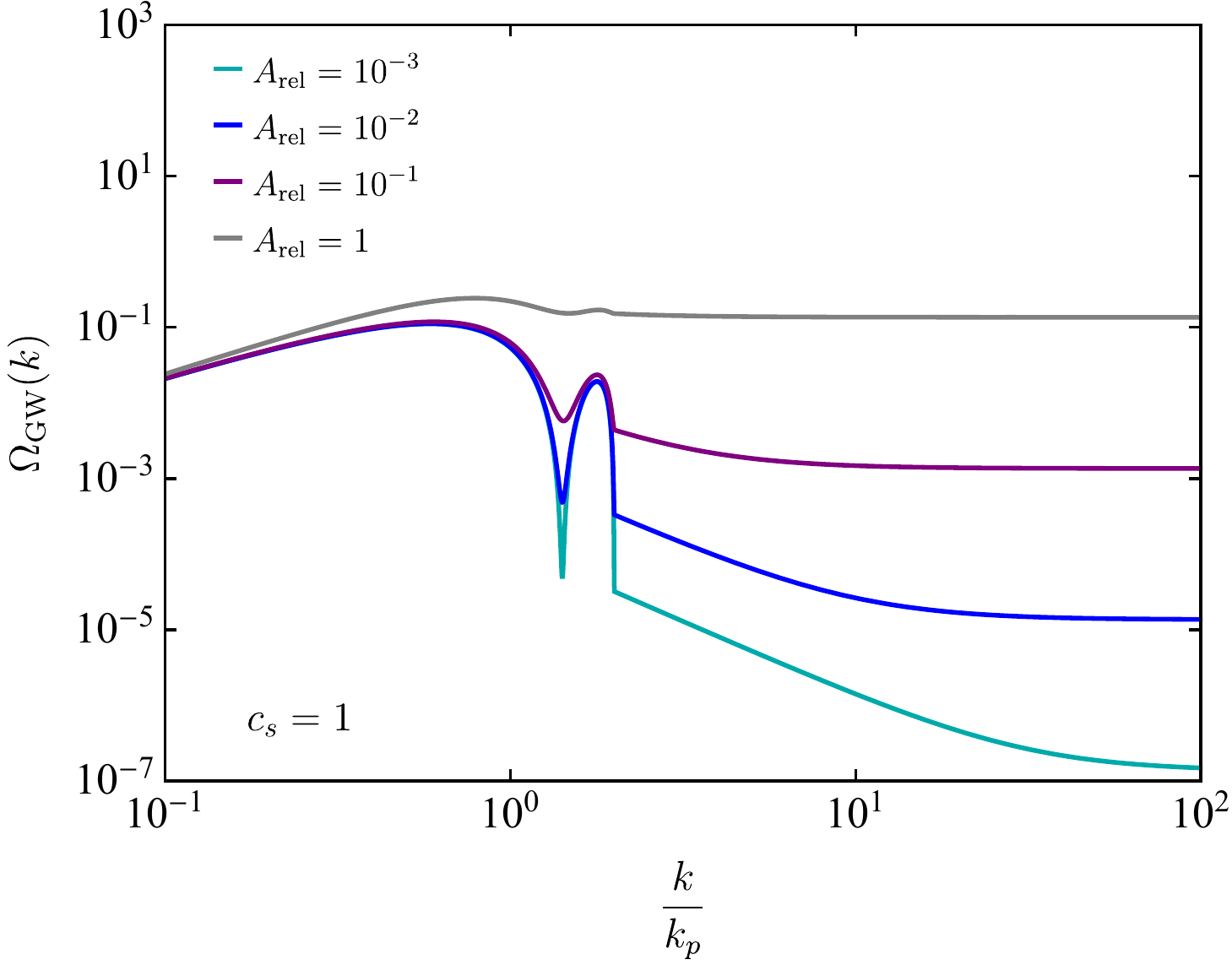}
\caption{The induced gravitational wave power spectrum $\Omega_{\textrm{GW}}$ is shown as a function of the dimensionless ratio $\frac{k}{k_p}$. The $\Omega_{\textrm{GW}}$ components are shown in the top panels and are separated by type: dirac delta peak, cross terms and smooth step. The bottom panels show the total summed $\Omega_{\textrm{GW}}$ as functions of the relative weight $A_{\rm rel}=[10^{-3},10^{-2},10^{-1},1]$ between the dirac delta peak and the smooth step amplitudes respectively. The left and right panels are for speeds of sound $c_s=\frac{1}{\sqrt{3}}$  and $c_s=1$ respectively.   \label{fig:diracdelta}}
\end{figure}

In Figure~\ref{fig:diracdelta} we numerically computed the different contributions to the total GW spectrum \eqref{eq:split} from a primordial spectrum given by the zero width limit, $\Delta\to0$, of the log-normal peak \eqref{eq:lognormal}. This case coincides with the Dirac delta case \eqref{eq:Pdelta} with the exception that ${\cal P}_{\cal R,\rm flat}=\Theta(k-k_p)$ instead of unity. This slight change of ${\cal P}_{\cal R,\rm flat}$ does not affect the UV regime at all nor the conclusions on the IR regime derived above. In Figure~\ref{fig:diracdelta}, we highlight how the cross term is substantially visible for $A_{\rm rel}<10^{-1}$ and it is particularly noticeable for the case when $c_s^2=1$. The range of $k$ where the 
cross term dominates the total GW spectrum increases with decreasing $A_{\rm rel}$. We also confirm that the second break point of the total GW spectrum is well described by \eqref{eq:break2}.\footnote{We also note that slightly before the second break point $k_{b2}$ the total GW spectrum starts to noticeably depart from a power-law. Finding such departure provides information about $k_{b2}$ and, therefore, about $A_{\rm rel}$, even in the case where the plateau is not detectable. This could be thought of as a consistency check of the size of the plateau.}

\subsection{Smooth model: Log-normal plus step primordial spectra \label{sec:lognormal}}

In more realistic situations, the peak in the primordial spectrum has a finite width and it is often well approximated by a log-normal peak \cite{Pi:2017gih,Chen:2019zza,Ashoorioon:2019xqc,Chen:2020uhe,Garcia-Bellido:1996mdl,Yokoyama:1998pt,Clesse:2015wea,Cheng:2016qzb,Espinosa:2017sgp,Kannike:2017bxn,Garcia-Bellido:2017mdw,Ando:2017veq,Cheng:2018yyr,Ando:2018nge,Espinosa:2018eve,Inomata:2018cht,Braglia:2020eai,Palma:2020ejf,Fumagalli:2020adf}. Thus, we consider the following smooth template
\begin{align}\label{eq:lognormal}
{\cal P}_{\cal R,\rm peak}=\frac{1}{\sqrt{2\pi}\Delta}e^{-\frac{1}{2\Delta^2}\ln^2\left(\frac{k}{k_p}\right)}\quad{\rm and}\quad {\cal P}_{\cal R,\rm flat}=\frac{1}{2}\left(1 + \tanh\left[\frac{2}{\Delta}\ln\left(\frac{k}{k_p}\right)\right]\right),
\end{align}
where $\Delta$ is the dimensionless width and we set ${\cal P}_{\cal R,\rm flat}$ to be a smooth step to represent the transition from the first to the second slow-roll plateau without significantly affecting the scales of the peak. In order to avoid several additional parameters, we choose the step in ${\cal P}_{\cal R,\rm flat}$ to be as gradual as the log-normal function and to share the same transition scale. We will be interested in the case where the peak is sharp and $\Delta<1$. It is worth noting that the particular shape of ${\cal P}_{\cal R,\rm flat}$ is not relevant for our results as long as the step happens faster or with a similar width to ${\cal P}_{\cal R,\rm peak}$. For instance, since we consider the case where ${\cal A}_{\rm peak}>{\cal A}_{\rm flat}$ in Eqs.~\eqref{eq:PR} and \eqref{eq:split}, we find that whenever $\Omega_{\rm GW,cross}$ or $\Omega_{\rm GW,flat}$ dominate the GW spectrum, the dominant contribution to the GW spectrum comes from the ${\cal P}_{\cal R,\rm flat}\approx 1$ region. This also implies that we can approximately use the results of \S\ref{sec:dirac} for the visibility of the cross term.

Before we compute the GW spectrum numerically by integrating \eqref{eq:inducedGWs} and \eqref{eq:inducedGWscross}, we investigate the analytical approximations for the IR \eqref{eq:inducedGWspeakIR}, and UV \eqref{eq:inducedGWspeakUVcs1/3}-\eqref{eq:inducedGWspeakUVcs1} to understand the asymptotic behaviour of the spectrum.\footnote{More sophisticated approximations for $\Omega_{\rm GW, peak}$ for $c_s^2=1/3$ are derived by Pi and Sasaki in Ref.~\cite{Pi:2020otn} which also includes the peak in the GW spectrum. We believe that one could perhaps derive similar formulas for $c_s^2=1$. However, the approximations of \cite{Pi:2020otn} for the GW peak do not work well for $\Delta\sim 0.1-0.4$ which is the case we consider here. In the IR and UV regimes, our approximations match those of \cite{Pi:2020otn} for $c_s^2=1/3$.} On one hand, it is easy to convince oneself from \eqref{eq:inducedGWspeakIR} that both the peak and cross contributions in the IR regime fall off as $k^3\ln^2k$. One can also show that $\Omega_{\rm GW, flat}$ also decays as $k^3\ln^2k$ in the IR. The reason is that the integrand in Eq.~\eqref{eq:inducedGWs} with ${\cal P}_{\cal R,\rm flat}$ given by the step in Eq.~\eqref{eq:lognormal} effectively behaves as a sharp peak for $k\ll k_p$, because the kernel \eqref{eq:kernelIR} decays as ${\cal T}(u,v,v_*\gg1)\sim v^{-4}$ and the region after the step becomes increasingly small for large $v$ (small $k$). Thus, the GW spectrum induced by the primordial spectrum \eqref{eq:lognormal} decays in the IR as $k^3\ln^2k$ and, since ${\cal A}_{\rm peak}\gg {\cal A}_{\rm flat}$ is dominated by the peak contribution $\Omega_{\rm GW, peak}$.

\begin{figure}
\includegraphics[width=0.45\columnwidth]{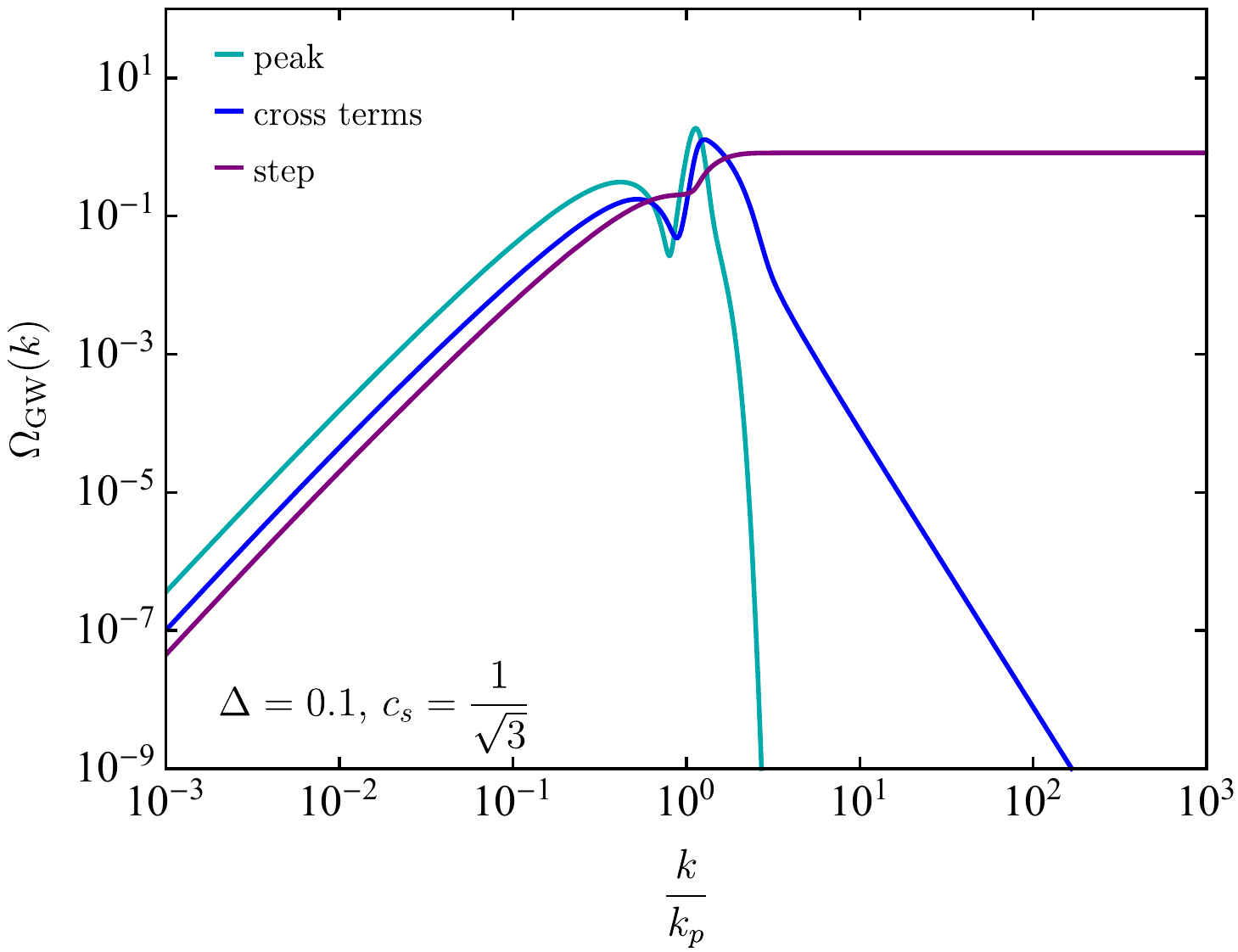}
\includegraphics[width=0.45\columnwidth]{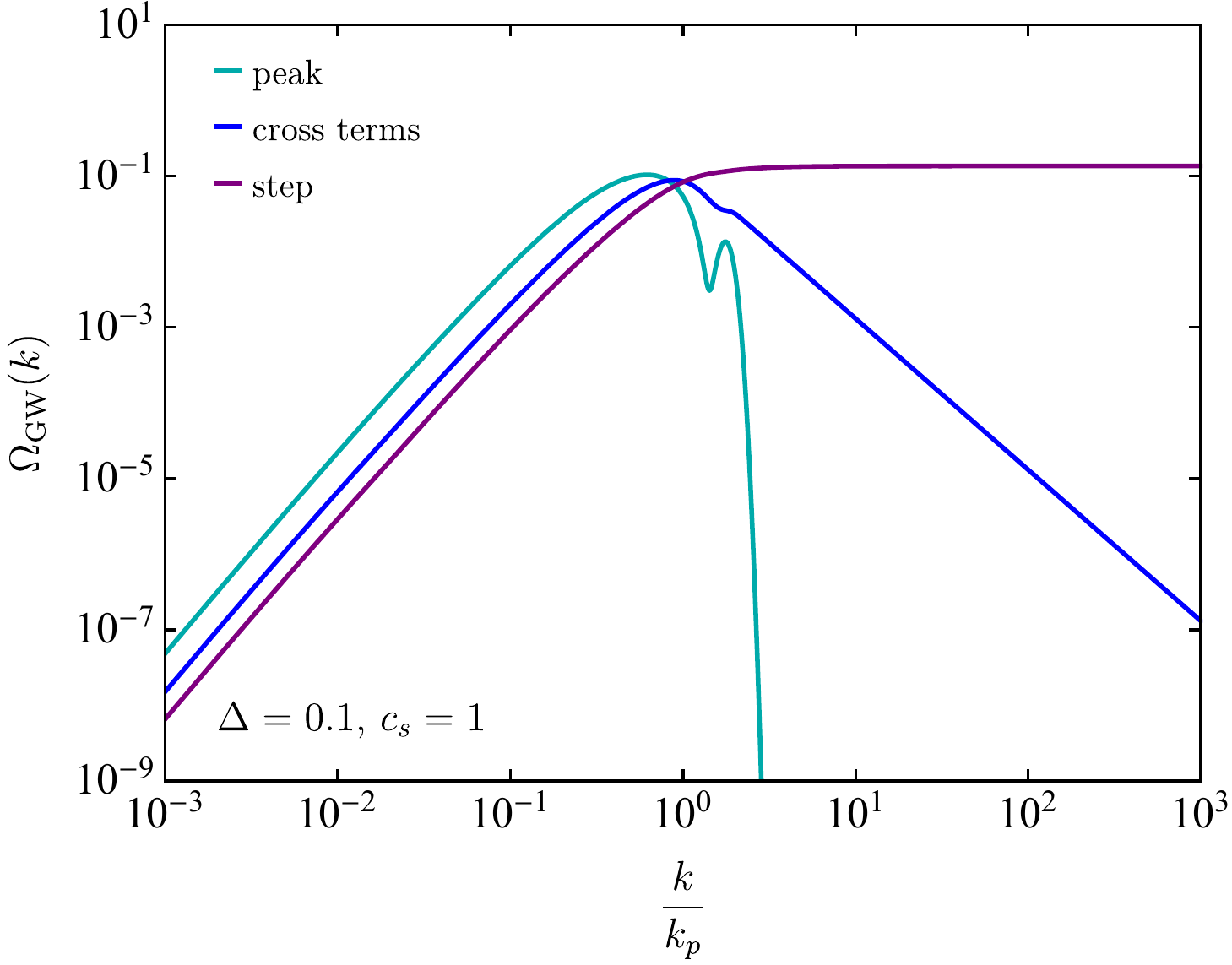} 
\includegraphics[width=0.45\columnwidth]{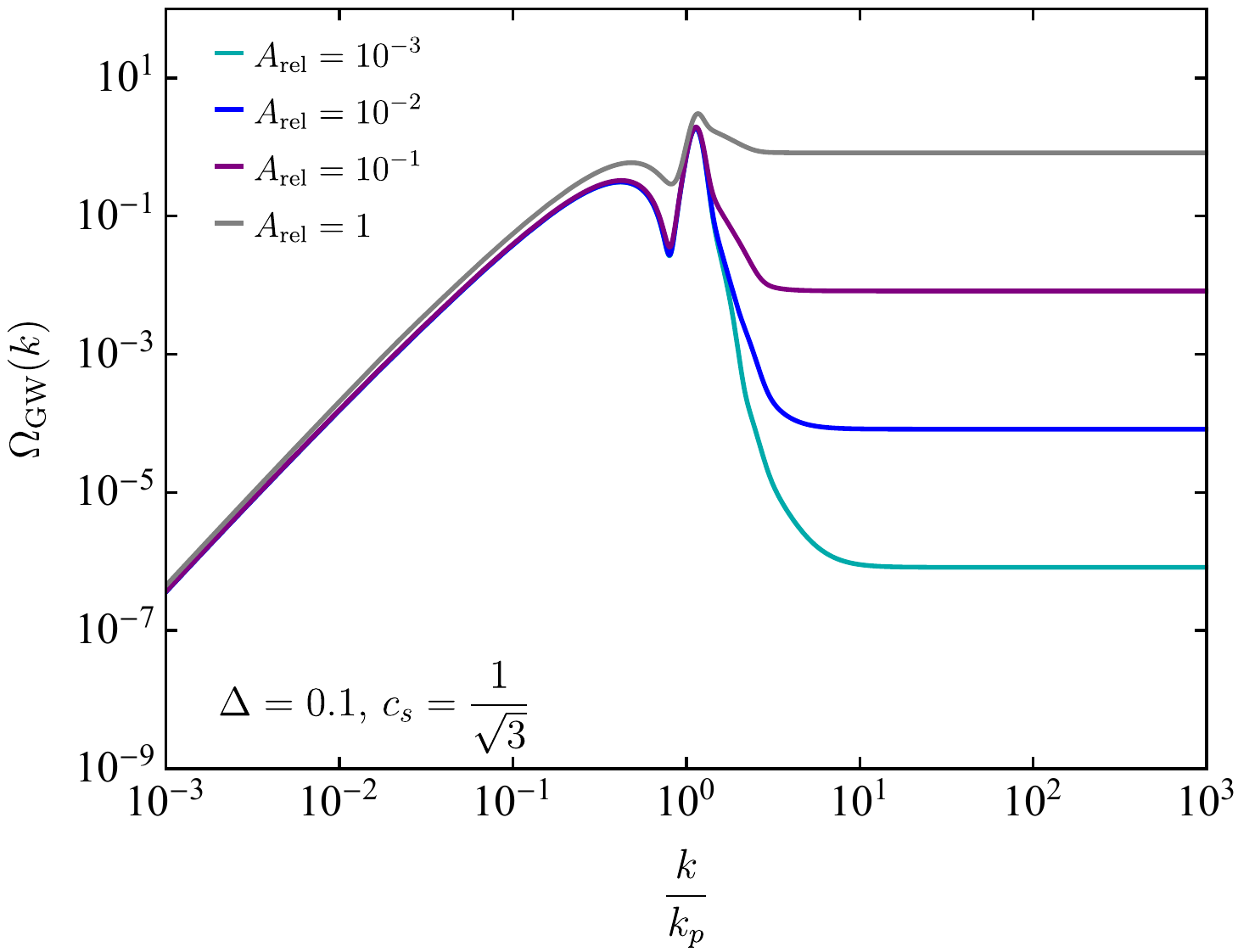}
\includegraphics[width=0.45\columnwidth]{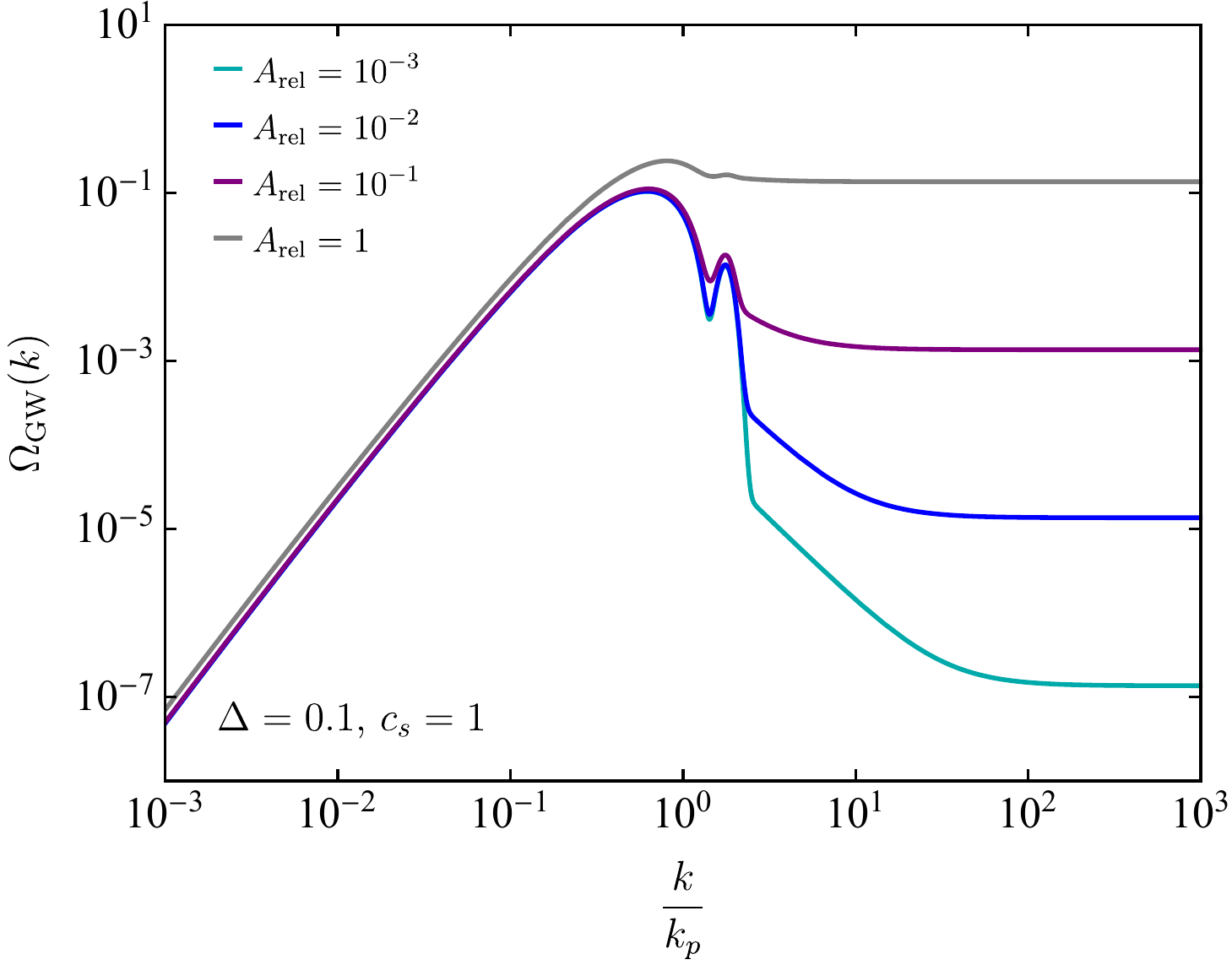}
\caption{The induced gravitational wave power spectrum $\Omega_{\textrm{GW}}$ is shown as a function of the dimensionless ratio $\frac{k}{k_p}$. The $\Omega_{\textrm{GW}}$ components are shown in the top panels and are separated by type: log-normal peak, cross term and smooth step. The bottom panels show the total summed $\Omega_{\textrm{GW}}$ as functions of the relative weight $A_{\rm rel}=[10^{-3},10^{-2},10^{-1},1]$ between the log-normal peak and the smooth step amplitudes respectively. The left and right panels are for speeds of sound $c_s=\frac{1}{\sqrt{3}}$  and $c_s=1$ respectively. In all cases, we fix the log-normal peak width to $\Delta=0.1$.   \label{fig:lognormal}}
\end{figure}

On the other hand, the UV regime is very similar to the one described by the Dirac delta case. First, the plateau is well approximated by $\Omega_{\rm GW, flat}$ in Eq.~\eqref{eq:flatOmega}. Second, the cross terms are also given by \eqref{eq:UVtaildelta} but for an additional factor ${\rm e}^{8\Delta^2}$ for the $c_s^2=1/3$ and ${\rm e}^{2\Delta^2}$ for $c_s^2=1$, which come from the definite integrals in Eqs.~\eqref{eq:inducedGWspeakUVcs1/3} and \eqref{eq:inducedGWspeakUVcs1} with a log-normal \eqref{eq:lognormal}. Thus, for $\Delta<1$ these additional numerical factors play no significant role. Third, in contrast to the Dirac delta case, the UV tail of $\Omega_{\rm GW, peak}$ does not terminate at $k=2k_p$. However, from \eqref{eq:inducedGWspeakUVcs1/3} we see that it decays exponentially as ${\cal P}_{\cal R,\rm peak}(k)$ in Eq.~\eqref{eq:lognormal}. Therefore, for $\Delta<1$, the effective cut-off scale is exponentially close to $k\sim2k_p$. We conclude that the approximations derived for the Dirac delta case, Eqs.~\eqref{eq:UVtaildelta} and \eqref{eq:break2}, work well in the UV regime of the log-normal case with $\Delta<1$. This implies that cross term dominates the total GW spectrum in the range of scales given by \eqref{eq:widthdelta}.

We numerically compute the different contributions to the total GW spectrum \eqref{eq:split} in Figure~\ref{fig:lognormal} for the log-normal spectrum \eqref{eq:lognormal} with $\Delta=0.1$. We choose $\Delta=0.1$ as an example of a log-normal peak which is sharp but sufficiently wide so as to differentiate it from a Dirac delta source. We will show that the results are qualitatively similar to the Dirac delta case shown in Figure~\ref{fig:diracdelta}, except that the peak of the GW spectrum is finite and smoother. In particular, we see that the 
cross term is already visible for $A_{\rm rel}<10^{-1}$ and that the visible width increases for decreasing $A_{\rm rel}$. In the case when $c_s^{2}=1$, the cross term decays much slower and, therefore, it is also more clearly visible.

\section{Probing the second inflationary phase with future detectors\label{sec:forecasts}}

The predictions for the second inflationary phase that we consider yields compelling GW phenomenology in the observational window of SKA, LISA, ET and DECIGO. The relevant GW bounds include those determined by studying the time of arrival from many pulsars in space. This includes data from the pulsar timing array (PTA) \cite{1979ApJ...234.1100D} which is comprised of three constituent projects, the European Pulsar Timing Array (EPTA) \cite{Desvignes:2016yex}, the Parkes Pulsar Timing Array
(PPTA) \cite{Hobbs:2013aka}, and the North American Observatory for Gravitational Waves (NANOGrav) \cite{McLaughlin:2013ira}, while the International Pulsar Timing Array (IPTA) \cite{IPTA2016} constraint comes from a combination of all three and covers the frequency band $10^{-9}$-$10^{-7}$ Hz. At higher frequency, we have a band that would be observable with the Einstein Telescope (ET) \cite{Maggiore:2019uih}, which would be sensitive to the range $10$--$10^3$ Hz. In between SKA and ET we expect LISA \cite{Barausse:2020rsu} and DECIGO \cite{Yagi:2011wg,Kawamura:2020pcg} to be applicable.
In order to compare theoretical predictions with experimental projections, we use the recently derived future sensitivity curves for these GW limits using the latest experimental design specification and peak integrated sensitivity curves computed in Ref.~\cite{Schmitz:2020syl}. We use the semi-analytical fit functions which enable quick and systematic comparison of theoretical predictions with experimental sensitivities. The data can be found in Ref.~\cite{schmitz_kai_2020_3689582}. The power signal-to-noise ratio (SNR) of the SGWB is generally given by
\begin{align}
\label{eq:snr}
    \rho = \sqrt{2t_\textrm{obs}\left[\int_{f_\textrm{min}}^{{f_\textrm{max}}}df\left(\frac{\Gamma^2_{IJ}(f) S^2_h(f)}{P_{nI}(f)P_{nJ}(f)}\right)\right]},
\end{align}
where $t_\textrm{obs}$ is the total observation time and $P_{nI}(f)$,$P_{nJ}(f)$ are the auto power spectral densities for the noise in detectors $I,J$. The frequency limits for integration $[f_\textrm{min}$,$f_\textrm{max}]$ define the bandwidth of the detector. This represents the total broadband SNR, integrated over both time and frequency. It can be calculated as the expected SNR of a filtered cross-correlation.  Here, in general one assumes that the SGWB is sufficiently described by a power-law of the from $\Omega_\textrm{GW}=\Omega_\beta \left(f/f_\textrm{ref}\right)^\beta$ where $\beta$ is the spectral index and $f_\textrm{ref}$ is the reference frequency which we set to $1\textrm{yr}^{-1}$ for PTA and $100$Hz for ground based detectors over the sensitivity region of interest. We set the observation times $T=1$yr in general and $20$yr for PTA observations. We can then use \eqref{eq:snr} to compute the value of GW amplitude required to reach a target SNR. In order to determine the detectability of the SGWB signal, we solve for the primordial amplitudes required to ensure an SNR of unity i.e. $\rho=1$.

\begin{figure}[!ht]
\includegraphics[width=0.45\columnwidth]{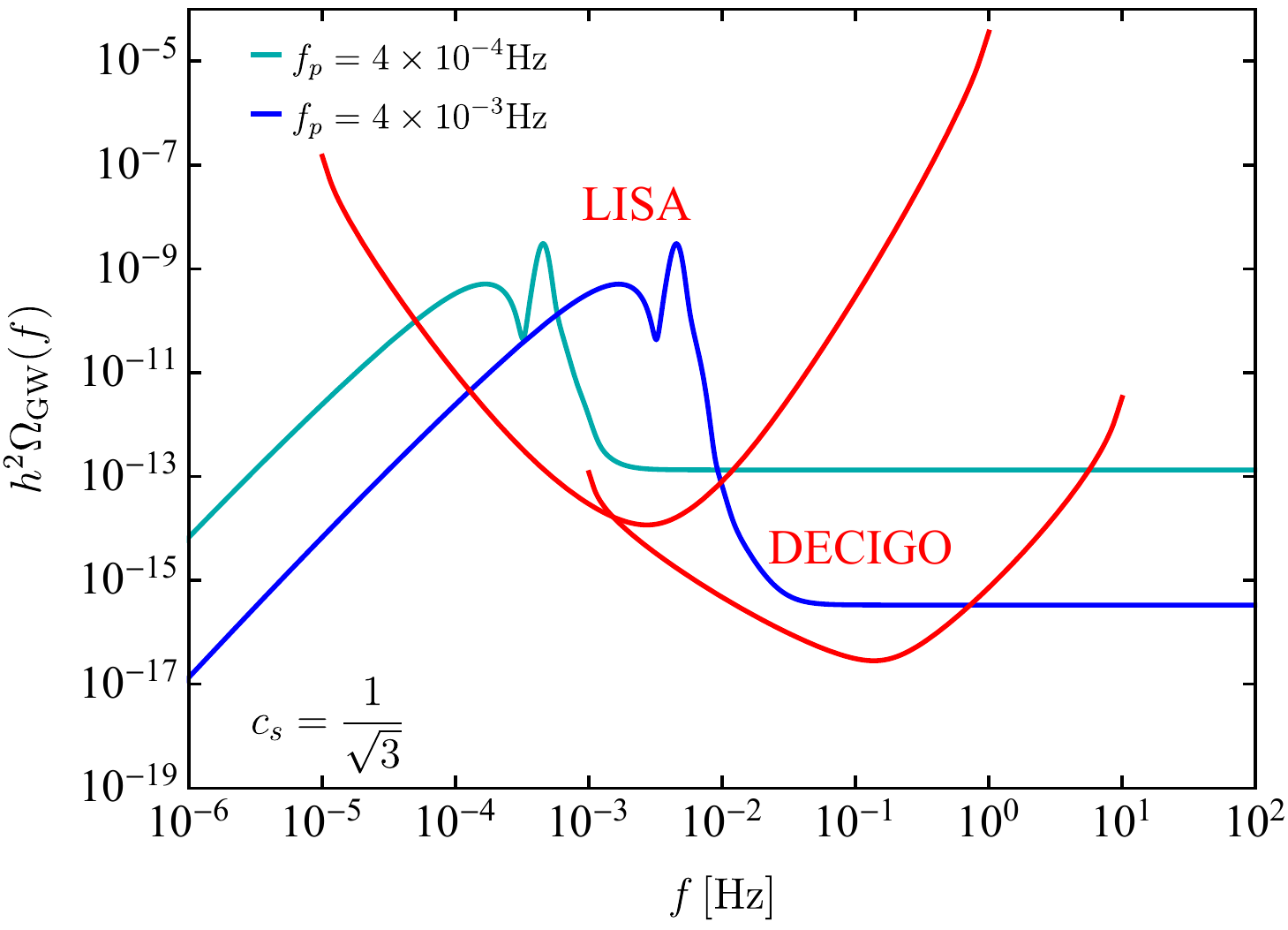}
\includegraphics[width=0.45\columnwidth]{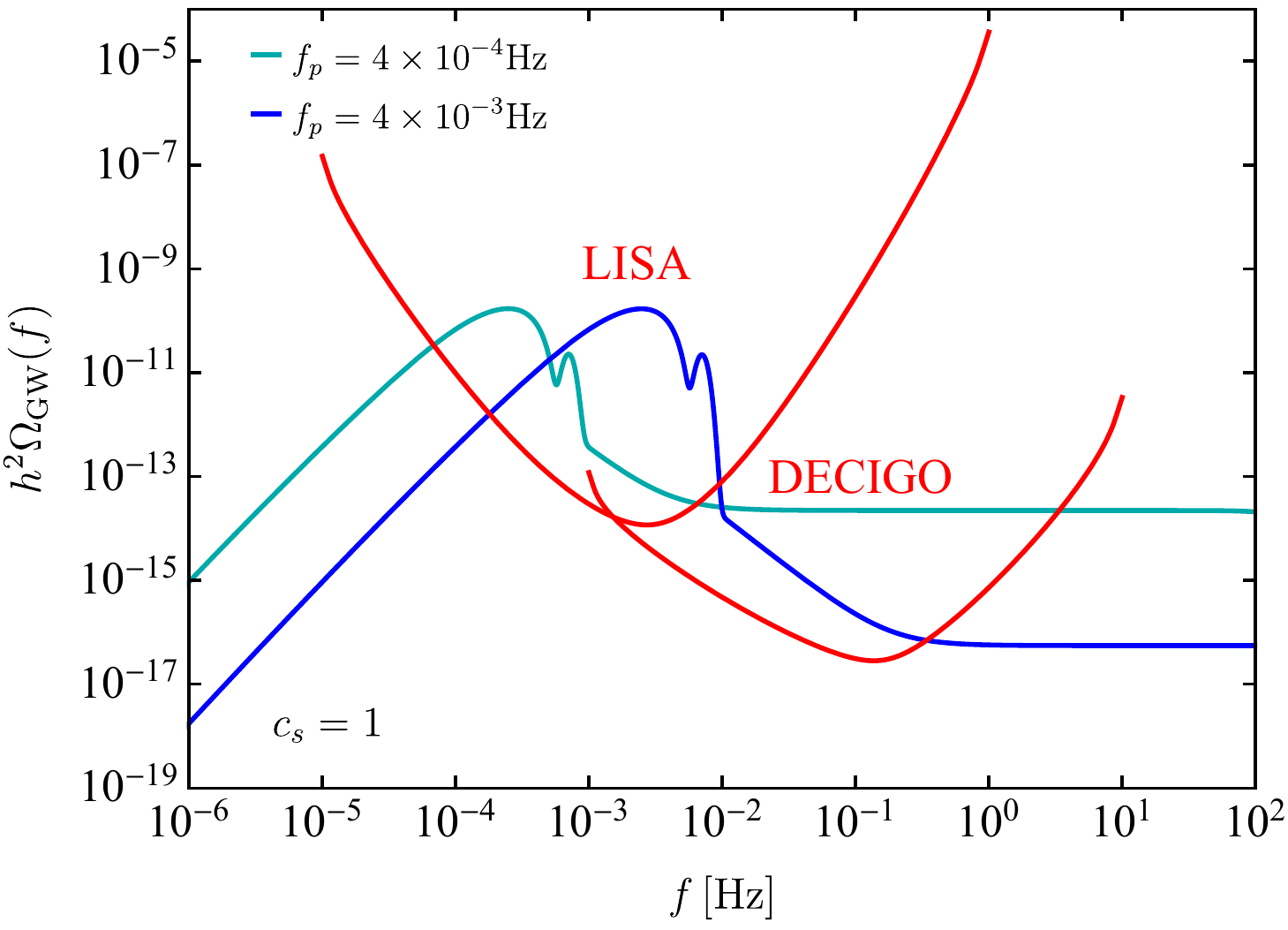}
\caption{The induced gravitational wave spectrum in the present universe $h^2\Omega_{\textrm{GW}}(f)$  is shown as a function of the frequency $f$ in Hz for speed of sound $c_s=\frac{1}{\sqrt{3}}$ (left panel) and $c_s=1$ (right panel) respectively. We consider log-normal primordial sources of width $\Delta=0.1$ and peak frequencies of $f_p=4\times10^{-4}$Hz and $f_p=4\times10^{-3}$Hz respectively. We fix ${\cal A}_\textrm{peak}=10^{-2}$ and ${\cal A}_\textrm{flat}=5\times10^{-6}$ for the blue curve and ${\cal A}_\textrm{flat}=5\times10^{-4}$ for the cyan curve for both panels. The LISA and DECIGO power-law integrated sensitivity curves \cite{Thrane:2013oya} are shown in red.\label{fig:examplecs}}
\end{figure}

We first provide an example of a theoretical prediction for the total GW spectrum produced for peak wave numbers corresponding to a peak frequency\footnote{The relation between frequency and wavenumber in reduced Planck units is $k_p=2\pi f_p$.} of $4\times 10^{-4}$Hz and $4\times 10^{-3}$Hz with the LISA and DECIGO sensitivity bands in Figure~\ref{fig:examplecs} for speed of sound $c_s=\frac{1}{\sqrt{3}}$ and $c_s=1$ in the left and right panels respectively. In this example, we consider a log-normal template with width $\Delta=0.1$ and fix ${\cal A}_\textrm{peak}=10^{-2}$ while considering ${\cal A}_\textrm{flat}=5\times10^{-6}$ for the blue curve and ${\cal A}_\textrm{flat}=5\times10^{-4}$ for the cyan curve for both panels. We see a more pronounced cross term in the case of $c_s=1$. We note in this example, that for $c_s=1$ and  $f_\textrm{peak}=4\times 10^{-4}$Hz, the peak is not observable by DECIGO but the cross term is. We note that this particular example with a relatively large ${\cal A}_{\rm peak}$ is just for illustration purposes. So although such a scenario might lead to a substantial production of PBHs, we do not consider them here. This is a simple example, however, it is important to formally consider the regions in which this happens for general $f_p$. 

In this vein, we now aim to reliably calculate the domains where either the peak from the log-normal GW source, the cross term or the step background can be observed by various experiments as a function of the log-normal peak wave number $k_p$ and the amplitude of the relevant term. To simplify the analysis we proceed in the following way. While the total GW spectrum is composed of the sum of $\Omega_{\rm GW, peak}$, $\Omega_{\rm GW, cross}$ and $\Omega_{\rm GW, flat}$ with their respective amplitudes given in Eq.~\eqref{eq:split}, here we shall separately calculate the SNR \eqref{eq:snr} for each contribution. By doing this, we reduce the parameter space to plot from three parameters ($k_p$, ${\cal A}_\textrm{peak}$, ${\cal A}_\textrm{flat}$) to two parameters ($k_p$ and amplitude). The resulting plot from using each spectrum separately is concise and informative, as we proceed to show.\footnote{An alternative, and perhaps more accurate procedure, would be to: $(i)$ fix a value of $A_{\rm rel}$, $(ii)$ compute the total GW spectrum which now has an overall amplitude of ${\cal A}_{\rm peak}$, $(iii)$ cut the total GW spectrum into three pieces at the frequency where each contribution dominates and $(iv)$ compute the SNR for each piece. This, however, does not reduce the parameter space and so it has to be done for the $A_{\rm rel}$ of interest. This is the reason why we decided to follow the simplest approach.} We follow the same convention here as preceding sections where the amplitude of each GW contribution is given by ${\cal A}_\textrm{peak},\sqrt{2{\cal A}_\textrm{peak}{\cal A}_\textrm{flat}}$ and ${\cal A}_\textrm{flat}$ for the peak, cross term and flat profiles respectively.

\begin{figure}[!h]
\includegraphics[width=0.45\columnwidth]{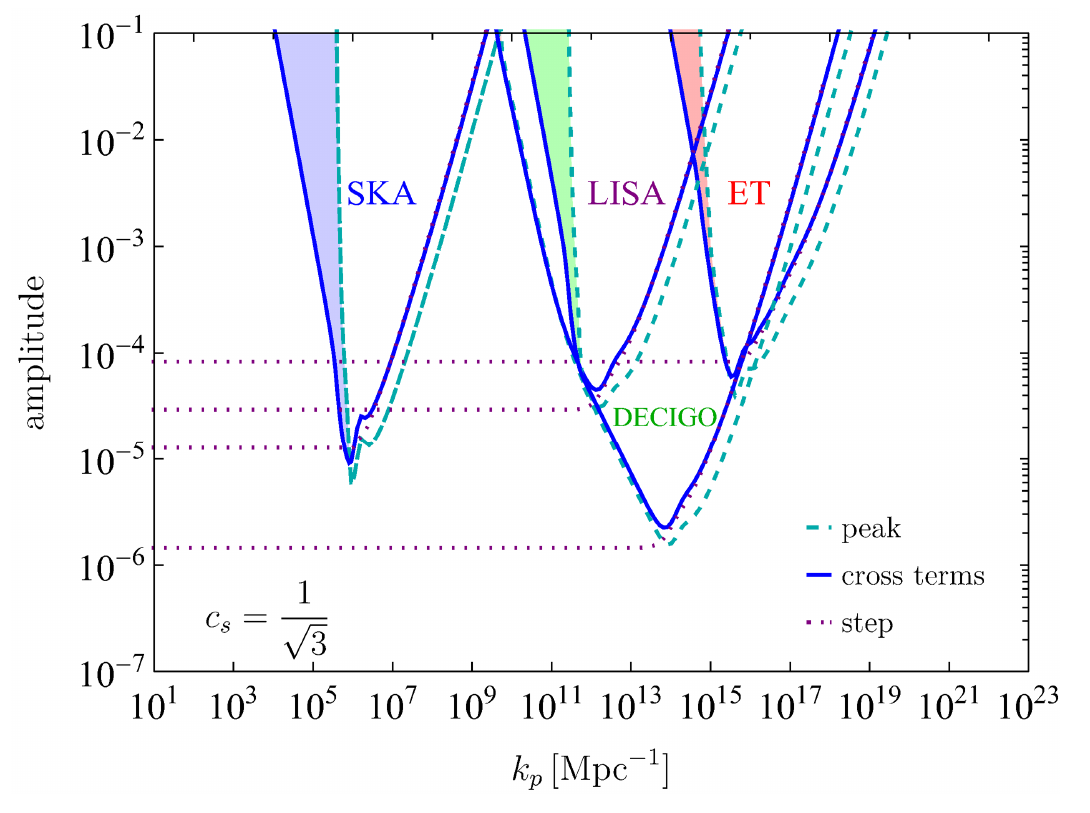}
\includegraphics[width=0.45\columnwidth]{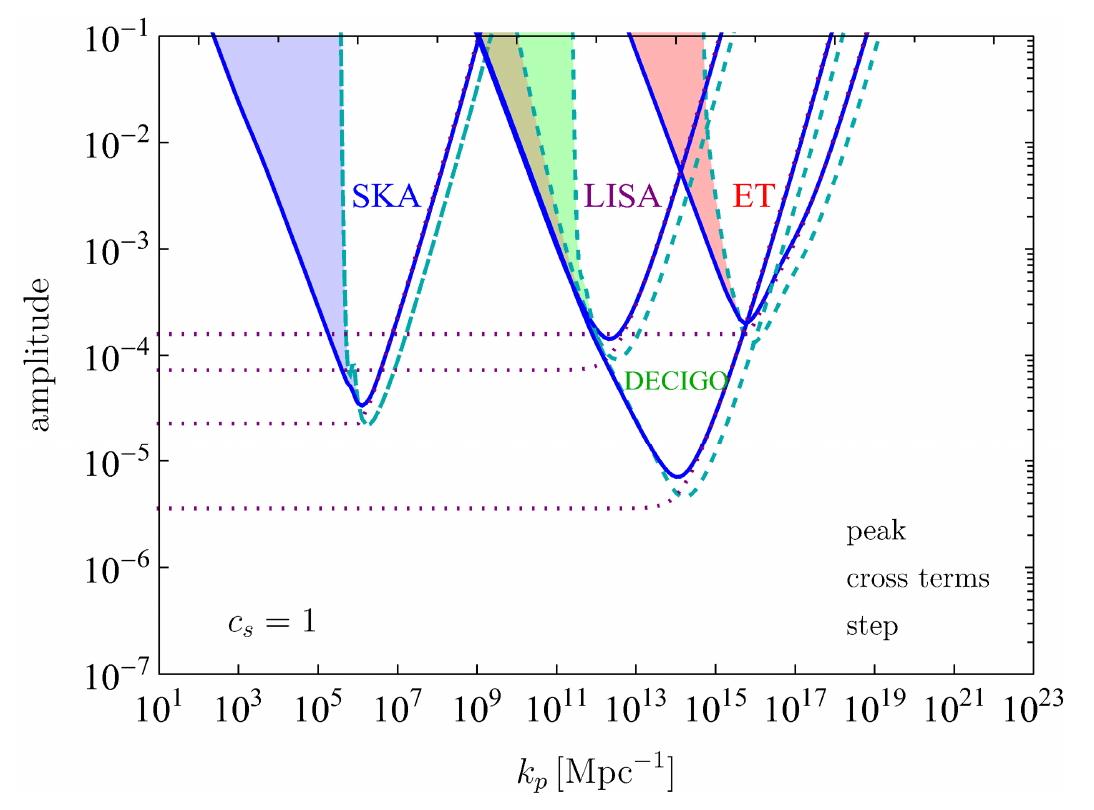}
\caption{Lower bounds on the generic amplitude as a function of peak wave number $k_p$. Limit on ${\cal A}$ correspond to limits on amplitudes ${\cal A}_\textrm{peak}$, $\sqrt{ 2{\cal A}_\textrm{peak}{\cal A}_\textrm{flat}}$ and ${\cal A}_\textrm{flat}$ for the log-normal peak (dashed), cross terms (solid) and smooth step (dotted) gravitational wave components respectively. The limits are set for $c_s=1$ (left panel) and $c_s=\frac{1}{\sqrt{3}}$ (right panel) using the sensitivity curves for SKA, LISA, DECIGO and ET and the peak integrated sensitivity curves of \cite{schmitz_kai_2020_3689582}. \label{fig:lognormallimits}}
\end{figure}

We show these results in Figure~\ref{fig:lognormallimits} for speed of sound $c_s=\frac{1}{\sqrt{3}}$ and $c_s=1$ on the left and right panels respectively. We note that we show amplitudes up to $0.1$, which are just for information purposes and excluded by over-abundance of PBHs \cite{Sasaki:2018dmp,Gow:2020bzo}. We also remind the reader that the choices of $c_s=1/\sqrt{3}$ and $c_s=1$ are two representative cases of an adiabatic perfect fluid and a canonical scalar field. Nevertheless, the same qualitative conclusions apply for general values of $c_s$. In particular, for $c_s^2\lesssim0.6$ we have the characteristic behaviour of $k^{-4}$ and for $c_s^2>0.6$ we see the appearance of the $k^{-2}$ slope distinctive of the $c_s^2=1$ case (see Appendix~\ref{app:generalcs}). The regions enclosed within the curves correspond to where particular terms in the GW spectrum are observable. Specifically, the blue, cyan and purple dashed boundaries represent where the cross, log-normal peak and step background terms are in principle observable, respectively. The shaded coloured region between the cyan and blue boundaries represents where the cross term enters the observational window but the peak does not for the specific experiment in question.\footnote{Note that if only the cross term is observed, then the induced GW signal is degenerate with the UV slope of GWs induced by a broken power-law with $n_{\rm UV}=1,2$ for $c_s^2=1,1/\sqrt{3}$, \textit{e.g.} see Eq.~\eqref{eq:OGWUVpeak}. In this case, the observation of the peak provides crucial information to break the degeneracy. We thank Shi Pi for pointing this out.} Also, although not directly obvious from Figure~\ref{fig:lognormallimits}, we find a substantial parameter space where the peak and the cross term are visible but not the plateau, which is clear from Figure~\ref{fig:examplecs}. This implies that while we might not in fact observe the plateau, we may extract ${\cal A}_{\rm peak}$ from the peak signal and then ${\cal A}_{\rm flat}$ from the cross term, effectively probing the second slow-roll phase of inflation.

To avoid any confusion we clarify how Figure~\ref{fig:lognormallimits} should be used with an illustrative example. First, choose a value for $k_p$. Second, pick up the value of interest for ${\cal A}_{\rm peak}$. Look at the dashed cyan line to know whether the peak falls within the observable range or not. Third, pick an amplitude for ${\cal A}_{\rm flat}$ and compute $\sqrt{2{\cal A}_{\rm peak}{\cal A}_{\rm flat}}$ and $k_{b2}$ \eqref{eq:break2}. Lastly, look at the solid blue line and purple dotted line and check whether the cross term or the plateau are observable. For the observability of the plateau is more accurate to use $k_{b2}$ instead of $k_p$. For example, for the cyan lines in Figure~\ref{fig:examplecs} we have $k_p=4\times 10^{10}\,{\rm Mpc}^{-1}$, ${\cal A}_{\rm peak}=10^{-2}$ and ${\cal A}_{\rm flat}=5\times10^{-4}$. This gives $\sqrt{2{\cal A}_{\rm peak}{\cal A}_{\rm flat}}\approx 3\times 10^{-3}$. So we see that this point falls above all lines for LISA in the left of Figure~\ref{fig:lognormallimits} and above the cyan and blue but below the purple line in the right of Figure~\ref{fig:lognormallimits}. This means that all components are visible by LISA for $c_s^2=1/3$ but for $c_s^2=1$ only the peak and the cross term are visible.

We note that in Figure~\ref{fig:lognormallimits}, the shaded region is particularly exaggerated for SKA due to the narrow shape of the SKA sensitivity window, characteristic for PTA experiments which operate on very large distance scales. For the discussion below, we shall consider $k_p$ observability windows for amplitudes of $\simeq 10^{-2}$, while bearing in mind previous remarks about exclusion due to PBH bounds. For $c_s=\frac{1}{\sqrt{3}}$, we find that amplitudes below $10^{-5}$ are too faint to be resolved by SKA and that the cross terms fall into the sensitivity band at amplitude for $k_p \gtrsim 4\times10^4 \textrm{Mpc}^{-1}$ while the peak only falls into SKA at $k_p\gtrsim 4\times 10^{5}\textrm{Mpc}^{-1}$. For LISA, the cross term and the peak becomes visible at an almost degenerate position of $k_p\simeq 1.6 \times 10^{10}\textrm{Mpc}^{-1}$, which explains why there is no shaded region visible. For ET, the cross term becomes visible at $k_p\gtrsim 3\times 10^{14}\textrm{Mpc}^{-1}$, while the peak at $k_p  \simeq 8\times 10^{14}\textrm{Mpc}^{-1}$. For DECIGO, there is a very tiny region where the cross term is visible and the peak is not since the cross term becomes visible at $k_p \simeq 7\times 10^{10} $ and the peak at $k_p\gtrsim 3\times 10^{11}$.   

For $c_s=1$, we see a much larger shaded region in Figure~\ref{fig:lognormallimits}, indicating more parameter space wherein the cross term falls into the sensitivity band at lower $k_p$ than the peak. We may once again consider $k_p$ bounds for a benchmark amplitude of $\simeq 10^{-2}$. For SKA, we find that the cross terms fall into the sensitivity band at $k_p \gtrsim 3\times10^3 \textrm{Mpc}^{-1}$ while the peak only falls into SKA at $k_p\gtrsim 4\times 10^{5}\textrm{Mpc}^{-1}$. For LISA, the cross term becomes visible at $k_p\gtrsim 10^{10}\textrm{Mpc}^{-1}$ while the peak at $k_p\simeq 4 \times 10^{10}\textrm{Mpc}^{-1}$, which is no longer degenerate and explains the small visible shaded region. For ET, the cross term becomes visible at $k_p\gtrsim 7\times 10^{13}\textrm{Mpc}^{-1}$, while the peak at $k_p  \gtrsim 7\times 10^{14}\textrm{Mpc}^{-1}$. For DECIGO, there is a also a region where the cross term is visible and the peak is not since the cross term becomes visible at $k_p \simeq 10^{10} $ and the peak at $k_p\gtrsim 3\times 10^{11}$.

In general, we see that there is a rich parameter region available to explore where the cross term contributions dominate the peak, and a connection between relative amplitudes between peaked and flat phases, speed of sound in the primordial plasma, cross term UV tail behaviour and observability with future experiments.
% \begin{align}
% \textrm{SNR}=\sqrt{T\int_{f_\textrm{min}}^{{f_\textrm{max}}}df\left(\frac{\Omega_\textrm{GW}(f)}{\Omega_s(f)}\right)^2}
% \end{align}
% where $\Omega_s(f)$ is the energy density computed from the power spectral density. Here, in general one assumes that the SGWB is sufficiently described ny a power-law of the from $\Omega_\textrm{GW}=A_* f^\alpha$ over the sensitivity region of interest and compute the value of amplitude $A$ required to give a targed SNR.
We also explain why in Figure~\ref{fig:lognormallimits}, there is no shaded region for LISA nor for the lower amplitudes of ET and DECIGO. This is related to how the sensitivity of the GW detectors behave at low frequencies compared to the slope of the cross terms. For instance, from \cite{Robson:2018ifk,Moore:2014lga,schmitz_kai_2020_3689582} we see that the acceleration component of the LISA effective noise power spectral density at low frequencies goes as $S_n(f)\propto f^{-6}$ and, therefore, the sensitivity curve in terms of spectral density goes as $\Omega_{\rm GW, LISA}\propto f^3S_n(f)\propto f^{-3}$. This means that if $\Omega_{\rm GW, cross}\propto f^{-4}$, as is the case for $c_s^2=1/3$, the cross term cannot enter the LISA range if the peak is already unobservable. Similar conclusions hold for ET. In contrast, in the case of $c_s^2=1$ we have that $\Omega_{\rm GW, cross}\propto f^{-2}$ and the cross term eventually enters the observable range even if the peak is unobservable. For DECIGO \cite{Yagi:2011wg} we have that $S_n(f)\propto f^{-4}$ so that for low frequencies $\Omega_{\rm GW, DECIGO}\propto f^{-1}$. Thus, for DECIGO in the frequency ranges of $10^{-3}$-$10\,{\rm Hz}$, the cross term is not visible if the peak is not visible.

\section{Conclusions\label{sec:conclusions}}

We considered GWs induced by a primordial spectrum of fluctuations with a large peak followed by a plateau, and a relative amplitude between them given by $A_{\rm rel}<1$ \eqref{eq:arel}. The large peak is produced by a special feature during inflation which includes, but is not limited to, bumps in the inflaton's potential, ultra-slow-roll phases, and sudden turns in multi-field space. Since inflation might not necessarily end immediately after the feature, we considered an additional generic plateau in the primordial spectrum which comes from a second phase of slow-roll. The resulting induced GW spectrum \eqref{eq:split} has contributions from the peak, the plateau and the cross terms between them. We also studied in detail the effects of the speed of sound $c_s$ on the shape of the cross terms and the total spectrum as well as the observable windows where the cross terms dominate for experiments such as SKA, LISA, DECIGO and ET. We mainly focused on the cases where $c_s^2=w=1/3$ and $c_s^2=1$, two representative values corresponding to standard radiation and a canonical scalar field, and we found that the UV tail of the GW spectrum is very different between these two cases. Our studies exemplify that there is a rich parameter region wherein the cross terms dominate the peak contributions in the presence of a flat slow-roll style background. We thoroughly explored the relationship between the relative amplitudes for the inflationary phases, the speed of sound in the primordial plasma and identify important features in the cross term spectrum.

Our main finding is that, for sharp peaks (in the main text a sharp log-normal \eqref{eq:lognormal}), the cross term is the dominant contribution of the total GW spectrum for $k>2 k_p$ \eqref{eq:UVtaildelta} until it is overcome by the plateau at around $k/k_p\sim A_{\rm rel}^{-1/\alpha}$ where $\alpha=\{4,2\}$ respectively for $c_s^2=\{1/3,1\}$ \eqref{eq:break2}. Furthermore, we found that the cross term has a characteristic slope given by $\Omega_{\rm GW,cross}\propto k^{-\alpha}$ \eqref{eq:UVtaildelta}. In Appendix~\ref{app:generalcs}, we showed that the transition between these two different regimes occurs at around $c_s^2\sim 0.6$ where a $k^{-2}$ behaviour appears after the peak and then transitions to $k^{-4}$. The $c_s^2=1$ is the limiting case where the $k^{-4}$ piece is pushed to $k\to\infty$. Nevertheless, we find that for $c_s^2=0.8$ there is already a substantial $k^{-2}$ contribution to the cross term. We then concluded that in the case with $c_s^2=1$, not only the UV tail of the cross term decays much more slowly as compared to $c_s^2=1/3$ but it also dominates for a wider range of $k$. In addition, while the plateau in the induced GW is proportional to $A_{\rm rel}^2$, the cross term instead is only suppressed by $A_{\rm rel}$ \eqref{eq:split}. This leaves an interesting region in the parameter space where the induced GWs from the peak and the cross term are observable but not the plateau (see Figure~\ref{fig:examplecs} for examples). We argue that even in this case, one might infer the amplitude of the primordial spectrum during the second slow-roll phase from the cross term. We show the observable parameter range for PTA, LISA, DECIGO and ET in Figure~\ref{fig:lognormallimits}.

In this work, we have studied cases where $c_s^2\sim{\cal O}(0.1$-$1)$. However, one may also wonder what occurs for $c_s^2\ll1$. The first change is that the position of the resonant peak moves to low wavenumbers as $k_{\rm res}=2c_sk_p$. The second and most important change is that there is a $c_s^{-4}$ enhancement of the amplitude of the induced GWs. This is clear from the expression of the kernel \eqref{eq:kernel}. On the other hand, the asymptotic behaviours in the IR and UV tail are similar to the $c_s^2=1/3$ case. For instance, in the UV the cross term dominates for $k>2k_p$ and decays as $k^{-4}$. Nevertheless, while the $c_s^{-4}$ enhancement for $c_s^2\ll1$ is an interesting possibility to boost the induced GW signal for low values of ${\cal A}_{\rm peak}$ and ${\cal A}_{\rm flat}$, it remains to be seen whether the approximation used to derive the kernel \eqref{eq:kernel} in Ref.~\cite{Domenech:2019quo,Domenech:2021ztg} is still valid for values such as $c_s^2\sim 0.01$. The main issue for the case $c_s^2\ll1$ is that the kernel \eqref{eq:kernel} assumes an instantaneous transition from $c_s^2\ll1$ to the standard radiation domination with $c_s^2=w=1/3$. However, for $c_s^2\ll1$ the curvature perturbation is almost constant and the sudden transition to oscillations for $c_s^2=w=1/3$ could be very important. For instance, in a transition from 
dust-domination ($c_s^2=w=0$) to radiation-domination ($c_s^2=w=1/3$), there is a huge production of induced GWs right after the transition, as shown by Ref.~\cite{Inomata:2019ivs}. For this reason, we leave the study of $c_s^2\ll1$ for future work.

Another interesting question that we leave for future work is how well a single GW detector would be able to reconstruct the induced GW signal \cite{Caprini:2019pxz}. That is, whether one may be able to reconstruct the peak and the cross term at the same time from a detection of a stochastic GW background.

\section*{Acknowledgments} 
G.D. would like to thank V.~Atal, A.~Ricciardone, M.~Sasaki and S.~Pi for useful discussions and S.~Passaglia for permission to use his python code to calculate some of the spectra of induced GWs. S.B. is supported by 
funding from the European Union's Horizon 2020 
research and innovation programme under grant 
agreement No 101002846 (ERC CoG ``CosmoChart'') 
as well as support from the Initiative Physique 
des Infinis (IPI), a research training program of the Idex SUPER at Sorbonne
Universit\'{e}. G.D. as a Fellini fellow was supported by the European Union's Horizon 2020 research and innovation programme under the Marie Sk{\l}odowska-Curie grant agreement No 754496.
	
\appendix

\section{Broken power-law plus plateau primordial spectrum \label{app:brokenpowerlaw}}
In this appendix, we consider the case of \S~\ref{sec:dirac} Eq.~\eqref{eq:Pdelta} but where the primordial spectrum is given by a broken power-law instead of a Dirac delta. This is often the case in single field inflation models \cite{Byrnes:2018txb,Atal:2018neu,Cole:2022xqc}. We show that in the broken power-law case, the cross terms of \eqref{eq:split} are often not important. The broken power-law primordial spectrum is given by
\begin{align}\label{eq:PRBPL}
{\cal P}_{{\cal R},\rm peak}(k)=\left\{
\begin{aligned}
&\left(\frac{k}{k_p}\right)^{n_{\rm IR}}\quad &k\leq k_p\\
&\left(\frac{k}{k_p}\right)^{-n_{\rm UV}}\quad &k\geq k_p
\end{aligned}
\right.\,,
\end{align}
where $n_{\rm IR},n_{\rm UV}>0$ are free parameters. 
We then follow the same type of approximations explained in \cite{Atal:2021jyo}, which also work for the cross term \eqref{eq:inducedGWscross}. In this appendix we only focus on $c_s^2=1/3$ but similar results also hold for $c_s^2=1$. We proceed to study the IR and UV regimes separately. For simplicity we only focus on the $k$-dependence of the approximations. We checked that doing so does not change the main conclusions of this appendix.

\subsection{infrared approximation}

For the IR approximation we present the results for the peak and the cross term separately, as they behave differently depending on the value of $n_{\rm IR}$. First, the IR tail of the induced GWs \eqref{eq:inducedGWs} from the peak \eqref{eq:PRBPL} is given by
\begin{align}\label{eq:OGWIRBPL}
\Omega^{\rm IR}_{\rm GW,peak}(n_{\rm IR}>3/2)\propto\left(\frac{k}{k_p}\right)^{3}\ln^2\left(\frac{k}{k_p}\right)\quad{\rm and}\quad
\Omega^{\rm IR}_{\rm GW,peak}(n_{\rm IR}<3/2)\propto \left(\frac{k}{k_p}\right)^{2n_{\rm IR}}\,,
\end{align}
Second, the IR tail of the cross term \eqref{eq:inducedGWscross} is given by
\begin{align}\label{eq:crossIRBPL}
\Omega^{\rm IR}_{\rm GW,cross}(n_{\rm IR}>3)\propto\left(\frac{k}{k_p}\right)^{3}\ln^2\left(\frac{k}{k_p}\right)\quad{\rm and}\quad
\Omega^{\rm IR}_{\rm GW,cross}(n_{\rm IR}<3)\propto \left(\frac{k}{k_p}\right)^{n_{\rm IR}}\,.
\end{align}
It should be noted, however, that the IR approximations of the cross term have been derived assuming a scale invariant spectrum for ${\cal P}_{{\cal R},\rm flat}$ as in \S~\ref{sec:dirac}. If instead we considered a step function for ${\cal P}_{{\cal R},\rm flat}$ as in \S~\ref{sec:lognormal}, the IR tail of the cross term always goes as $\Omega^{\rm IR}_{\rm GW,cross}\propto k^3\ln^2k$ independent of the value of $n_{\rm IR}$, which we also confirmed numerically. Thus, we conclude that in general the cross term does not dominate the IR tail of the total GW spectrum \eqref{eq:split}.

\subsection{ultraviolet approximation}

In the UV regime, the integrand of the peak and cross contributions behave similarly for a given value of $n_{\rm UV}$. And, as shown in Ref.~\cite{Atal:2021jyo}, for $c_s^2=1/3$ the GW spectrum behaves differently for $n_{\rm UV}<4$ and $n_{\rm UV}>4$. Thus, we treat these two $n_{\rm UV}$ cases separately. Note that if we had considered $c_s^2=1$ the limiting value of $n_{\rm UV}$ would be $2$ instead of $4$. This means that the following analysis for the case $c_s^2=1/3$ is also applicable to $c_s^2=1$ by replacing the factors $4$ by factors $2$. Then, proceeding with the $c_s^2=1/3$ case we find, on one hand, for $n_{\rm UV}<4$ that 
\begin{align}\label{eq:OGWUVpeak}
\Omega^{\rm UV}_{\rm GW,peak}(n_{\rm UV}<4)\propto\left(\frac{k}{k_p}\right)^{-2n_{\rm UV}}\quad{\rm and}\quad \Omega^{\rm UV}_{\rm GW,cross}(k\gg k_p) \propto\left(\frac{k}{k_p}\right)^{-n_{\rm UV}}\,.
\end{align}
On the other hand, when $n_{\rm UV}>4$ we obtain
\begin{align}\label{eq:OGWUV2}
\Omega^{\rm UV}_{\rm GW,\rm peak}(n_{\rm UV}>4)\propto\left(\frac{k}{k_p}\right)^{-4-n_{\rm UV}}\quad{\rm and}\quad
\Omega^{\rm UV}_{\rm GW,cross}(n_{\rm UV}>4)\propto \left(\frac{k}{k_p}\right)^{-4}\,.
\end{align}

\begin{figure}
\includegraphics[width=0.45\columnwidth]{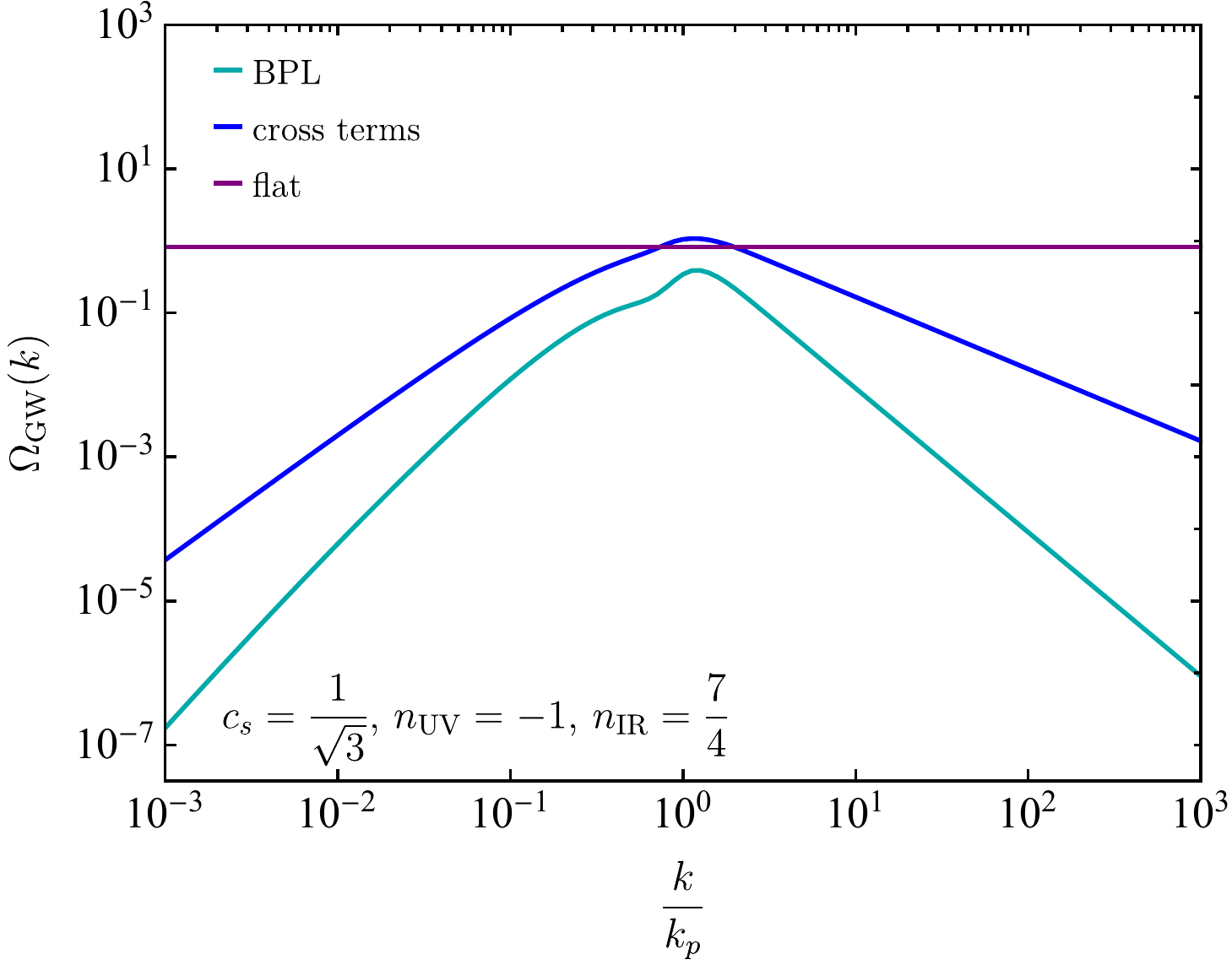}
\includegraphics[width=0.45\columnwidth]{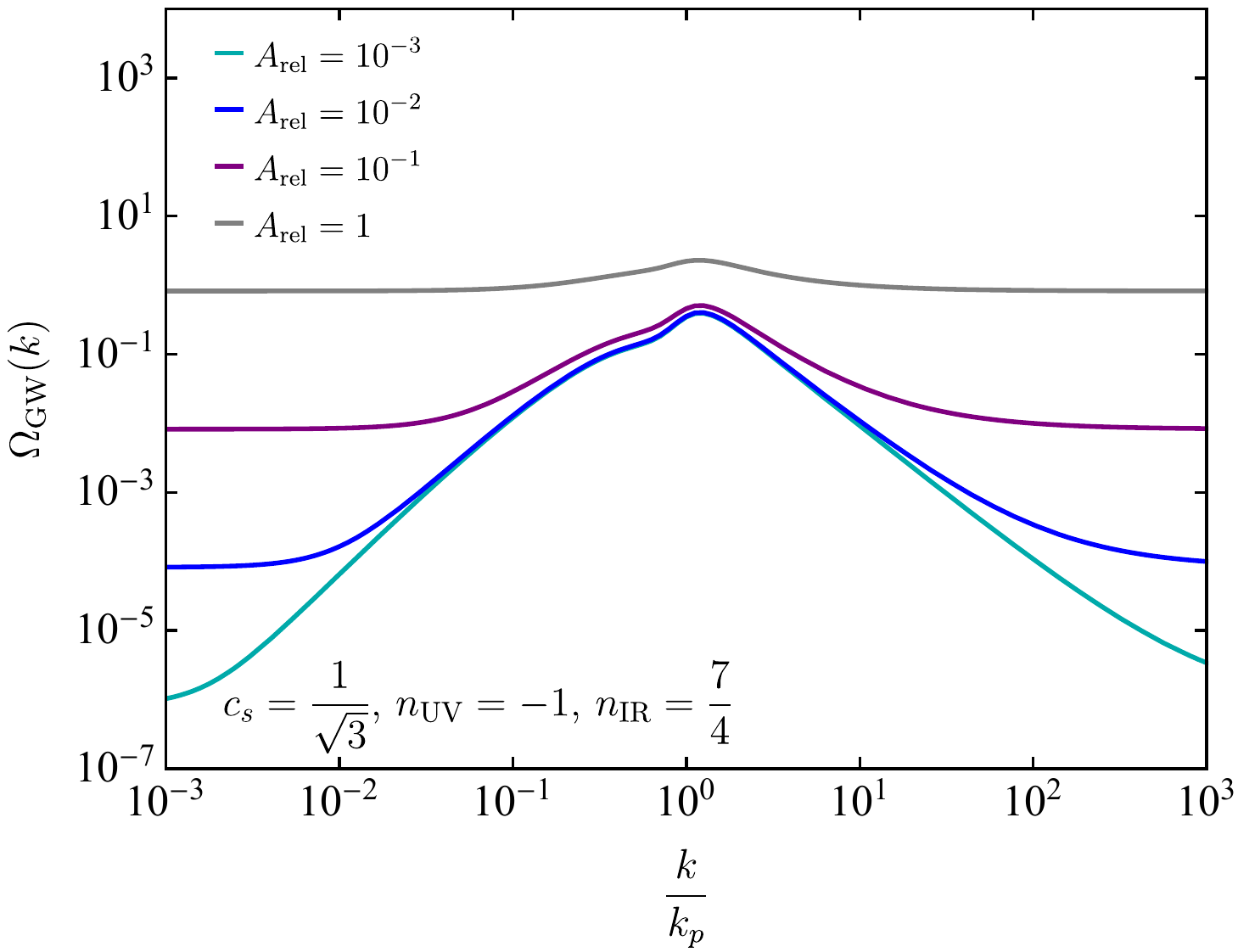}
\caption{The induced gravitational wave power spectrum $\Omega_{\textrm{GW}}$ is shown as a function of the dimensionless ratio $\frac{k}{k_p}$. The $\Omega_{\textrm{GW}}$ components are shown. They are separated by type: broken power-law, cross term and scale invariant flat background (left panel) and shown together as functions of the relative weight $A_{\rm rel}=[10^{-3},10^{-2},10^{-1},1]$ between the broken power-law and flat background amplitudes respectively. We consider a speed of sound $c_s=\frac{1}{\sqrt{3}}$ and in all we fix the IR and UV slopes to be $n_\textrm{IR}=\frac{7}{4}$ and $n_\textrm{UV}=-1$ respectively. \label{fig:brokenpowerlaw}}
\end{figure}

Let us show that, in the broken power-law case, the cross term is not as pronounced as in the sharp peak case studied in the main text \S~\ref{sec:template}. We will do so by finding the breaking points in the UV of the total GW spectrum \eqref{eq:split}. Some simple algebra leads us to
\begin{align}\label{eq:breaknuvl4}
\frac{k_{b1}}{k_p}\sim \frac{k_{b2}}{k_p}\sim A_{\rm rel}^{-1/n_{\rm UV}}\qquad (n_{\rm UV}<4)
\end{align}
and
\begin{align}\label{eq:breaknuvg4}
\frac{k_{b1}}{k_p}\sim A_{\rm rel}^{-1/n_{\rm UV}}\quad,\quad \frac{k_{b2}}{k_p}\sim A_{\rm rel}^{-1/4}\qquad (n_{\rm UV}>4)\,.
\end{align}
From \eqref{eq:breaknuvl4} it is easy to convince oneself that the contribution of the cross term in the total GW spectrum is often negligible.

First, for $n_{\rm UV}<4$, we find that the amplitude of the GW spectrum when the cross term might dominate is already of the order of the flat contribution as, that is $\Omega_{\rm GW}\sim {\cal A}_{\rm flat}^2$. Then, we also have that the visible width of the cross term is given by 
\begin{align}
\frac{\Delta k}{k_{b1}}\equiv \frac{k_{b1}-k_{b2}}{k_{b1}}\lesssim O(1)
\end{align}
Thus, we conclude that for $n_{\rm UV}<4$ the total GW spectrum is very close to the sum of $\Omega_{\rm GW, peak}$ and $\Omega_{\rm GW, flat}$. We show an example of this case in Figure~\ref{fig:brokenpowerlaw}. A similar calculation for $n_{\rm UV}>4$ yields
\begin{align}
\frac{\Delta k}{k_{b1}}\equiv \frac{k_{b1}-k_{b2}}{k_{b1}}\lesssim A_{\rm rel}^{\tfrac{1}{n_{\rm UV}}-\tfrac{1}{4}}-1\,.
\end{align}
We also see that when the cross term starts to dominate the total GW spectrum has an amplitude $\Omega_{\rm GW}\sim {\cal A}_{\rm peak}^2A_{\rm rel}^{1+\tfrac{4}{n_{\rm UV}}}$. Thus, the observationally interesting case is when $n_{\rm UV}\gg1$ which practically behaves as a very sharp peak which we studied in \S~\ref{sec:template}. For this reason, we did not consider a broken power-law primordial spectrum in the main text.

\section{Heuristic derivation of the power-law behaviour \label{app:derivation}}

Here we derive the power-law behaviour of the cross term by directly looking at the source term, which provides more physical insight. We will do so in an heuristic way. A mathematically rigorous derivation is provided in the main text.

The simplest way to understand the power-law behaviour of the cross term  is to look at the source term of tensor modes induced by a scalar field in the spatially flat gauge. This is roughly given by \cite{Domenech:2021ztg}
\begin{align}\label{eq:testt}
    h''_\mathbf{k}+2{\cal H}h_\mathbf{k}'+k^2h_\mathbf{k}\sim \int d^3q \, e^{ij}(k)q_iq_j \delta\varphi_\mathbf{q} \delta\varphi_{|\mathbf{k}-\mathbf{q}|}\,.
\end{align}
Then, if one of the fluctuations is a Dirac delta at $k=k_p$ and the other has a scale invariant spectrum,\footnote{The case when the two fluctuations come from a Dirac delta at $k=k_p$ is discussed in \cite{Domenech:2021ztg}} Eq.~\eqref{eq:testt} approximates to
\begin{align}\label{eq:test2}
    h''_\mathbf{k}+2{\cal H}h_\mathbf{k}'+k^2h_\mathbf{k}\sim \frac{k_p^{3/2}}{|k-k_p|^{3/2}} k_p^2\delta\varphi_{\rm peak} \delta\varphi_{\rm flat}{\cal T}_{\delta\varphi}(k_p\tau){\cal T}_{\delta\varphi}(|k-k_p|\tau)\,,
\end{align}
where $\delta\varphi_{\rm peak}$ and $ \delta\varphi_{\rm flat}$ are respectively the amplitude of the peak and the plateau, ${\cal T}_{\delta\varphi}(q\tau)$ is the transfer function of the scalar mode $q$ and since the flat contribution is scale invariant we use that $\delta\varphi_{\rm flat}(q)\propto (q/k_p)^{-3/2}$ with the pivot scale set at $k_p$.
We have that ${\cal T}_{\delta\varphi}(q\tau)$ is first constant on superhorizon scales ($q\tau<1$) and then it decays on subhorizon scales ($q\tau>1$) as $a^{-1}$ and oscillates as $e^{-ic_sq\tau}$. This means that for $k>k_p$ \eqref{eq:test2} has a constant source for $k\tau<1$ and it follows that $h_k\sim k^{-3/2}(k_p\tau)^2\delta\varphi_{\rm peak}\delta\varphi_{\rm flat}$. Evaluating at horizon crossing of the tensor mode (at $k\tau=1$) we find that
\begin{align}\label{eq:test4}
    \Omega_{\rm GW,cross}\propto k^3\langle h^2\rangle\propto \left(\frac{k}{k_p}\right)^{-4}\,.
\end{align}
However, for $k>k_p$ when the scalar mode with $|k-k_p|$ enters the horizon it oscillates as $e^{-ic_s|k-k_p|\tau}\sim e^{-ic_sk\tau}$. This means that for $k_p\tau<1$, \textit{i.e.} when $k_p$ is still superhorizon, the source term goes as
\begin{align}\label{eq:test3}
    h''_\mathbf{k}+2{\cal H}h_\mathbf{k}'+k^2h_\mathbf{k}\sim \frac{k_p^{3/2}}{|k-k_p|^{3/2}} k_p^2 \delta\varphi_{\rm peak}\delta\varphi_{\rm flat} a^{-1}e^{-ic_sk\tau}\,,
\end{align}
which has a resonance for $c_s=1$. This resonance stops at $\tau=1/k_p$, when the scalar mode with $k=k_p$ enters the horizon and decays. To see this resonance, we show the particular solution to \eqref{eq:test3} which at leading order in $k\tau$ is given by
\begin{align}\label{eq:test33}
    h_\mathbf{k}(c_s\neq1,k_p\tau<1)\sim \frac{k_p^{7/2}}{k^{7/2}}\frac{ e^{i {c_s} k \tau }}{(1-c_s^2)k\tau }\delta\varphi_{\rm peak}\delta\varphi_{\rm flat}\,,
\end{align}
 for $c_s\neq1$ and 
\begin{align}
h_\mathbf{k}(c_s=1,k_p\tau<1)\sim e^{i k \tau } \frac{k_p^{7/2}}{k^{7/2}}\delta\varphi_{\rm peak}\delta\varphi_{\rm flat}\,,
\end{align}
for $c_s=1$. Matching the above solutions at horizon crossing of the peak, that is at $\tau=1/k_p$, to the subhorizon homogeneous solutions of \eqref{eq:testt} we find for $c_s\neq1$ that, at leading order in $k/k_p$,
\begin{align}
h_\mathbf{k}(c_s\neq1,k_p\tau>1)\sim a^{-1} e^{i k \tau } \frac{k_p^{7/2}}{k^{7/2}}\delta\varphi_{\rm peak}\delta\varphi_{\rm flat}\,,
\end{align}
and for $c_s=1$
\begin{align}
h_\mathbf{k}(c_s=1,k_p\tau>1)\sim a^{-1} e^{i k \tau } \frac{k_p^{5/2}}{k^{5/2}}\delta\varphi_{\rm peak}\delta\varphi_{\rm flat}\,.
\end{align}
Proceeding as in Eq.~\eqref{eq:test4} we conclude that
\begin{align}\label{eq:test5}
    \Omega_{\rm GW,cross}(c_s\neq1, k>k_p)\propto \left(\frac{k}{k_p}\right)^{-4}\,\quad {\rm and}\quad \Omega_{\rm GW,cross}(c_s=1, k>k_p)\propto \left(\frac{k}{k_p}\right)^{-2}\,,
\end{align}
as we have explicitly shown in the main text.

\section{High-frequency behaviour for general propagation speed \label{app:generalcs}}

In this appendix, we derive the approximations  for the transfer function ${\cal T}(u,v,c_s)$ \eqref{eq:kernel} used in the in the UV limit of the GW spectrum, Eqs.~\eqref{eq:inducedGWspeakUVcs1} and \eqref{eq:inducedGWspeakUVcs1/3}, for a peaked primordial spectrum. Let us first present two distinct limits of the transfer function and then turn to intermediate cases. First, expanding $y$ \eqref{eq:y} for $v\ll 1$ and $u\sim 1$ we have, as given in the main text
\begin{align}
y=\frac{1-c_s^{-2}}{2v} + \frac{v}{2} +\mathcal{O}(v^2)\,.
\end{align}
We see that for $v\ll1$ we may have $|y|\gg1$ and $|y|\ll1$ depending on the value of $c_s^2$. In fact, the transition between these two regimes, i.e. at roughly $|y|\sim1$ occurs at
\begin{align}
v_*=\frac{c_s^{-2}-1}{2}\,,
\end{align}
where since $v>0$ we took the positive solution. 
Note that this point $v_*$ also corresponds to the point where the resonant line $u=c_s^{-2}-v_*$ meets the upper boundary $u=1+v_*$. This also means that for $c_s^{2}<1/2$, which implies $v_{*}>1/2$, the change in the behaviour of ${\cal T}(u,v,c_s)$ is well inside the UV regime, i.e. at $v\ll1$ and $u\sim 1$. If the primordial spectrum ${\cal P}_{{\cal R},\rm peak}$ is peaked, we should see a break in the GW spectrum at
\begin{align}\label{eq:ks}
\frac{k_{*}}{k_p}\approx\frac{2}{c_s^{-2}-1}\,.
\end{align}
For $k>k_*$ we are in the $|y|\gg 1$ regime, and we find that
\begin{align}\label{eq:ivlvs}
{\cal I}(v<v_*)\equiv\int_{|1-v|}^{1+v}{\cal T}(u\sim 1, v \gg 1,c_s)\approx \frac{8}{27}\frac{v^3}{(1-c_s^2)^2}\,,
\end{align}
where we expanded the integral for $v\ll1$ and we took $u=1$.

\begin{figure}
\includegraphics[width=0.45\columnwidth]{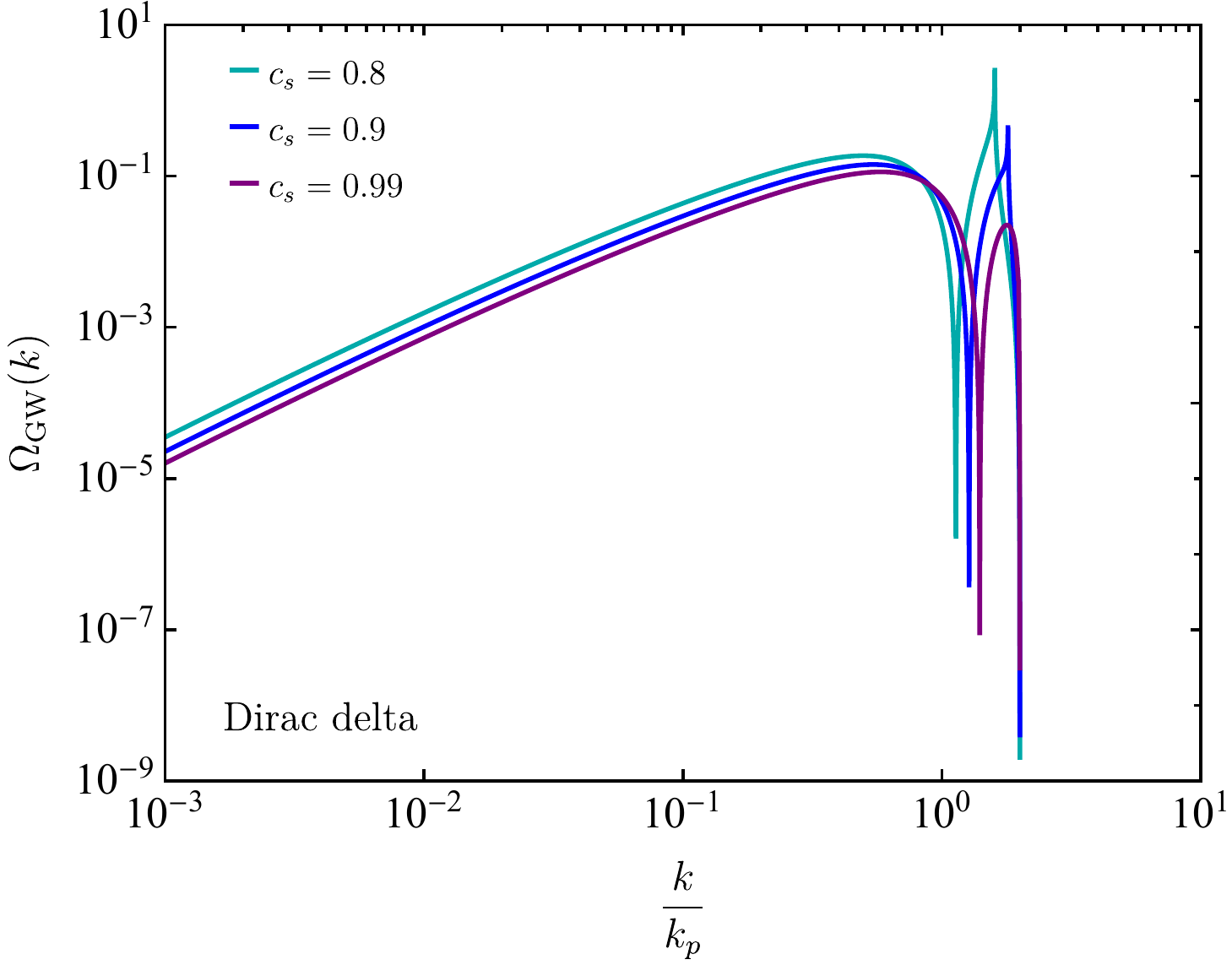}
\includegraphics[width=0.45\columnwidth]{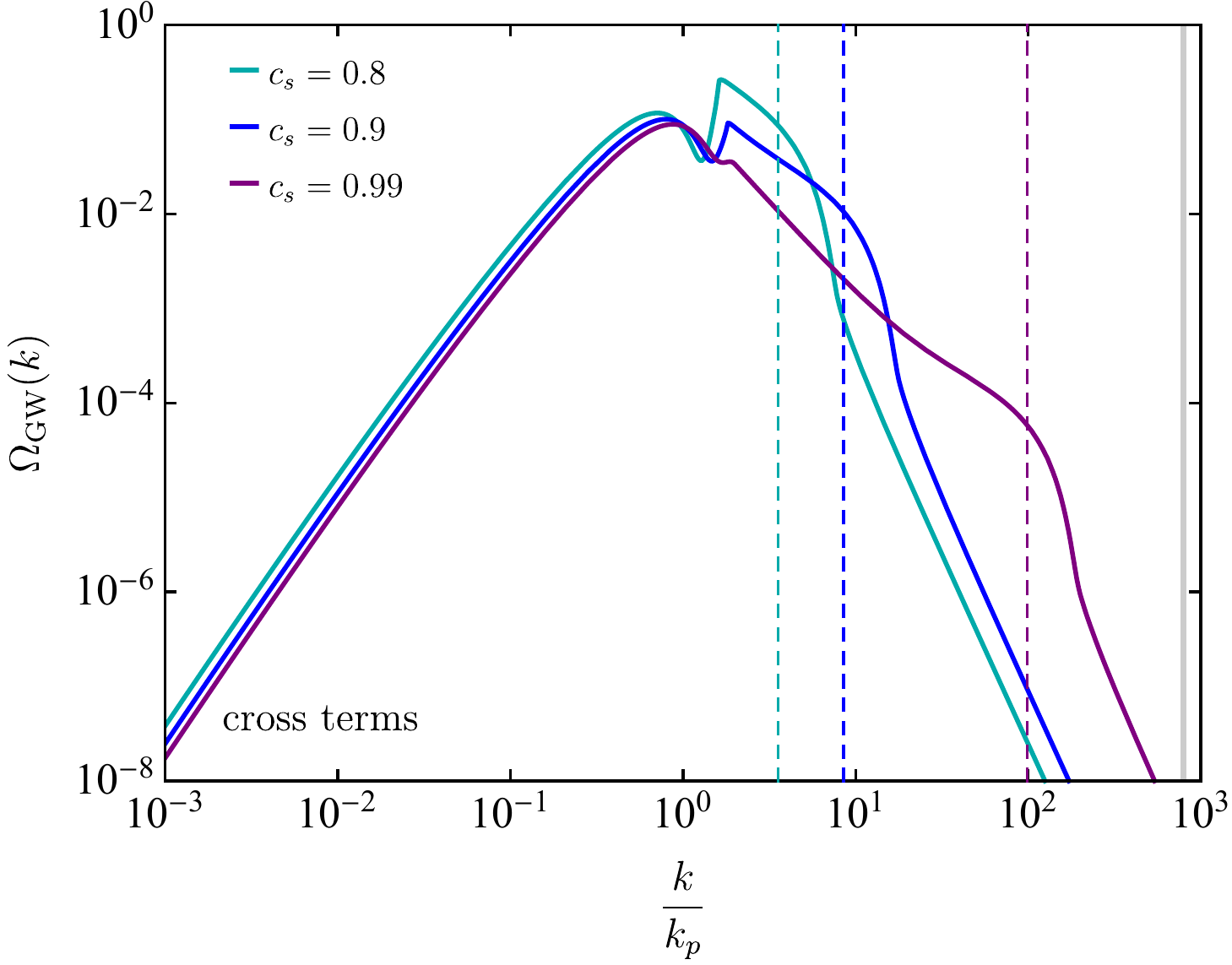}
\caption{The induced gravitational wave power spectrum $\Omega_{\textrm{GW}}$  is shown as a function of the dimensionless ratio $\frac{k}{k_p}$ for speed of sound $c_s=0.8, 0.9, 0.99$ respectively. $\Omega_{\textrm{GW}}$ is shown for a Dirac delta peak (left panel) and for the cross terms (right panel). The dashed lines in the right panel correspond to the break points defined in Eq.~\eqref{eq:ks}.  \label{fig:diracdeltacs}}
\end{figure}

For $k<k_*$ we are in the $|y|\lesssim1$ regime. This case is more subtle than the previous one as the integrand vanishes for $u=1$, so that we have to consider an extended region of integration. However, we can proceed as follows. Since we are in a regime where the resonance is cancelled and since we also know that we have to recover the $c_s^2=1$ limit, let us just take the case $c_s^2=1$. Then, by continuity, the GW spectrum should for $k<k_*$ should be well approximated by that limiting case. Interestingly, when $c_s^2=1$ we can write the integrand as a function of $y$ only as
\begin{align}
{\cal T}(u\sim 1, v \gg 1,c_s^2=1)=\frac{y^2}{3}\left(1-y^2\right)^2\left(\frac{\pi^2y^2}{4}+\left(1-\frac{y}{2}\ln\left|\frac{1+y}{1-y}\right|\right)^2\right)\,.
\end{align}
Note that since $y$ is quadratic in $u$ and $v$, there are two solutions of $u$ in terms of $y$ and $v$. Nevertheless, they yield the same integrand. Now, considering $v$ to be fixed, because we are assuming a peaked primordial spectrum in the $v$ variable, the variable transformation of $u$ for $y$ yields
\begin{align}
du=\left(v\pm\frac{v^2y}{\sqrt{1-v^2(1-y^2)}}\right)dy\approx v dy\,,
\end{align}
where in the last step we took the leading order when $v\ll1$. The integration range for $y$ is $-1<y<1$. Then we have that
\begin{align}\label{eq:ivgvs}
{\cal I}(v>v_*)\approx 2v\int_0^1dy\frac{y^2}{3}\left(1-y^2\right)^2\left(\frac{\pi^2y^2}{4}+\left(1-\frac{y}{2}\ln\left|\frac{1+y}{1-y}\right|\right)^2\right)=
2v \frac{35+24\pi^2}{8505}\,.
\end{align}
In Figure~\ref{fig:diracdeltacs} we show these approximations for the cross term $\Omega_{\rm GW, cross}$ in the Dirac delta case studied in Section~\ref{fig:diracdeltacs}. We find that the approximations \eqref{eq:ivlvs} and \eqref{eq:ivgvs} provide a good fit to the numerical integration and that the breaking point $k_*$ is indeed well approximated by \eqref{eq:ks}.

\bibliography{bibliographyIGWs.bib} 

\begin{thebibliography}{141}%
\makeatletter
\providecommand \@ifxundefined [1]{%
 \@ifx{#1\undefined}
}%
\providecommand \@ifnum [1]{%
 \ifnum #1\expandafter \@firstoftwo
 \else \expandafter \@secondoftwo
 \fi
}%
\providecommand \@ifx [1]{%
 \ifx #1\expandafter \@firstoftwo
 \else \expandafter \@secondoftwo
 \fi
}%
\providecommand \natexlab [1]{#1}%
\providecommand \enquote  [1]{``#1''}%
\providecommand \bibnamefont  [1]{#1}%
\providecommand \bibfnamefont [1]{#1}%
\providecommand \citenamefont [1]{#1}%
\providecommand \href@noop [0]{\@secondoftwo}%
\providecommand \href [0]{\begingroup \@sanitize@url \@href}%
\providecommand \@href[1]{\@@startlink{#1}\@@href}%
\providecommand \@@href[1]{\endgroup#1\@@endlink}%
\providecommand \@sanitize@url [0]{\catcode `\\12\catcode `\$12\catcode
  `\&12\catcode `\#12\catcode `\^12\catcode `\_12\catcode `\%12\relax}%
\providecommand \@@startlink[1]{}%
\providecommand \@@endlink[0]{}%
\providecommand \url  [0]{\begingroup\@sanitize@url \@url }%
\providecommand \@url [1]{\endgroup\@href {#1}{\urlprefix }}%
\providecommand \urlprefix  [0]{URL }%
\providecommand \Eprint [0]{\href }%
\providecommand \doibase [0]{http://dx.doi.org/}%
\providecommand \selectlanguage [0]{\@gobble}%
\providecommand \bibinfo  [0]{\@secondoftwo}%
\providecommand \bibfield  [0]{\@secondoftwo}%
\providecommand \translation [1]{[#1]}%
\providecommand \BibitemOpen [0]{}%
\providecommand \bibitemStop [0]{}%
\providecommand \bibitemNoStop [0]{.\EOS\space}%
\providecommand \EOS [0]{\spacefactor3000\relax}%
\providecommand \BibitemShut  [1]{\csname bibitem#1\endcsname}%
\let\auto@bib@innerbib\@empty
%</preamble>
\bibitem [{\citenamefont {Guzzetti}\ \emph {et~al.}(2016)\citenamefont
  {Guzzetti}, \citenamefont {Bartolo}, \citenamefont {Liguori},\ and\
  \citenamefont {Matarrese}}]{Guzzetti:2016mkm}%
  \BibitemOpen
  \bibfield  {author} {\bibinfo {author} {\bibfnamefont {M.~C.}\ \bibnamefont
  {Guzzetti}}, \bibinfo {author} {\bibfnamefont {N.}~\bibnamefont {Bartolo}},
  \bibinfo {author} {\bibfnamefont {M.}~\bibnamefont {Liguori}}, \ and\
  \bibinfo {author} {\bibfnamefont {S.}~\bibnamefont {Matarrese}},\ }\href
  {\doibase 10.1393/ncr/i2016-10127-1} {\bibfield  {journal} {\bibinfo
  {journal} {Riv. Nuovo Cim.}\ }\textbf {\bibinfo {volume} {39}},\ \bibinfo
  {pages} {399} (\bibinfo {year} {2016})},\ \Eprint
  {http://arxiv.org/abs/1605.01615} {arXiv:1605.01615 [astro-ph.CO]}
  \BibitemShut {NoStop}%
\bibitem [{\citenamefont {Caprini}\ and\ \citenamefont
  {Figueroa}(2018)}]{Caprini:2018mtu}%
  \BibitemOpen
  \bibfield  {author} {\bibinfo {author} {\bibfnamefont {C.}~\bibnamefont
  {Caprini}}\ and\ \bibinfo {author} {\bibfnamefont {D.~G.}\ \bibnamefont
  {Figueroa}},\ }\href {\doibase 10.1088/1361-6382/aac608} {\bibfield
  {journal} {\bibinfo  {journal} {Class. Quant. Grav.}\ }\textbf {\bibinfo
  {volume} {35}},\ \bibinfo {pages} {163001} (\bibinfo {year} {2018})},\
  \Eprint {http://arxiv.org/abs/1801.04268} {arXiv:1801.04268 [astro-ph.CO]}
  \BibitemShut {NoStop}%
\bibitem [{\citenamefont {Tomita}(1967)}]{Tomita}%
  \BibitemOpen
  \bibfield  {author} {\bibinfo {author} {\bibfnamefont {K.}~\bibnamefont
  {Tomita}},\ }\href {\doibase 10.1143/PTP.37.831} {\bibfield  {journal}
  {\bibinfo  {journal} {Progress of Theoretical Physics}\ }\textbf {\bibinfo
  {volume} {37}},\ \bibinfo {pages} {831} (\bibinfo {year} {1967})},\ \Eprint
  {http://arxiv.org/abs/https://academic.oup.com/ptp/article-pdf/37/5/831/5234391/37-5-831.pdf}
  {https://academic.oup.com/ptp/article-pdf/37/5/831/5234391/37-5-831.pdf}
  \BibitemShut {NoStop}%
\bibitem [{\citenamefont {Matarrese}\ \emph {et~al.}(1993)\citenamefont
  {Matarrese}, \citenamefont {Pantano},\ and\ \citenamefont
  {Saez}}]{Matarrese:1992rp}%
  \BibitemOpen
  \bibfield  {author} {\bibinfo {author} {\bibfnamefont {S.}~\bibnamefont
  {Matarrese}}, \bibinfo {author} {\bibfnamefont {O.}~\bibnamefont {Pantano}},
  \ and\ \bibinfo {author} {\bibfnamefont {D.}~\bibnamefont {Saez}},\ }\href
  {\doibase 10.1103/PhysRevD.47.1311} {\bibfield  {journal} {\bibinfo
  {journal} {Phys. Rev. D}\ }\textbf {\bibinfo {volume} {47}},\ \bibinfo
  {pages} {1311} (\bibinfo {year} {1993})}\BibitemShut {NoStop}%
\bibitem [{\citenamefont {Matarrese}\ \emph {et~al.}(1994)\citenamefont
  {Matarrese}, \citenamefont {Pantano},\ and\ \citenamefont
  {Saez}}]{Matarrese:1993zf}%
  \BibitemOpen
  \bibfield  {author} {\bibinfo {author} {\bibfnamefont {S.}~\bibnamefont
  {Matarrese}}, \bibinfo {author} {\bibfnamefont {O.}~\bibnamefont {Pantano}},
  \ and\ \bibinfo {author} {\bibfnamefont {D.}~\bibnamefont {Saez}},\ }\href
  {\doibase 10.1103/PhysRevLett.72.320} {\bibfield  {journal} {\bibinfo
  {journal} {Phys. Rev. Lett.}\ }\textbf {\bibinfo {volume} {72}},\ \bibinfo
  {pages} {320} (\bibinfo {year} {1994})},\ \Eprint
  {http://arxiv.org/abs/astro-ph/9310036} {arXiv:astro-ph/9310036} \BibitemShut
  {NoStop}%
\bibitem [{\citenamefont {Ananda}\ \emph {et~al.}(2007)\citenamefont {Ananda},
  \citenamefont {Clarkson},\ and\ \citenamefont {Wands}}]{Ananda:2006af}%
  \BibitemOpen
  \bibfield  {author} {\bibinfo {author} {\bibfnamefont {K.~N.}\ \bibnamefont
  {Ananda}}, \bibinfo {author} {\bibfnamefont {C.}~\bibnamefont {Clarkson}}, \
  and\ \bibinfo {author} {\bibfnamefont {D.}~\bibnamefont {Wands}},\ }\href
  {\doibase 10.1103/PhysRevD.75.123518} {\bibfield  {journal} {\bibinfo
  {journal} {Phys. Rev. D}\ }\textbf {\bibinfo {volume} {75}},\ \bibinfo
  {pages} {123518} (\bibinfo {year} {2007})},\ \Eprint
  {http://arxiv.org/abs/gr-qc/0612013} {arXiv:gr-qc/0612013} \BibitemShut
  {NoStop}%
\bibitem [{\citenamefont {Baumann}\ \emph {et~al.}(2007)\citenamefont
  {Baumann}, \citenamefont {Steinhardt}, \citenamefont {Takahashi},\ and\
  \citenamefont {Ichiki}}]{Baumann:2007zm}%
  \BibitemOpen
  \bibfield  {author} {\bibinfo {author} {\bibfnamefont {D.}~\bibnamefont
  {Baumann}}, \bibinfo {author} {\bibfnamefont {P.~J.}\ \bibnamefont
  {Steinhardt}}, \bibinfo {author} {\bibfnamefont {K.}~\bibnamefont
  {Takahashi}}, \ and\ \bibinfo {author} {\bibfnamefont {K.}~\bibnamefont
  {Ichiki}},\ }\href {\doibase 10.1103/PhysRevD.76.084019} {\bibfield
  {journal} {\bibinfo  {journal} {Phys. Rev. D}\ }\textbf {\bibinfo {volume}
  {76}},\ \bibinfo {pages} {084019} (\bibinfo {year} {2007})},\ \Eprint
  {http://arxiv.org/abs/hep-th/0703290} {arXiv:hep-th/0703290} \BibitemShut
  {NoStop}%
\bibitem [{\citenamefont {Saito}\ and\ \citenamefont
  {Yokoyama}(2009)}]{Saito:2008jc}%
  \BibitemOpen
  \bibfield  {author} {\bibinfo {author} {\bibfnamefont {R.}~\bibnamefont
  {Saito}}\ and\ \bibinfo {author} {\bibfnamefont {J.}~\bibnamefont
  {Yokoyama}},\ }\href {\doibase 10.1103/PhysRevLett.102.161101} {\bibfield
  {journal} {\bibinfo  {journal} {Phys. Rev. Lett.}\ }\textbf {\bibinfo
  {volume} {102}},\ \bibinfo {pages} {161101} (\bibinfo {year} {2009})},\
  \bibinfo {note} {[Erratum: Phys.Rev.Lett. 107, 069901 (2011)]},\ \Eprint
  {http://arxiv.org/abs/0812.4339} {arXiv:0812.4339 [astro-ph]} \BibitemShut
  {NoStop}%
\bibitem [{\citenamefont {Saito}\ and\ \citenamefont
  {Yokoyama}(2010)}]{Saito:2009jt}%
  \BibitemOpen
  \bibfield  {author} {\bibinfo {author} {\bibfnamefont {R.}~\bibnamefont
  {Saito}}\ and\ \bibinfo {author} {\bibfnamefont {J.}~\bibnamefont
  {Yokoyama}},\ }\href {\doibase 10.1143/PTP.126.351} {\bibfield  {journal}
  {\bibinfo  {journal} {Prog. Theor. Phys.}\ }\textbf {\bibinfo {volume}
  {123}},\ \bibinfo {pages} {867} (\bibinfo {year} {2010})},\ \bibinfo {note}
  {[Erratum: Prog.Theor.Phys. 126, 351--352 (2011)]},\ \Eprint
  {http://arxiv.org/abs/0912.5317} {arXiv:0912.5317 [astro-ph.CO]} \BibitemShut
  {NoStop}%
\bibitem [{\citenamefont {Yuan}\ and\ \citenamefont
  {Huang}(2021)}]{Yuan:2021qgz}%
  \BibitemOpen
  \bibfield  {author} {\bibinfo {author} {\bibfnamefont {C.}~\bibnamefont
  {Yuan}}\ and\ \bibinfo {author} {\bibfnamefont {Q.-G.}\ \bibnamefont
  {Huang}},\ }\href@noop {} {\  (\bibinfo {year} {2021})},\ \Eprint
  {http://arxiv.org/abs/2103.04739} {arXiv:2103.04739 [astro-ph.GA]}
  \BibitemShut {NoStop}%
\bibitem [{\citenamefont {Dom\`enech}(2021)}]{Domenech:2021ztg}%
  \BibitemOpen
  \bibfield  {author} {\bibinfo {author} {\bibfnamefont {G.}~\bibnamefont
  {Dom\`enech}},\ }\href {\doibase 10.3390/universe7110398} {\bibfield
  {journal} {\bibinfo  {journal} {Universe}\ }\textbf {\bibinfo {volume} {7}},\
  \bibinfo {pages} {398} (\bibinfo {year} {2021})},\ \Eprint
  {http://arxiv.org/abs/2109.01398} {arXiv:2109.01398 [gr-qc]} \BibitemShut
  {NoStop}%
\bibitem [{\citenamefont {Akrami}\ \emph {et~al.}(2020)\citenamefont {Akrami}
  \emph {et~al.}}]{Akrami:2018odb}%
  \BibitemOpen
  \bibfield  {author} {\bibinfo {author} {\bibfnamefont {Y.}~\bibnamefont
  {Akrami}} \emph {et~al.} (\bibinfo {collaboration} {Planck}),\ }\href
  {\doibase 10.1051/0004-6361/201833887} {\bibfield  {journal} {\bibinfo
  {journal} {Astron. Astrophys.}\ }\textbf {\bibinfo {volume} {641}},\ \bibinfo
  {pages} {A10} (\bibinfo {year} {2020})},\ \Eprint
  {http://arxiv.org/abs/1807.06211} {arXiv:1807.06211 [astro-ph.CO]}
  \BibitemShut {NoStop}%
\bibitem [{\citenamefont {Brandenberger}\ and\ \citenamefont
  {Peter}(2017)}]{Brandenberger:2016vhg}%
  \BibitemOpen
  \bibfield  {author} {\bibinfo {author} {\bibfnamefont {R.}~\bibnamefont
  {Brandenberger}}\ and\ \bibinfo {author} {\bibfnamefont {P.}~\bibnamefont
  {Peter}},\ }\href {\doibase 10.1007/s10701-016-0057-0} {\bibfield  {journal}
  {\bibinfo  {journal} {Found. Phys.}\ }\textbf {\bibinfo {volume} {47}},\
  \bibinfo {pages} {797} (\bibinfo {year} {2017})},\ \Eprint
  {http://arxiv.org/abs/1603.05834} {arXiv:1603.05834 [hep-th]} \BibitemShut
  {NoStop}%
\bibitem [{\citenamefont {Garcia-Bellido}\ \emph {et~al.}(2016)\citenamefont
  {Garcia-Bellido}, \citenamefont {Peloso},\ and\ \citenamefont
  {Unal}}]{Garcia-Bellido:2016dkw}%
  \BibitemOpen
  \bibfield  {author} {\bibinfo {author} {\bibfnamefont {J.}~\bibnamefont
  {Garcia-Bellido}}, \bibinfo {author} {\bibfnamefont {M.}~\bibnamefont
  {Peloso}}, \ and\ \bibinfo {author} {\bibfnamefont {C.}~\bibnamefont
  {Unal}},\ }\href {\doibase 10.1088/1475-7516/2016/12/031} {\bibfield
  {journal} {\bibinfo  {journal} {JCAP}\ }\textbf {\bibinfo {volume} {12}},\
  \bibinfo {pages} {031} (\bibinfo {year} {2016})},\ \Eprint
  {http://arxiv.org/abs/1610.03763} {arXiv:1610.03763 [astro-ph.CO]}
  \BibitemShut {NoStop}%
\bibitem [{\citenamefont {Di}\ and\ \citenamefont {Gong}(2018)}]{Gong:2017qlj}%
  \BibitemOpen
  \bibfield  {author} {\bibinfo {author} {\bibfnamefont {H.}~\bibnamefont
  {Di}}\ and\ \bibinfo {author} {\bibfnamefont {Y.}~\bibnamefont {Gong}},\
  }\href {\doibase 10.1088/1475-7516/2018/07/007} {\bibfield  {journal}
  {\bibinfo  {journal} {JCAP}\ }\textbf {\bibinfo {volume} {07}},\ \bibinfo
  {pages} {007} (\bibinfo {year} {2018})},\ \Eprint
  {http://arxiv.org/abs/1707.09578} {arXiv:1707.09578 [astro-ph.CO]}
  \BibitemShut {NoStop}%
\bibitem [{\citenamefont {Ando}\ \emph
  {et~al.}(2018{\natexlab{a}})\citenamefont {Ando}, \citenamefont {Kawasaki},\
  and\ \citenamefont {Nakatsuka}}]{Ando:2018nge}%
  \BibitemOpen
  \bibfield  {author} {\bibinfo {author} {\bibfnamefont {K.}~\bibnamefont
  {Ando}}, \bibinfo {author} {\bibfnamefont {M.}~\bibnamefont {Kawasaki}}, \
  and\ \bibinfo {author} {\bibfnamefont {H.}~\bibnamefont {Nakatsuka}},\ }\href
  {\doibase 10.1103/PhysRevD.98.083508} {\bibfield  {journal} {\bibinfo
  {journal} {Phys. Rev. D}\ }\textbf {\bibinfo {volume} {98}},\ \bibinfo
  {pages} {083508} (\bibinfo {year} {2018}{\natexlab{a}})},\ \Eprint
  {http://arxiv.org/abs/1805.07757} {arXiv:1805.07757 [astro-ph.CO]}
  \BibitemShut {NoStop}%
\bibitem [{\citenamefont {Byrnes}\ \emph {et~al.}(2019)\citenamefont {Byrnes},
  \citenamefont {Cole},\ and\ \citenamefont {Patil}}]{Byrnes:2018txb}%
  \BibitemOpen
  \bibfield  {author} {\bibinfo {author} {\bibfnamefont {C.~T.}\ \bibnamefont
  {Byrnes}}, \bibinfo {author} {\bibfnamefont {P.~S.}\ \bibnamefont {Cole}}, \
  and\ \bibinfo {author} {\bibfnamefont {S.~P.}\ \bibnamefont {Patil}},\ }\href
  {\doibase 10.1088/1475-7516/2019/06/028} {\bibfield  {journal} {\bibinfo
  {journal} {JCAP}\ }\textbf {\bibinfo {volume} {06}},\ \bibinfo {pages} {028}
  (\bibinfo {year} {2019})},\ \Eprint {http://arxiv.org/abs/1811.11158}
  {arXiv:1811.11158 [astro-ph.CO]} \BibitemShut {NoStop}%
\bibitem [{\citenamefont {Gao}\ and\ \citenamefont {Yang}(2019)}]{Gao:2019kto}%
  \BibitemOpen
  \bibfield  {author} {\bibinfo {author} {\bibfnamefont {T.-J.}\ \bibnamefont
  {Gao}}\ and\ \bibinfo {author} {\bibfnamefont {X.-Y.}\ \bibnamefont {Yang}},\
  }\href {\doibase 10.1142/S0217751X19502130} {\bibfield  {journal} {\bibinfo
  {journal} {Int. J. Mod. Phys. A}\ }\textbf {\bibinfo {volume} {34}},\
  \bibinfo {pages} {1950213} (\bibinfo {year} {2019})}\BibitemShut {NoStop}%
\bibitem [{\citenamefont {Xu}\ \emph {et~al.}(2020)\citenamefont {Xu},
  \citenamefont {Liu}, \citenamefont {Gao},\ and\ \citenamefont
  {Guo}}]{Xu:2019bdp}%
  \BibitemOpen
  \bibfield  {author} {\bibinfo {author} {\bibfnamefont {W.-T.}\ \bibnamefont
  {Xu}}, \bibinfo {author} {\bibfnamefont {J.}~\bibnamefont {Liu}}, \bibinfo
  {author} {\bibfnamefont {T.-J.}\ \bibnamefont {Gao}}, \ and\ \bibinfo
  {author} {\bibfnamefont {Z.-K.}\ \bibnamefont {Guo}},\ }\href {\doibase
  10.1103/PhysRevD.101.023505} {\bibfield  {journal} {\bibinfo  {journal}
  {Phys. Rev. D}\ }\textbf {\bibinfo {volume} {101}},\ \bibinfo {pages}
  {023505} (\bibinfo {year} {2020})},\ \Eprint
  {http://arxiv.org/abs/1907.05213} {arXiv:1907.05213 [astro-ph.CO]}
  \BibitemShut {NoStop}%
\bibitem [{\citenamefont {Liu}\ \emph {et~al.}(2020)\citenamefont {Liu},
  \citenamefont {Guo},\ and\ \citenamefont {Cai}}]{Liu:2020oqe}%
  \BibitemOpen
  \bibfield  {author} {\bibinfo {author} {\bibfnamefont {J.}~\bibnamefont
  {Liu}}, \bibinfo {author} {\bibfnamefont {Z.-K.}\ \bibnamefont {Guo}}, \ and\
  \bibinfo {author} {\bibfnamefont {R.-G.}\ \bibnamefont {Cai}},\ }\href
  {\doibase 10.1103/PhysRevD.101.083535} {\bibfield  {journal} {\bibinfo
  {journal} {Phys. Rev. D}\ }\textbf {\bibinfo {volume} {101}},\ \bibinfo
  {pages} {083535} (\bibinfo {year} {2020})},\ \Eprint
  {http://arxiv.org/abs/2003.02075} {arXiv:2003.02075 [astro-ph.CO]}
  \BibitemShut {NoStop}%
\bibitem [{\citenamefont {Cai}\ \emph {et~al.}(2019{\natexlab{a}})\citenamefont
  {Cai}, \citenamefont {Pi}, \citenamefont {Wang},\ and\ \citenamefont
  {Yang}}]{Cai:2019amo}%
  \BibitemOpen
  \bibfield  {author} {\bibinfo {author} {\bibfnamefont {R.-G.}\ \bibnamefont
  {Cai}}, \bibinfo {author} {\bibfnamefont {S.}~\bibnamefont {Pi}}, \bibinfo
  {author} {\bibfnamefont {S.-J.}\ \bibnamefont {Wang}}, \ and\ \bibinfo
  {author} {\bibfnamefont {X.-Y.}\ \bibnamefont {Yang}},\ }\href {\doibase
  10.1088/1475-7516/2019/05/013} {\bibfield  {journal} {\bibinfo  {journal}
  {JCAP}\ }\textbf {\bibinfo {volume} {05}},\ \bibinfo {pages} {013} (\bibinfo
  {year} {2019}{\natexlab{a}})},\ \Eprint {http://arxiv.org/abs/1901.10152}
  {arXiv:1901.10152 [astro-ph.CO]} \BibitemShut {NoStop}%
\bibitem [{\citenamefont {\"Ozsoy}\ and\ \citenamefont
  {Tasinato}(2020)}]{Ozsoy:2019lyy}%
  \BibitemOpen
  \bibfield  {author} {\bibinfo {author} {\bibfnamefont {O.}~\bibnamefont
  {\"Ozsoy}}\ and\ \bibinfo {author} {\bibfnamefont {G.}~\bibnamefont
  {Tasinato}},\ }\href {\doibase 10.1088/1475-7516/2020/04/048} {\bibfield
  {journal} {\bibinfo  {journal} {JCAP}\ }\textbf {\bibinfo {volume} {04}},\
  \bibinfo {pages} {048} (\bibinfo {year} {2020})},\ \Eprint
  {http://arxiv.org/abs/1912.01061} {arXiv:1912.01061 [astro-ph.CO]}
  \BibitemShut {NoStop}%
\bibitem [{\citenamefont {\"Ozsoy}\ and\ \citenamefont
  {Lalak}(2021)}]{Ozsoy:2020kat}%
  \BibitemOpen
  \bibfield  {author} {\bibinfo {author} {\bibfnamefont {O.}~\bibnamefont
  {\"Ozsoy}}\ and\ \bibinfo {author} {\bibfnamefont {Z.}~\bibnamefont
  {Lalak}},\ }\href {\doibase 10.1088/1475-7516/2021/01/040} {\bibfield
  {journal} {\bibinfo  {journal} {JCAP}\ }\textbf {\bibinfo {volume} {01}},\
  \bibinfo {pages} {040} (\bibinfo {year} {2021})},\ \Eprint
  {http://arxiv.org/abs/2008.07549} {arXiv:2008.07549 [astro-ph.CO]}
  \BibitemShut {NoStop}%
\bibitem [{\citenamefont {Ragavendra}\ \emph {et~al.}(2020)\citenamefont
  {Ragavendra}, \citenamefont {Saha}, \citenamefont {Sriramkumar},\ and\
  \citenamefont {Silk}}]{Ragavendra:2020sop}%
  \BibitemOpen
  \bibfield  {author} {\bibinfo {author} {\bibfnamefont {H.~V.}\ \bibnamefont
  {Ragavendra}}, \bibinfo {author} {\bibfnamefont {P.}~\bibnamefont {Saha}},
  \bibinfo {author} {\bibfnamefont {L.}~\bibnamefont {Sriramkumar}}, \ and\
  \bibinfo {author} {\bibfnamefont {J.}~\bibnamefont {Silk}},\ }\href@noop {}
  {\  (\bibinfo {year} {2020})},\ \Eprint {http://arxiv.org/abs/2008.12202}
  {arXiv:2008.12202 [astro-ph.CO]} \BibitemShut {NoStop}%
\bibitem [{\citenamefont {Fumagalli}\ \emph
  {et~al.}(2020{\natexlab{a}})\citenamefont {Fumagalli}, \citenamefont
  {Renaux-Petel},\ and\ \citenamefont {Witkowski}}]{Fumagalli:2020nvq}%
  \BibitemOpen
  \bibfield  {author} {\bibinfo {author} {\bibfnamefont {J.}~\bibnamefont
  {Fumagalli}}, \bibinfo {author} {\bibfnamefont {S.}~\bibnamefont
  {Renaux-Petel}}, \ and\ \bibinfo {author} {\bibfnamefont {L.~T.}\
  \bibnamefont {Witkowski}},\ }\href@noop {} {\  (\bibinfo {year}
  {2020}{\natexlab{a}})},\ \Eprint {http://arxiv.org/abs/2012.02761}
  {arXiv:2012.02761 [astro-ph.CO]} \BibitemShut {NoStop}%
\bibitem [{\citenamefont {Braglia}\ \emph
  {et~al.}(2020{\natexlab{a}})\citenamefont {Braglia}, \citenamefont {Hazra},
  \citenamefont {Finelli}, \citenamefont {Smoot}, \citenamefont {Sriramkumar},\
  and\ \citenamefont {Starobinsky}}]{Braglia:2020eai}%
  \BibitemOpen
  \bibfield  {author} {\bibinfo {author} {\bibfnamefont {M.}~\bibnamefont
  {Braglia}}, \bibinfo {author} {\bibfnamefont {D.~K.}\ \bibnamefont {Hazra}},
  \bibinfo {author} {\bibfnamefont {F.}~\bibnamefont {Finelli}}, \bibinfo
  {author} {\bibfnamefont {G.~F.}\ \bibnamefont {Smoot}}, \bibinfo {author}
  {\bibfnamefont {L.}~\bibnamefont {Sriramkumar}}, \ and\ \bibinfo {author}
  {\bibfnamefont {A.~A.}\ \bibnamefont {Starobinsky}},\ }\href {\doibase
  10.1088/1475-7516/2020/08/001} {\bibfield  {journal} {\bibinfo  {journal}
  {JCAP}\ }\textbf {\bibinfo {volume} {08}},\ \bibinfo {pages} {001} (\bibinfo
  {year} {2020}{\natexlab{a}})},\ \Eprint {http://arxiv.org/abs/2005.02895}
  {arXiv:2005.02895 [astro-ph.CO]} \BibitemShut {NoStop}%
\bibitem [{\citenamefont {Atal}\ and\ \citenamefont
  {Dom\`enech}(2021)}]{Atal:2021jyo}%
  \BibitemOpen
  \bibfield  {author} {\bibinfo {author} {\bibfnamefont {V.}~\bibnamefont
  {Atal}}\ and\ \bibinfo {author} {\bibfnamefont {G.}~\bibnamefont
  {Dom\`enech}},\ }\href {\doibase 10.1088/1475-7516/2021/06/001} {\bibfield
  {journal} {\bibinfo  {journal} {JCAP}\ }\textbf {\bibinfo {volume} {06}},\
  \bibinfo {pages} {001} (\bibinfo {year} {2021})},\ \Eprint
  {http://arxiv.org/abs/2103.01056} {arXiv:2103.01056 [astro-ph.CO]}
  \BibitemShut {NoStop}%
\bibitem [{\citenamefont {Braglia}\ \emph
  {et~al.}(2020{\natexlab{b}})\citenamefont {Braglia}, \citenamefont {Chen},\
  and\ \citenamefont {Hazra}}]{Braglia:2020taf}%
  \BibitemOpen
  \bibfield  {author} {\bibinfo {author} {\bibfnamefont {M.}~\bibnamefont
  {Braglia}}, \bibinfo {author} {\bibfnamefont {X.}~\bibnamefont {Chen}}, \
  and\ \bibinfo {author} {\bibfnamefont {D.~K.}\ \bibnamefont {Hazra}},\
  }\href@noop {} {\  (\bibinfo {year} {2020}{\natexlab{b}})},\ \Eprint
  {http://arxiv.org/abs/2012.05821} {arXiv:2012.05821 [astro-ph.CO]}
  \BibitemShut {NoStop}%
\bibitem [{\citenamefont {Fumagalli}\ \emph
  {et~al.}(2021{\natexlab{a}})\citenamefont {Fumagalli}, \citenamefont
  {Renaux-Petel},\ and\ \citenamefont {Witkowski}}]{Fumagalli:2021cel}%
  \BibitemOpen
  \bibfield  {author} {\bibinfo {author} {\bibfnamefont {J.}~\bibnamefont
  {Fumagalli}}, \bibinfo {author} {\bibfnamefont {S.}~\bibnamefont
  {Renaux-Petel}}, \ and\ \bibinfo {author} {\bibfnamefont {L.~T.}\
  \bibnamefont {Witkowski}},\ }\href@noop {} {\  (\bibinfo {year}
  {2021}{\natexlab{a}})},\ \Eprint {http://arxiv.org/abs/2105.06481}
  {arXiv:2105.06481 [astro-ph.CO]} \BibitemShut {NoStop}%
\bibitem [{\citenamefont {Bastero-Gil}\ and\ \citenamefont
  {D\'\i{}az-Blanco}(2021)}]{Bastero-Gil:2021fac}%
  \BibitemOpen
  \bibfield  {author} {\bibinfo {author} {\bibfnamefont {M.}~\bibnamefont
  {Bastero-Gil}}\ and\ \bibinfo {author} {\bibfnamefont {M.~S.}\ \bibnamefont
  {D\'\i{}az-Blanco}},\ }\href@noop {} {\  (\bibinfo {year} {2021})},\ \Eprint
  {http://arxiv.org/abs/2105.08045} {arXiv:2105.08045 [hep-ph]} \BibitemShut
  {NoStop}%
\bibitem [{\citenamefont {Fumagalli}\ \emph {et~al.}(2022)\citenamefont
  {Fumagalli}, \citenamefont {Palma}, \citenamefont {Renaux-Petel},
  \citenamefont {Sypsas}, \citenamefont {Witkowski},\ and\ \citenamefont
  {Zenteno}}]{Fumagalli:2021mpc}%
  \BibitemOpen
  \bibfield  {author} {\bibinfo {author} {\bibfnamefont {J.}~\bibnamefont
  {Fumagalli}}, \bibinfo {author} {\bibfnamefont {G.~A.}\ \bibnamefont
  {Palma}}, \bibinfo {author} {\bibfnamefont {S.}~\bibnamefont {Renaux-Petel}},
  \bibinfo {author} {\bibfnamefont {S.}~\bibnamefont {Sypsas}}, \bibinfo
  {author} {\bibfnamefont {L.~T.}\ \bibnamefont {Witkowski}}, \ and\ \bibinfo
  {author} {\bibfnamefont {C.}~\bibnamefont {Zenteno}},\ }\href {\doibase
  10.1007/JHEP03(2022)196} {\bibfield  {journal} {\bibinfo  {journal} {JHEP}\
  }\textbf {\bibinfo {volume} {03}},\ \bibinfo {pages} {196} (\bibinfo {year}
  {2022})},\ \Eprint {http://arxiv.org/abs/2111.14664} {arXiv:2111.14664
  [astro-ph.CO]} \BibitemShut {NoStop}%
\bibitem [{\citenamefont {Fumagalli}\ \emph
  {et~al.}(2021{\natexlab{b}})\citenamefont {Fumagalli}, \citenamefont
  {Pieroni}, \citenamefont {Renaux-Petel},\ and\ \citenamefont
  {Witkowski}}]{Fumagalli:2021dtd}%
  \BibitemOpen
  \bibfield  {author} {\bibinfo {author} {\bibfnamefont {J.}~\bibnamefont
  {Fumagalli}}, \bibinfo {author} {\bibfnamefont {M.}~\bibnamefont {Pieroni}},
  \bibinfo {author} {\bibfnamefont {S.}~\bibnamefont {Renaux-Petel}}, \ and\
  \bibinfo {author} {\bibfnamefont {L.~T.}\ \bibnamefont {Witkowski}},\
  }\href@noop {} {\  (\bibinfo {year} {2021}{\natexlab{b}})},\ \Eprint
  {http://arxiv.org/abs/2112.06903} {arXiv:2112.06903 [astro-ph.CO]}
  \BibitemShut {NoStop}%
\bibitem [{\citenamefont {Saikawa}\ and\ \citenamefont
  {Shirai}(2018)}]{Saikawa:2018rcs}%
  \BibitemOpen
  \bibfield  {author} {\bibinfo {author} {\bibfnamefont {K.}~\bibnamefont
  {Saikawa}}\ and\ \bibinfo {author} {\bibfnamefont {S.}~\bibnamefont
  {Shirai}},\ }\href {\doibase 10.1088/1475-7516/2018/05/035} {\bibfield
  {journal} {\bibinfo  {journal} {JCAP}\ }\textbf {\bibinfo {volume} {05}},\
  \bibinfo {pages} {035} (\bibinfo {year} {2018})},\ \Eprint
  {http://arxiv.org/abs/1803.01038} {arXiv:1803.01038 [hep-ph]} \BibitemShut
  {NoStop}%
\bibitem [{\citenamefont {Pi}\ and\ \citenamefont {Sasaki}(2021)}]{Pi:2021dft}%
  \BibitemOpen
  \bibfield  {author} {\bibinfo {author} {\bibfnamefont {S.}~\bibnamefont
  {Pi}}\ and\ \bibinfo {author} {\bibfnamefont {M.}~\bibnamefont {Sasaki}},\
  }\href@noop {} {\  (\bibinfo {year} {2021})},\ \Eprint
  {http://arxiv.org/abs/2112.12680} {arXiv:2112.12680 [astro-ph.CO]}
  \BibitemShut {NoStop}%
\bibitem [{\citenamefont {Fujita}\ \emph {et~al.}(2022)\citenamefont {Fujita},
  \citenamefont {Nakatsuka}, \citenamefont {Obata},\ and\ \citenamefont
  {Young}}]{Fujita:2022ait}%
  \BibitemOpen
  \bibfield  {author} {\bibinfo {author} {\bibfnamefont {T.}~\bibnamefont
  {Fujita}}, \bibinfo {author} {\bibfnamefont {H.}~\bibnamefont {Nakatsuka}},
  \bibinfo {author} {\bibfnamefont {I.}~\bibnamefont {Obata}}, \ and\ \bibinfo
  {author} {\bibfnamefont {S.}~\bibnamefont {Young}},\ }\href@noop {} {\
  (\bibinfo {year} {2022})},\ \Eprint {http://arxiv.org/abs/2202.02401}
  {arXiv:2202.02401 [astro-ph.CO]} \BibitemShut {NoStop}%
\bibitem [{\citenamefont {Lozanov}\ and\ \citenamefont
  {Takhistov}(2022)}]{Lozanov:2022yoy}%
  \BibitemOpen
  \bibfield  {author} {\bibinfo {author} {\bibfnamefont {K.~D.}\ \bibnamefont
  {Lozanov}}\ and\ \bibinfo {author} {\bibfnamefont {V.}~\bibnamefont
  {Takhistov}},\ }\href@noop {} {\  (\bibinfo {year} {2022})},\ \Eprint
  {http://arxiv.org/abs/2204.07152} {arXiv:2204.07152 [astro-ph.CO]}
  \BibitemShut {NoStop}%
\bibitem [{\citenamefont {Inomata}(2022)}]{Inomata:2022ydj}%
  \BibitemOpen
  \bibfield  {author} {\bibinfo {author} {\bibfnamefont {K.}~\bibnamefont
  {Inomata}},\ }\href@noop {} {\  (\bibinfo {year} {2022})},\ \Eprint
  {http://arxiv.org/abs/2203.04974} {arXiv:2203.04974 [astro-ph.CO]}
  \BibitemShut {NoStop}%
\bibitem [{\citenamefont {Balaji}\ \emph {et~al.}(2022)\citenamefont {Balaji},
  \citenamefont {Silk},\ and\ \citenamefont {Wu}}]{Balaji:2022rsy}%
  \BibitemOpen
  \bibfield  {author} {\bibinfo {author} {\bibfnamefont {S.}~\bibnamefont
  {Balaji}}, \bibinfo {author} {\bibfnamefont {J.}~\bibnamefont {Silk}}, \ and\
  \bibinfo {author} {\bibfnamefont {Y.-P.}\ \bibnamefont {Wu}},\ }\href@noop {}
  {\  (\bibinfo {year} {2022})},\ \Eprint {http://arxiv.org/abs/2202.00700}
  {arXiv:2202.00700 [astro-ph.CO]} \BibitemShut {NoStop}%
\bibitem [{\citenamefont {Dom\`enech}\ \emph {et~al.}(2022)\citenamefont
  {Dom\`enech}, \citenamefont {Passaglia},\ and\ \citenamefont
  {Renaux-Petel}}]{Domenech:2021and}%
  \BibitemOpen
  \bibfield  {author} {\bibinfo {author} {\bibfnamefont {G.}~\bibnamefont
  {Dom\`enech}}, \bibinfo {author} {\bibfnamefont {S.}~\bibnamefont
  {Passaglia}}, \ and\ \bibinfo {author} {\bibfnamefont {S.}~\bibnamefont
  {Renaux-Petel}},\ }\href {\doibase 10.1088/1475-7516/2022/03/023} {\bibfield
  {journal} {\bibinfo  {journal} {JCAP}\ }\textbf {\bibinfo {volume} {03}},\
  \bibinfo {pages} {023} (\bibinfo {year} {2022})},\ \Eprint
  {http://arxiv.org/abs/2112.10163} {arXiv:2112.10163 [astro-ph.CO]}
  \BibitemShut {NoStop}%
\bibitem [{\citenamefont {Addazi}\ \emph {et~al.}(2022)\citenamefont {Addazi},
  \citenamefont {Capozziello},\ and\ \citenamefont {Gan}}]{Addazi:2022ukh}%
  \BibitemOpen
  \bibfield  {author} {\bibinfo {author} {\bibfnamefont {A.}~\bibnamefont
  {Addazi}}, \bibinfo {author} {\bibfnamefont {S.}~\bibnamefont {Capozziello}},
  \ and\ \bibinfo {author} {\bibfnamefont {Q.}~\bibnamefont {Gan}},\
  }\href@noop {} {\  (\bibinfo {year} {2022})},\ \Eprint
  {http://arxiv.org/abs/2204.07668} {arXiv:2204.07668 [astro-ph.CO]}
  \BibitemShut {NoStop}%
\bibitem [{\citenamefont {Balaji}\ \emph {et~al.}(2021)\citenamefont {Balaji},
  \citenamefont {Spannowsky},\ and\ \citenamefont {Tamarit}}]{Balaji:2020yrx}%
  \BibitemOpen
  \bibfield  {author} {\bibinfo {author} {\bibfnamefont {S.}~\bibnamefont
  {Balaji}}, \bibinfo {author} {\bibfnamefont {M.}~\bibnamefont {Spannowsky}},
  \ and\ \bibinfo {author} {\bibfnamefont {C.}~\bibnamefont {Tamarit}},\ }\href
  {\doibase 10.1088/1475-7516/2021/03/051} {\bibfield  {journal} {\bibinfo
  {journal} {JCAP}\ }\textbf {\bibinfo {volume} {03}},\ \bibinfo {pages} {051}
  (\bibinfo {year} {2021})},\ \Eprint {http://arxiv.org/abs/2010.08013}
  {arXiv:2010.08013 [hep-ph]} \BibitemShut {NoStop}%
\bibitem [{\citenamefont {Balaji}\ and\ \citenamefont
  {Kobakhidze}(2018)}]{Balaji:2018qyo}%
  \BibitemOpen
  \bibfield  {author} {\bibinfo {author} {\bibfnamefont {S.}~\bibnamefont
  {Balaji}}\ and\ \bibinfo {author} {\bibfnamefont {A.}~\bibnamefont
  {Kobakhidze}},\ }\href@noop {} {\  (\bibinfo {year} {2018})},\ \Eprint
  {http://arxiv.org/abs/1812.10914} {arXiv:1812.10914 [hep-ph]} \BibitemShut
  {NoStop}%
\bibitem [{\citenamefont {Assadullahi}\ and\ \citenamefont
  {Wands}(2009)}]{Assadullahi:2009nf}%
  \BibitemOpen
  \bibfield  {author} {\bibinfo {author} {\bibfnamefont {H.}~\bibnamefont
  {Assadullahi}}\ and\ \bibinfo {author} {\bibfnamefont {D.}~\bibnamefont
  {Wands}},\ }\href {\doibase 10.1103/PhysRevD.79.083511} {\bibfield  {journal}
  {\bibinfo  {journal} {Phys. Rev. D}\ }\textbf {\bibinfo {volume} {79}},\
  \bibinfo {pages} {083511} (\bibinfo {year} {2009})},\ \Eprint
  {http://arxiv.org/abs/0901.0989} {arXiv:0901.0989 [astro-ph.CO]} \BibitemShut
  {NoStop}%
\bibitem [{\citenamefont {Inomata}\ \emph
  {et~al.}(2019{\natexlab{a}})\citenamefont {Inomata}, \citenamefont {Kohri},
  \citenamefont {Nakama},\ and\ \citenamefont {Terada}}]{Inomata:2019zqy}%
  \BibitemOpen
  \bibfield  {author} {\bibinfo {author} {\bibfnamefont {K.}~\bibnamefont
  {Inomata}}, \bibinfo {author} {\bibfnamefont {K.}~\bibnamefont {Kohri}},
  \bibinfo {author} {\bibfnamefont {T.}~\bibnamefont {Nakama}}, \ and\ \bibinfo
  {author} {\bibfnamefont {T.}~\bibnamefont {Terada}},\ }\href {\doibase
  10.1088/1475-7516/2019/10/071} {\bibfield  {journal} {\bibinfo  {journal}
  {JCAP}\ }\textbf {\bibinfo {volume} {10}},\ \bibinfo {pages} {071} (\bibinfo
  {year} {2019}{\natexlab{a}})},\ \Eprint {http://arxiv.org/abs/1904.12878}
  {arXiv:1904.12878 [astro-ph.CO]} \BibitemShut {NoStop}%
\bibitem [{\citenamefont {Inomata}\ \emph
  {et~al.}(2019{\natexlab{b}})\citenamefont {Inomata}, \citenamefont {Kohri},
  \citenamefont {Nakama},\ and\ \citenamefont {Terada}}]{Inomata:2019ivs}%
  \BibitemOpen
  \bibfield  {author} {\bibinfo {author} {\bibfnamefont {K.}~\bibnamefont
  {Inomata}}, \bibinfo {author} {\bibfnamefont {K.}~\bibnamefont {Kohri}},
  \bibinfo {author} {\bibfnamefont {T.}~\bibnamefont {Nakama}}, \ and\ \bibinfo
  {author} {\bibfnamefont {T.}~\bibnamefont {Terada}},\ }\href {\doibase
  10.1103/PhysRevD.100.043532} {\bibfield  {journal} {\bibinfo  {journal}
  {Phys. Rev. D}\ }\textbf {\bibinfo {volume} {100}},\ \bibinfo {pages}
  {043532} (\bibinfo {year} {2019}{\natexlab{b}})},\ \Eprint
  {http://arxiv.org/abs/1904.12879} {arXiv:1904.12879 [astro-ph.CO]}
  \BibitemShut {NoStop}%
\bibitem [{\citenamefont {Inomata}\ \emph {et~al.}(2020)\citenamefont
  {Inomata}, \citenamefont {Kawasaki}, \citenamefont {Mukaida}, \citenamefont
  {Terada},\ and\ \citenamefont {Yanagida}}]{Inomata:2020lmk}%
  \BibitemOpen
  \bibfield  {author} {\bibinfo {author} {\bibfnamefont {K.}~\bibnamefont
  {Inomata}}, \bibinfo {author} {\bibfnamefont {M.}~\bibnamefont {Kawasaki}},
  \bibinfo {author} {\bibfnamefont {K.}~\bibnamefont {Mukaida}}, \bibinfo
  {author} {\bibfnamefont {T.}~\bibnamefont {Terada}}, \ and\ \bibinfo {author}
  {\bibfnamefont {T.~T.}\ \bibnamefont {Yanagida}},\ }\href {\doibase
  10.1103/PhysRevD.101.123533} {\bibfield  {journal} {\bibinfo  {journal}
  {Phys. Rev. D}\ }\textbf {\bibinfo {volume} {101}},\ \bibinfo {pages}
  {123533} (\bibinfo {year} {2020})},\ \Eprint
  {http://arxiv.org/abs/2003.10455} {arXiv:2003.10455 [astro-ph.CO]}
  \BibitemShut {NoStop}%
\bibitem [{\citenamefont {Papanikolaou}\ \emph {et~al.}(2020)\citenamefont
  {Papanikolaou}, \citenamefont {Vennin},\ and\ \citenamefont
  {Langlois}}]{Papanikolaou:2020qtd}%
  \BibitemOpen
  \bibfield  {author} {\bibinfo {author} {\bibfnamefont {T.}~\bibnamefont
  {Papanikolaou}}, \bibinfo {author} {\bibfnamefont {V.}~\bibnamefont
  {Vennin}}, \ and\ \bibinfo {author} {\bibfnamefont {D.}~\bibnamefont
  {Langlois}},\ }\href@noop {} {\  (\bibinfo {year} {2020})},\ \Eprint
  {http://arxiv.org/abs/2010.11573} {arXiv:2010.11573 [astro-ph.CO]}
  \BibitemShut {NoStop}%
\bibitem [{\citenamefont {Dom\`enech}\ \emph
  {et~al.}(2021{\natexlab{a}})\citenamefont {Dom\`enech}, \citenamefont {Lin},\
  and\ \citenamefont {Sasaki}}]{Domenech:2020ssp}%
  \BibitemOpen
  \bibfield  {author} {\bibinfo {author} {\bibfnamefont {G.}~\bibnamefont
  {Dom\`enech}}, \bibinfo {author} {\bibfnamefont {C.}~\bibnamefont {Lin}}, \
  and\ \bibinfo {author} {\bibfnamefont {M.}~\bibnamefont {Sasaki}},\ }\href
  {\doibase 10.1088/1475-7516/2021/11/E01} {\bibfield  {journal} {\bibinfo
  {journal} {JCAP}\ }\textbf {\bibinfo {volume} {11}},\ \bibinfo {pages} {E01}
  (\bibinfo {year} {2021}{\natexlab{a}})},\ \Eprint
  {http://arxiv.org/abs/2012.08151} {arXiv:2012.08151 [gr-qc]} \BibitemShut
  {NoStop}%
\bibitem [{\citenamefont {Dom\`enech}\ \emph
  {et~al.}(2021{\natexlab{b}})\citenamefont {Dom\`enech}, \citenamefont
  {Takhistov},\ and\ \citenamefont {Sasaki}}]{Domenech:2021wkk}%
  \BibitemOpen
  \bibfield  {author} {\bibinfo {author} {\bibfnamefont {G.}~\bibnamefont
  {Dom\`enech}}, \bibinfo {author} {\bibfnamefont {V.}~\bibnamefont
  {Takhistov}}, \ and\ \bibinfo {author} {\bibfnamefont {M.}~\bibnamefont
  {Sasaki}},\ }\href {\doibase 10.1016/j.physletb.2021.136722} {\bibfield
  {journal} {\bibinfo  {journal} {Phys. Lett. B}\ }\textbf {\bibinfo {volume}
  {823}},\ \bibinfo {pages} {136722} (\bibinfo {year} {2021}{\natexlab{b}})},\
  \Eprint {http://arxiv.org/abs/2105.06816} {arXiv:2105.06816 [astro-ph.CO]}
  \BibitemShut {NoStop}%
\bibitem [{\citenamefont {Dalianis}\ and\ \citenamefont
  {Kouvaris}(2020)}]{Dalianis:2020gup}%
  \BibitemOpen
  \bibfield  {author} {\bibinfo {author} {\bibfnamefont {I.}~\bibnamefont
  {Dalianis}}\ and\ \bibinfo {author} {\bibfnamefont {C.}~\bibnamefont
  {Kouvaris}},\ }\href@noop {} {\  (\bibinfo {year} {2020})},\ \Eprint
  {http://arxiv.org/abs/2012.09255} {arXiv:2012.09255 [astro-ph.CO]}
  \BibitemShut {NoStop}%
\bibitem [{\citenamefont {Hajkarim}\ and\ \citenamefont
  {Schaffner-Bielich}(2020)}]{Hajkarim:2019nbx}%
  \BibitemOpen
  \bibfield  {author} {\bibinfo {author} {\bibfnamefont {F.}~\bibnamefont
  {Hajkarim}}\ and\ \bibinfo {author} {\bibfnamefont {J.}~\bibnamefont
  {Schaffner-Bielich}},\ }\href {\doibase 10.1103/PhysRevD.101.043522}
  {\bibfield  {journal} {\bibinfo  {journal} {Phys. Rev. D}\ }\textbf {\bibinfo
  {volume} {101}},\ \bibinfo {pages} {043522} (\bibinfo {year} {2020})},\
  \Eprint {http://arxiv.org/abs/1910.12357} {arXiv:1910.12357 [hep-ph]}
  \BibitemShut {NoStop}%
\bibitem [{\citenamefont {Bhattacharya}\ \emph {et~al.}(2020)\citenamefont
  {Bhattacharya}, \citenamefont {Mohanty},\ and\ \citenamefont
  {Parashari}}]{Bhattacharya:2019bvk}%
  \BibitemOpen
  \bibfield  {author} {\bibinfo {author} {\bibfnamefont {S.}~\bibnamefont
  {Bhattacharya}}, \bibinfo {author} {\bibfnamefont {S.}~\bibnamefont
  {Mohanty}}, \ and\ \bibinfo {author} {\bibfnamefont {P.}~\bibnamefont
  {Parashari}},\ }\href {\doibase 10.1103/PhysRevD.102.043522} {\bibfield
  {journal} {\bibinfo  {journal} {Phys. Rev. D}\ }\textbf {\bibinfo {volume}
  {102}},\ \bibinfo {pages} {043522} (\bibinfo {year} {2020})},\ \Eprint
  {http://arxiv.org/abs/1912.01653} {arXiv:1912.01653 [astro-ph.CO]}
  \BibitemShut {NoStop}%
\bibitem [{\citenamefont {Dom\`enech}(2020)}]{Domenech:2019quo}%
  \BibitemOpen
  \bibfield  {author} {\bibinfo {author} {\bibfnamefont {G.}~\bibnamefont
  {Dom\`enech}},\ }\href {\doibase 10.1142/S0218271820500285} {\bibfield
  {journal} {\bibinfo  {journal} {Int. J. Mod. Phys. D}\ }\textbf {\bibinfo
  {volume} {29}},\ \bibinfo {pages} {2050028} (\bibinfo {year} {2020})},\
  \Eprint {http://arxiv.org/abs/1912.05583} {arXiv:1912.05583 [gr-qc]}
  \BibitemShut {NoStop}%
\bibitem [{\citenamefont {Dom\`enech}\ \emph {et~al.}(2020)\citenamefont
  {Dom\`enech}, \citenamefont {Pi},\ and\ \citenamefont
  {Sasaki}}]{Domenech:2020kqm}%
  \BibitemOpen
  \bibfield  {author} {\bibinfo {author} {\bibfnamefont {G.}~\bibnamefont
  {Dom\`enech}}, \bibinfo {author} {\bibfnamefont {S.}~\bibnamefont {Pi}}, \
  and\ \bibinfo {author} {\bibfnamefont {M.}~\bibnamefont {Sasaki}},\ }\href
  {\doibase 10.1088/1475-7516/2020/08/017} {\bibfield  {journal} {\bibinfo
  {journal} {JCAP}\ }\textbf {\bibinfo {volume} {08}},\ \bibinfo {pages} {017}
  (\bibinfo {year} {2020})},\ \Eprint {http://arxiv.org/abs/2005.12314}
  {arXiv:2005.12314 [gr-qc]} \BibitemShut {NoStop}%
\bibitem [{\citenamefont {Dalianis}\ and\ \citenamefont
  {Kritos}(2021)}]{Dalianis:2020cla}%
  \BibitemOpen
  \bibfield  {author} {\bibinfo {author} {\bibfnamefont {I.}~\bibnamefont
  {Dalianis}}\ and\ \bibinfo {author} {\bibfnamefont {K.}~\bibnamefont
  {Kritos}},\ }\href {\doibase 10.1103/PhysRevD.103.023505} {\bibfield
  {journal} {\bibinfo  {journal} {Phys. Rev. D}\ }\textbf {\bibinfo {volume}
  {103}},\ \bibinfo {pages} {023505} (\bibinfo {year} {2021})},\ \Eprint
  {http://arxiv.org/abs/2007.07915} {arXiv:2007.07915 [astro-ph.CO]}
  \BibitemShut {NoStop}%
\bibitem [{\citenamefont {Abe}\ \emph {et~al.}(2020)\citenamefont {Abe},
  \citenamefont {Tada},\ and\ \citenamefont {Ueda}}]{Abe:2020sqb}%
  \BibitemOpen
  \bibfield  {author} {\bibinfo {author} {\bibfnamefont {K.~T.}\ \bibnamefont
  {Abe}}, \bibinfo {author} {\bibfnamefont {Y.}~\bibnamefont {Tada}}, \ and\
  \bibinfo {author} {\bibfnamefont {I.}~\bibnamefont {Ueda}},\ }\href@noop {}
  {\  (\bibinfo {year} {2020})},\ \Eprint {http://arxiv.org/abs/2010.06193}
  {arXiv:2010.06193 [astro-ph.CO]} \BibitemShut {NoStop}%
\bibitem [{\citenamefont {Witkowski}\ \emph {et~al.}(2021)\citenamefont
  {Witkowski}, \citenamefont {Dom\`enech}, \citenamefont {Fumagalli},\ and\
  \citenamefont {Renaux-Petel}}]{Witkowski:2021raz}%
  \BibitemOpen
  \bibfield  {author} {\bibinfo {author} {\bibfnamefont {L.~T.}\ \bibnamefont
  {Witkowski}}, \bibinfo {author} {\bibfnamefont {G.}~\bibnamefont
  {Dom\`enech}}, \bibinfo {author} {\bibfnamefont {J.}~\bibnamefont
  {Fumagalli}}, \ and\ \bibinfo {author} {\bibfnamefont {S.}~\bibnamefont
  {Renaux-Petel}},\ }\href@noop {} {\  (\bibinfo {year} {2021})},\ \Eprint
  {http://arxiv.org/abs/2110.09480} {arXiv:2110.09480 [astro-ph.CO]}
  \BibitemShut {NoStop}%
\bibitem [{\citenamefont {Assadullahi}\ and\ \citenamefont
  {Wands}(2010)}]{Assadullahi:2009jc}%
  \BibitemOpen
  \bibfield  {author} {\bibinfo {author} {\bibfnamefont {H.}~\bibnamefont
  {Assadullahi}}\ and\ \bibinfo {author} {\bibfnamefont {D.}~\bibnamefont
  {Wands}},\ }\href {\doibase 10.1103/PhysRevD.81.023527} {\bibfield  {journal}
  {\bibinfo  {journal} {Phys. Rev. D}\ }\textbf {\bibinfo {volume} {81}},\
  \bibinfo {pages} {023527} (\bibinfo {year} {2010})},\ \Eprint
  {http://arxiv.org/abs/0907.4073} {arXiv:0907.4073 [astro-ph.CO]} \BibitemShut
  {NoStop}%
\bibitem [{\citenamefont {Bugaev}\ and\ \citenamefont
  {Klimai}(2010{\natexlab{a}})}]{Bugaev:2009zh}%
  \BibitemOpen
  \bibfield  {author} {\bibinfo {author} {\bibfnamefont {E.}~\bibnamefont
  {Bugaev}}\ and\ \bibinfo {author} {\bibfnamefont {P.}~\bibnamefont
  {Klimai}},\ }\href {\doibase 10.1103/PhysRevD.81.023517} {\bibfield
  {journal} {\bibinfo  {journal} {Phys. Rev. D}\ }\textbf {\bibinfo {volume}
  {81}},\ \bibinfo {pages} {023517} (\bibinfo {year} {2010}{\natexlab{a}})},\
  \Eprint {http://arxiv.org/abs/0908.0664} {arXiv:0908.0664 [astro-ph.CO]}
  \BibitemShut {NoStop}%
\bibitem [{\citenamefont {Bugaev}\ and\ \citenamefont
  {Klimai}(2010{\natexlab{b}})}]{Bugaev:2009kq}%
  \BibitemOpen
  \bibfield  {author} {\bibinfo {author} {\bibfnamefont {E.~V.}\ \bibnamefont
  {Bugaev}}\ and\ \bibinfo {author} {\bibfnamefont {P.~A.}\ \bibnamefont
  {Klimai}},\ }\href {\doibase 10.1134/S0021364010010017} {\bibfield  {journal}
  {\bibinfo  {journal} {JETP Lett.}\ }\textbf {\bibinfo {volume} {91}},\
  \bibinfo {pages} {1} (\bibinfo {year} {2010}{\natexlab{b}})},\ \Eprint
  {http://arxiv.org/abs/0911.0611} {arXiv:0911.0611 [astro-ph.CO]} \BibitemShut
  {NoStop}%
\bibitem [{\citenamefont {Bugaev}\ and\ \citenamefont
  {Klimai}(2011)}]{Bugaev:2010bb}%
  \BibitemOpen
  \bibfield  {author} {\bibinfo {author} {\bibfnamefont {E.}~\bibnamefont
  {Bugaev}}\ and\ \bibinfo {author} {\bibfnamefont {P.}~\bibnamefont
  {Klimai}},\ }\href {\doibase 10.1103/PhysRevD.83.083521} {\bibfield
  {journal} {\bibinfo  {journal} {Phys. Rev. D}\ }\textbf {\bibinfo {volume}
  {83}},\ \bibinfo {pages} {083521} (\bibinfo {year} {2011})},\ \Eprint
  {http://arxiv.org/abs/1012.4697} {arXiv:1012.4697 [astro-ph.CO]} \BibitemShut
  {NoStop}%
\bibitem [{\citenamefont {Inomata}\ and\ \citenamefont
  {Nakama}(2019)}]{Inomata:2018epa}%
  \BibitemOpen
  \bibfield  {author} {\bibinfo {author} {\bibfnamefont {K.}~\bibnamefont
  {Inomata}}\ and\ \bibinfo {author} {\bibfnamefont {T.}~\bibnamefont
  {Nakama}},\ }\href {\doibase 10.1103/PhysRevD.99.043511} {\bibfield
  {journal} {\bibinfo  {journal} {Phys. Rev. D}\ }\textbf {\bibinfo {volume}
  {99}},\ \bibinfo {pages} {043511} (\bibinfo {year} {2019})},\ \Eprint
  {http://arxiv.org/abs/1812.00674} {arXiv:1812.00674 [astro-ph.CO]}
  \BibitemShut {NoStop}%
\bibitem [{\citenamefont {Alabidi}\ \emph {et~al.}(2013)\citenamefont
  {Alabidi}, \citenamefont {Kohri}, \citenamefont {Sasaki},\ and\ \citenamefont
  {Sendouda}}]{Alabidi:2013lya}%
  \BibitemOpen
  \bibfield  {author} {\bibinfo {author} {\bibfnamefont {L.}~\bibnamefont
  {Alabidi}}, \bibinfo {author} {\bibfnamefont {K.}~\bibnamefont {Kohri}},
  \bibinfo {author} {\bibfnamefont {M.}~\bibnamefont {Sasaki}}, \ and\ \bibinfo
  {author} {\bibfnamefont {Y.}~\bibnamefont {Sendouda}},\ }\href {\doibase
  10.1088/1475-7516/2013/05/033} {\bibfield  {journal} {\bibinfo  {journal}
  {JCAP}\ }\textbf {\bibinfo {volume} {05}},\ \bibinfo {pages} {033} (\bibinfo
  {year} {2013})},\ \Eprint {http://arxiv.org/abs/1303.4519} {arXiv:1303.4519
  [astro-ph.CO]} \BibitemShut {NoStop}%
\bibitem [{\citenamefont {Zel'dovich}(1967)}]{Zeldovich:1967lct}%
  \BibitemOpen
  \bibfield  {author} {\bibinfo {author} {\bibfnamefont {I.~D.}\ \bibnamefont
  {Zel'dovich}, \bibfnamefont {Ya.B.;~Novikov}},\ }\href@noop {} {\bibfield
  {journal} {\bibinfo  {journal} {Soviet Astron. AJ (Engl. Transl. ),}\
  }\textbf {\bibinfo {volume} {10}},\ \bibinfo {pages} {602} (\bibinfo {year}
  {1967})}\BibitemShut {NoStop}%
\bibitem [{\citenamefont {Hawking}(1971)}]{Hawking:1971ei}%
  \BibitemOpen
  \bibfield  {author} {\bibinfo {author} {\bibfnamefont {S.}~\bibnamefont
  {Hawking}},\ }\href@noop {} {\bibfield  {journal} {\bibinfo  {journal} {Mon.
  Not. Roy. Astron. Soc.}\ }\textbf {\bibinfo {volume} {152}},\ \bibinfo
  {pages} {75} (\bibinfo {year} {1971})}\BibitemShut {NoStop}%
\bibitem [{\citenamefont {Carr}\ and\ \citenamefont
  {Hawking}(1974)}]{Carr:1974nx}%
  \BibitemOpen
  \bibfield  {author} {\bibinfo {author} {\bibfnamefont {B.~J.}\ \bibnamefont
  {Carr}}\ and\ \bibinfo {author} {\bibfnamefont {S.}~\bibnamefont {Hawking}},\
  }\href@noop {} {\bibfield  {journal} {\bibinfo  {journal} {Mon. Not. Roy.
  Astron. Soc.}\ }\textbf {\bibinfo {volume} {168}},\ \bibinfo {pages} {399}
  (\bibinfo {year} {1974})}\BibitemShut {NoStop}%
\bibitem [{\citenamefont {Meszaros}(1974)}]{Meszaros:1974tb}%
  \BibitemOpen
  \bibfield  {author} {\bibinfo {author} {\bibfnamefont {P.}~\bibnamefont
  {Meszaros}},\ }\href@noop {} {\bibfield  {journal} {\bibinfo  {journal}
  {Astron. Astrophys.}\ }\textbf {\bibinfo {volume} {37}},\ \bibinfo {pages}
  {225} (\bibinfo {year} {1974})}\BibitemShut {NoStop}%
\bibitem [{\citenamefont {Carr}(1975)}]{Carr:1975qj}%
  \BibitemOpen
  \bibfield  {author} {\bibinfo {author} {\bibfnamefont {B.~J.}\ \bibnamefont
  {Carr}},\ }\href {\doibase 10.1086/153853} {\bibfield  {journal} {\bibinfo
  {journal} {Astrophys. J.}\ }\textbf {\bibinfo {volume} {201}},\ \bibinfo
  {pages} {1} (\bibinfo {year} {1975})}\BibitemShut {NoStop}%
\bibitem [{\citenamefont {Khlopov}\ \emph {et~al.}(1985)\citenamefont
  {Khlopov}, \citenamefont {Malomed},\ and\ \citenamefont
  {Zeldovich}}]{Khlopov:1985jw}%
  \BibitemOpen
  \bibfield  {author} {\bibinfo {author} {\bibfnamefont {M.}~\bibnamefont
  {Khlopov}}, \bibinfo {author} {\bibfnamefont {B.}~\bibnamefont {Malomed}}, \
  and\ \bibinfo {author} {\bibfnamefont {I.}~\bibnamefont {Zeldovich}},\
  }\href@noop {} {\bibfield  {journal} {\bibinfo  {journal} {Mon. Not. Roy.
  Astron. Soc.}\ }\textbf {\bibinfo {volume} {215}},\ \bibinfo {pages} {575}
  (\bibinfo {year} {1985})}\BibitemShut {NoStop}%
\bibitem [{\citenamefont {Niemeyer}\ and\ \citenamefont
  {Jedamzik}(1999)}]{Niemeyer:1999ak}%
  \BibitemOpen
  \bibfield  {author} {\bibinfo {author} {\bibfnamefont {J.~C.}\ \bibnamefont
  {Niemeyer}}\ and\ \bibinfo {author} {\bibfnamefont {K.}~\bibnamefont
  {Jedamzik}},\ }\href {\doibase 10.1103/PhysRevD.59.124013} {\bibfield
  {journal} {\bibinfo  {journal} {Phys. Rev. D}\ }\textbf {\bibinfo {volume}
  {59}},\ \bibinfo {pages} {124013} (\bibinfo {year} {1999})},\ \Eprint
  {http://arxiv.org/abs/astro-ph/9901292} {arXiv:astro-ph/9901292} \BibitemShut
  {NoStop}%
\bibitem [{\citenamefont {Inomata}\ \emph
  {et~al.}(2017{\natexlab{a}})\citenamefont {Inomata}, \citenamefont
  {Kawasaki}, \citenamefont {Mukaida}, \citenamefont {Tada},\ and\
  \citenamefont {Yanagida}}]{Inomata:2016rbd}%
  \BibitemOpen
  \bibfield  {author} {\bibinfo {author} {\bibfnamefont {K.}~\bibnamefont
  {Inomata}}, \bibinfo {author} {\bibfnamefont {M.}~\bibnamefont {Kawasaki}},
  \bibinfo {author} {\bibfnamefont {K.}~\bibnamefont {Mukaida}}, \bibinfo
  {author} {\bibfnamefont {Y.}~\bibnamefont {Tada}}, \ and\ \bibinfo {author}
  {\bibfnamefont {T.~T.}\ \bibnamefont {Yanagida}},\ }\href {\doibase
  10.1103/PhysRevD.95.123510} {\bibfield  {journal} {\bibinfo  {journal} {Phys.
  Rev. D}\ }\textbf {\bibinfo {volume} {95}},\ \bibinfo {pages} {123510}
  (\bibinfo {year} {2017}{\natexlab{a}})},\ \Eprint
  {http://arxiv.org/abs/1611.06130} {arXiv:1611.06130 [astro-ph.CO]}
  \BibitemShut {NoStop}%
\bibitem [{\citenamefont {Nakama}\ \emph {et~al.}(2017)\citenamefont {Nakama},
  \citenamefont {Silk},\ and\ \citenamefont {Kamionkowski}}]{Nakama:2016gzw}%
  \BibitemOpen
  \bibfield  {author} {\bibinfo {author} {\bibfnamefont {T.}~\bibnamefont
  {Nakama}}, \bibinfo {author} {\bibfnamefont {J.}~\bibnamefont {Silk}}, \ and\
  \bibinfo {author} {\bibfnamefont {M.}~\bibnamefont {Kamionkowski}},\ }\href
  {\doibase 10.1103/PhysRevD.95.043511} {\bibfield  {journal} {\bibinfo
  {journal} {Phys. Rev. D}\ }\textbf {\bibinfo {volume} {95}},\ \bibinfo
  {pages} {043511} (\bibinfo {year} {2017})},\ \Eprint
  {http://arxiv.org/abs/1612.06264} {arXiv:1612.06264 [astro-ph.CO]}
  \BibitemShut {NoStop}%
\bibitem [{\citenamefont {Ando}\ \emph
  {et~al.}(2018{\natexlab{b}})\citenamefont {Ando}, \citenamefont {Inomata},
  \citenamefont {Kawasaki}, \citenamefont {Mukaida},\ and\ \citenamefont
  {Yanagida}}]{Ando:2017veq}%
  \BibitemOpen
  \bibfield  {author} {\bibinfo {author} {\bibfnamefont {K.}~\bibnamefont
  {Ando}}, \bibinfo {author} {\bibfnamefont {K.}~\bibnamefont {Inomata}},
  \bibinfo {author} {\bibfnamefont {M.}~\bibnamefont {Kawasaki}}, \bibinfo
  {author} {\bibfnamefont {K.}~\bibnamefont {Mukaida}}, \ and\ \bibinfo
  {author} {\bibfnamefont {T.~T.}\ \bibnamefont {Yanagida}},\ }\href {\doibase
  10.1103/PhysRevD.97.123512} {\bibfield  {journal} {\bibinfo  {journal} {Phys.
  Rev. D}\ }\textbf {\bibinfo {volume} {97}},\ \bibinfo {pages} {123512}
  (\bibinfo {year} {2018}{\natexlab{b}})},\ \Eprint
  {http://arxiv.org/abs/1711.08956} {arXiv:1711.08956 [astro-ph.CO]}
  \BibitemShut {NoStop}%
\bibitem [{\citenamefont {Clesse}\ and\ \citenamefont
  {Garc\'\i{}a-Bellido}(2018)}]{Clesse:2017bsw}%
  \BibitemOpen
  \bibfield  {author} {\bibinfo {author} {\bibfnamefont {S.}~\bibnamefont
  {Clesse}}\ and\ \bibinfo {author} {\bibfnamefont {J.}~\bibnamefont
  {Garc\'\i{}a-Bellido}},\ }\href {\doibase 10.1016/j.dark.2018.08.004}
  {\bibfield  {journal} {\bibinfo  {journal} {Phys. Dark Univ.}\ }\textbf
  {\bibinfo {volume} {22}},\ \bibinfo {pages} {137} (\bibinfo {year} {2018})},\
  \Eprint {http://arxiv.org/abs/1711.10458} {arXiv:1711.10458 [astro-ph.CO]}
  \BibitemShut {NoStop}%
\bibitem [{\citenamefont {Kohri}\ and\ \citenamefont
  {Terada}(2018{\natexlab{a}})}]{Kohri:2018qtx}%
  \BibitemOpen
  \bibfield  {author} {\bibinfo {author} {\bibfnamefont {K.}~\bibnamefont
  {Kohri}}\ and\ \bibinfo {author} {\bibfnamefont {T.}~\bibnamefont {Terada}},\
  }\href {\doibase 10.1088/1361-6382/aaea18} {\bibfield  {journal} {\bibinfo
  {journal} {Class. Quant. Grav.}\ }\textbf {\bibinfo {volume} {35}},\ \bibinfo
  {pages} {235017} (\bibinfo {year} {2018}{\natexlab{a}})},\ \Eprint
  {http://arxiv.org/abs/1802.06785} {arXiv:1802.06785 [astro-ph.CO]}
  \BibitemShut {NoStop}%
\bibitem [{\citenamefont {Garc\'\i{}a-Bellido}\ \emph
  {et~al.}(2021)\citenamefont {Garc\'\i{}a-Bellido}, \citenamefont {Nu\~no
  Siles},\ and\ \citenamefont {Ruiz~Morales}}]{Garcia-Bellido:2020pwq}%
  \BibitemOpen
  \bibfield  {author} {\bibinfo {author} {\bibfnamefont {J.}~\bibnamefont
  {Garc\'\i{}a-Bellido}}, \bibinfo {author} {\bibfnamefont {J.~F.}\
  \bibnamefont {Nu\~no Siles}}, \ and\ \bibinfo {author} {\bibfnamefont
  {E.}~\bibnamefont {Ruiz~Morales}},\ }\href {\doibase
  10.1016/j.dark.2021.100791} {\bibfield  {journal} {\bibinfo  {journal} {Phys.
  Dark Univ.}\ }\textbf {\bibinfo {volume} {31}},\ \bibinfo {pages} {100791}
  (\bibinfo {year} {2021})},\ \Eprint {http://arxiv.org/abs/2010.13811}
  {arXiv:2010.13811 [astro-ph.CO]} \BibitemShut {NoStop}%
\bibitem [{\citenamefont {Franciolini}\ \emph {et~al.}(2021)\citenamefont
  {Franciolini}, \citenamefont {Baibhav}, \citenamefont {De~Luca},
  \citenamefont {Ng}, \citenamefont {Wong}, \citenamefont {Berti},
  \citenamefont {Pani}, \citenamefont {Riotto},\ and\ \citenamefont
  {Vitale}}]{Franciolini:2021tla}%
  \BibitemOpen
  \bibfield  {author} {\bibinfo {author} {\bibfnamefont {G.}~\bibnamefont
  {Franciolini}}, \bibinfo {author} {\bibfnamefont {V.}~\bibnamefont
  {Baibhav}}, \bibinfo {author} {\bibfnamefont {V.}~\bibnamefont {De~Luca}},
  \bibinfo {author} {\bibfnamefont {K.~K.~Y.}\ \bibnamefont {Ng}}, \bibinfo
  {author} {\bibfnamefont {K.~W.~K.}\ \bibnamefont {Wong}}, \bibinfo {author}
  {\bibfnamefont {E.}~\bibnamefont {Berti}}, \bibinfo {author} {\bibfnamefont
  {P.}~\bibnamefont {Pani}}, \bibinfo {author} {\bibfnamefont {A.}~\bibnamefont
  {Riotto}}, \ and\ \bibinfo {author} {\bibfnamefont {S.}~\bibnamefont
  {Vitale}},\ }\href@noop {} {\  (\bibinfo {year} {2021})},\ \Eprint
  {http://arxiv.org/abs/2105.03349} {arXiv:2105.03349 [gr-qc]} \BibitemShut
  {NoStop}%
\bibitem [{\citenamefont {Khlopov}(2010)}]{Khlopov:2008qy}%
  \BibitemOpen
  \bibfield  {author} {\bibinfo {author} {\bibfnamefont {M.~Y.}\ \bibnamefont
  {Khlopov}},\ }\href {\doibase 10.1088/1674-4527/10/6/001} {\bibfield
  {journal} {\bibinfo  {journal} {Res. Astron. Astrophys.}\ }\textbf {\bibinfo
  {volume} {10}},\ \bibinfo {pages} {495} (\bibinfo {year} {2010})},\ \Eprint
  {http://arxiv.org/abs/0801.0116} {arXiv:0801.0116 [astro-ph]} \BibitemShut
  {NoStop}%
\bibitem [{\citenamefont {Sasaki}\ \emph {et~al.}(2018)\citenamefont {Sasaki},
  \citenamefont {Suyama}, \citenamefont {Tanaka},\ and\ \citenamefont
  {Yokoyama}}]{Sasaki:2018dmp}%
  \BibitemOpen
  \bibfield  {author} {\bibinfo {author} {\bibfnamefont {M.}~\bibnamefont
  {Sasaki}}, \bibinfo {author} {\bibfnamefont {T.}~\bibnamefont {Suyama}},
  \bibinfo {author} {\bibfnamefont {T.}~\bibnamefont {Tanaka}}, \ and\ \bibinfo
  {author} {\bibfnamefont {S.}~\bibnamefont {Yokoyama}},\ }\href {\doibase
  10.1088/1361-6382/aaa7b4} {\bibfield  {journal} {\bibinfo  {journal} {Class.
  Quant. Grav.}\ }\textbf {\bibinfo {volume} {35}},\ \bibinfo {pages} {063001}
  (\bibinfo {year} {2018})},\ \Eprint {http://arxiv.org/abs/1801.05235}
  {arXiv:1801.05235 [astro-ph.CO]} \BibitemShut {NoStop}%
\bibitem [{\citenamefont {Carr}\ \emph {et~al.}(2020)\citenamefont {Carr},
  \citenamefont {Kohri}, \citenamefont {Sendouda},\ and\ \citenamefont
  {Yokoyama}}]{Carr:2020gox}%
  \BibitemOpen
  \bibfield  {author} {\bibinfo {author} {\bibfnamefont {B.}~\bibnamefont
  {Carr}}, \bibinfo {author} {\bibfnamefont {K.}~\bibnamefont {Kohri}},
  \bibinfo {author} {\bibfnamefont {Y.}~\bibnamefont {Sendouda}}, \ and\
  \bibinfo {author} {\bibfnamefont {J.}~\bibnamefont {Yokoyama}},\ }\href@noop
  {} {\  (\bibinfo {year} {2020})},\ \Eprint {http://arxiv.org/abs/2002.12778}
  {arXiv:2002.12778 [astro-ph.CO]} \BibitemShut {NoStop}%
\bibitem [{\citenamefont {Carr}\ and\ \citenamefont
  {Kuhnel}(2020)}]{Carr:2020xqk}%
  \BibitemOpen
  \bibfield  {author} {\bibinfo {author} {\bibfnamefont {B.}~\bibnamefont
  {Carr}}\ and\ \bibinfo {author} {\bibfnamefont {F.}~\bibnamefont {Kuhnel}},\
  }\href {\doibase 10.1146/annurev-nucl-050520-125911} {\bibfield  {journal}
  {\bibinfo  {journal} {Ann. Rev. Nucl. Part. Sci.}\ }\textbf {\bibinfo
  {volume} {70}},\ \bibinfo {pages} {355} (\bibinfo {year} {2020})},\ \Eprint
  {http://arxiv.org/abs/2006.02838} {arXiv:2006.02838 [astro-ph.CO]}
  \BibitemShut {NoStop}%
\bibitem [{\citenamefont {Green}\ and\ \citenamefont
  {Kavanagh}(2021)}]{Green:2020jor}%
  \BibitemOpen
  \bibfield  {author} {\bibinfo {author} {\bibfnamefont {A.~M.}\ \bibnamefont
  {Green}}\ and\ \bibinfo {author} {\bibfnamefont {B.~J.}\ \bibnamefont
  {Kavanagh}},\ }\href {\doibase 10.1088/1361-6471/abc534} {\bibfield
  {journal} {\bibinfo  {journal} {J. Phys. G}\ }\textbf {\bibinfo {volume}
  {48}},\ \bibinfo {pages} {4} (\bibinfo {year} {2021})},\ \Eprint
  {http://arxiv.org/abs/2007.10722} {arXiv:2007.10722 [astro-ph.CO]}
  \BibitemShut {NoStop}%
\bibitem [{\citenamefont {Escriv\`a}(2021)}]{Escriva:2021aeh}%
  \BibitemOpen
  \bibfield  {author} {\bibinfo {author} {\bibfnamefont {A.}~\bibnamefont
  {Escriv\`a}},\ }\href@noop {} {\  (\bibinfo {year} {2021})},\ \Eprint
  {http://arxiv.org/abs/2111.12693} {arXiv:2111.12693 [gr-qc]} \BibitemShut
  {NoStop}%
\bibitem [{\citenamefont {Kawasaki}\ \emph {et~al.}(1998)\citenamefont
  {Kawasaki}, \citenamefont {Sugiyama},\ and\ \citenamefont
  {Yanagida}}]{Kawasaki:1997ju}%
  \BibitemOpen
  \bibfield  {author} {\bibinfo {author} {\bibfnamefont {M.}~\bibnamefont
  {Kawasaki}}, \bibinfo {author} {\bibfnamefont {N.}~\bibnamefont {Sugiyama}},
  \ and\ \bibinfo {author} {\bibfnamefont {T.}~\bibnamefont {Yanagida}},\
  }\href {\doibase 10.1103/PhysRevD.57.6050} {\bibfield  {journal} {\bibinfo
  {journal} {Phys. Rev. D}\ }\textbf {\bibinfo {volume} {57}},\ \bibinfo
  {pages} {6050} (\bibinfo {year} {1998})},\ \Eprint
  {http://arxiv.org/abs/hep-ph/9710259} {arXiv:hep-ph/9710259} \BibitemShut
  {NoStop}%
\bibitem [{\citenamefont {Frampton}\ \emph {et~al.}(2010)\citenamefont
  {Frampton}, \citenamefont {Kawasaki}, \citenamefont {Takahashi},\ and\
  \citenamefont {Yanagida}}]{Frampton:2010sw}%
  \BibitemOpen
  \bibfield  {author} {\bibinfo {author} {\bibfnamefont {P.~H.}\ \bibnamefont
  {Frampton}}, \bibinfo {author} {\bibfnamefont {M.}~\bibnamefont {Kawasaki}},
  \bibinfo {author} {\bibfnamefont {F.}~\bibnamefont {Takahashi}}, \ and\
  \bibinfo {author} {\bibfnamefont {T.~T.}\ \bibnamefont {Yanagida}},\ }\href
  {\doibase 10.1088/1475-7516/2010/04/023} {\bibfield  {journal} {\bibinfo
  {journal} {JCAP}\ }\textbf {\bibinfo {volume} {04}},\ \bibinfo {pages} {023}
  (\bibinfo {year} {2010})},\ \Eprint {http://arxiv.org/abs/1001.2308}
  {arXiv:1001.2308 [hep-ph]} \BibitemShut {NoStop}%
\bibitem [{\citenamefont {Kawasaki}\ \emph {et~al.}(2013)\citenamefont
  {Kawasaki}, \citenamefont {Kitajima},\ and\ \citenamefont
  {Yanagida}}]{Kawasaki:2012wr}%
  \BibitemOpen
  \bibfield  {author} {\bibinfo {author} {\bibfnamefont {M.}~\bibnamefont
  {Kawasaki}}, \bibinfo {author} {\bibfnamefont {N.}~\bibnamefont {Kitajima}},
  \ and\ \bibinfo {author} {\bibfnamefont {T.~T.}\ \bibnamefont {Yanagida}},\
  }\href {\doibase 10.1103/PhysRevD.87.063519} {\bibfield  {journal} {\bibinfo
  {journal} {Phys. Rev. D}\ }\textbf {\bibinfo {volume} {87}},\ \bibinfo
  {pages} {063519} (\bibinfo {year} {2013})},\ \Eprint
  {http://arxiv.org/abs/1207.2550} {arXiv:1207.2550 [hep-ph]} \BibitemShut
  {NoStop}%
\bibitem [{\citenamefont {Inomata}\ \emph
  {et~al.}(2017{\natexlab{b}})\citenamefont {Inomata}, \citenamefont
  {Kawasaki}, \citenamefont {Mukaida}, \citenamefont {Tada},\ and\
  \citenamefont {Yanagida}}]{Inomata:2017okj}%
  \BibitemOpen
  \bibfield  {author} {\bibinfo {author} {\bibfnamefont {K.}~\bibnamefont
  {Inomata}}, \bibinfo {author} {\bibfnamefont {M.}~\bibnamefont {Kawasaki}},
  \bibinfo {author} {\bibfnamefont {K.}~\bibnamefont {Mukaida}}, \bibinfo
  {author} {\bibfnamefont {Y.}~\bibnamefont {Tada}}, \ and\ \bibinfo {author}
  {\bibfnamefont {T.~T.}\ \bibnamefont {Yanagida}},\ }\href {\doibase
  10.1103/PhysRevD.96.043504} {\bibfield  {journal} {\bibinfo  {journal} {Phys.
  Rev. D}\ }\textbf {\bibinfo {volume} {96}},\ \bibinfo {pages} {043504}
  (\bibinfo {year} {2017}{\natexlab{b}})},\ \Eprint
  {http://arxiv.org/abs/1701.02544} {arXiv:1701.02544 [astro-ph.CO]}
  \BibitemShut {NoStop}%
\bibitem [{\citenamefont {Pi}\ \emph {et~al.}(2018)\citenamefont {Pi},
  \citenamefont {Zhang}, \citenamefont {Huang},\ and\ \citenamefont
  {Sasaki}}]{Pi:2017gih}%
  \BibitemOpen
  \bibfield  {author} {\bibinfo {author} {\bibfnamefont {S.}~\bibnamefont
  {Pi}}, \bibinfo {author} {\bibfnamefont {Y.-l.}\ \bibnamefont {Zhang}},
  \bibinfo {author} {\bibfnamefont {Q.-G.}\ \bibnamefont {Huang}}, \ and\
  \bibinfo {author} {\bibfnamefont {M.}~\bibnamefont {Sasaki}},\ }\href
  {\doibase 10.1088/1475-7516/2018/05/042} {\bibfield  {journal} {\bibinfo
  {journal} {JCAP}\ }\textbf {\bibinfo {volume} {05}},\ \bibinfo {pages} {042}
  (\bibinfo {year} {2018})},\ \Eprint {http://arxiv.org/abs/1712.09896}
  {arXiv:1712.09896 [astro-ph.CO]} \BibitemShut {NoStop}%
\bibitem [{\citenamefont {Cai}\ \emph {et~al.}(2018)\citenamefont {Cai},
  \citenamefont {Tong}, \citenamefont {Wang},\ and\ \citenamefont
  {Yan}}]{Cai:2018tuh}%
  \BibitemOpen
  \bibfield  {author} {\bibinfo {author} {\bibfnamefont {Y.-F.}\ \bibnamefont
  {Cai}}, \bibinfo {author} {\bibfnamefont {X.}~\bibnamefont {Tong}}, \bibinfo
  {author} {\bibfnamefont {D.-G.}\ \bibnamefont {Wang}}, \ and\ \bibinfo
  {author} {\bibfnamefont {S.-F.}\ \bibnamefont {Yan}},\ }\href {\doibase
  10.1103/PhysRevLett.121.081306} {\bibfield  {journal} {\bibinfo  {journal}
  {Phys. Rev. Lett.}\ }\textbf {\bibinfo {volume} {121}},\ \bibinfo {pages}
  {081306} (\bibinfo {year} {2018})},\ \Eprint
  {http://arxiv.org/abs/1805.03639} {arXiv:1805.03639 [astro-ph.CO]}
  \BibitemShut {NoStop}%
\bibitem [{\citenamefont {Cai}\ \emph {et~al.}(2019{\natexlab{b}})\citenamefont
  {Cai}, \citenamefont {Chen}, \citenamefont {Tong}, \citenamefont {Wang},\
  and\ \citenamefont {Yan}}]{Cai:2019jah}%
  \BibitemOpen
  \bibfield  {author} {\bibinfo {author} {\bibfnamefont {Y.-F.}\ \bibnamefont
  {Cai}}, \bibinfo {author} {\bibfnamefont {C.}~\bibnamefont {Chen}}, \bibinfo
  {author} {\bibfnamefont {X.}~\bibnamefont {Tong}}, \bibinfo {author}
  {\bibfnamefont {D.-G.}\ \bibnamefont {Wang}}, \ and\ \bibinfo {author}
  {\bibfnamefont {S.-F.}\ \bibnamefont {Yan}},\ }\href {\doibase
  10.1103/PhysRevD.100.043518} {\bibfield  {journal} {\bibinfo  {journal}
  {Phys. Rev. D}\ }\textbf {\bibinfo {volume} {100}},\ \bibinfo {pages}
  {043518} (\bibinfo {year} {2019}{\natexlab{b}})},\ \Eprint
  {http://arxiv.org/abs/1902.08187} {arXiv:1902.08187 [astro-ph.CO]}
  \BibitemShut {NoStop}%
\bibitem [{\citenamefont {Chen}\ and\ \citenamefont
  {Cai}(2019)}]{Chen:2019zza}%
  \BibitemOpen
  \bibfield  {author} {\bibinfo {author} {\bibfnamefont {C.}~\bibnamefont
  {Chen}}\ and\ \bibinfo {author} {\bibfnamefont {Y.-F.}\ \bibnamefont {Cai}},\
  }\href {\doibase 10.1088/1475-7516/2019/10/068} {\bibfield  {journal}
  {\bibinfo  {journal} {JCAP}\ }\textbf {\bibinfo {volume} {10}},\ \bibinfo
  {pages} {068} (\bibinfo {year} {2019})},\ \Eprint
  {http://arxiv.org/abs/1908.03942} {arXiv:1908.03942 [astro-ph.CO]}
  \BibitemShut {NoStop}%
\bibitem [{\citenamefont {Ashoorioon}\ \emph {et~al.}(2021)\citenamefont
  {Ashoorioon}, \citenamefont {Rostami},\ and\ \citenamefont
  {Firouzjaee}}]{Ashoorioon:2019xqc}%
  \BibitemOpen
  \bibfield  {author} {\bibinfo {author} {\bibfnamefont {A.}~\bibnamefont
  {Ashoorioon}}, \bibinfo {author} {\bibfnamefont {A.}~\bibnamefont {Rostami}},
  \ and\ \bibinfo {author} {\bibfnamefont {J.~T.}\ \bibnamefont {Firouzjaee}},\
  }\href {\doibase 10.1007/JHEP07(2021)087} {\bibfield  {journal} {\bibinfo
  {journal} {JHEP}\ }\textbf {\bibinfo {volume} {07}},\ \bibinfo {pages} {087}
  (\bibinfo {year} {2021})},\ \Eprint {http://arxiv.org/abs/1912.13326}
  {arXiv:1912.13326 [astro-ph.CO]} \BibitemShut {NoStop}%
\bibitem [{\citenamefont {Chen}\ \emph {et~al.}(2020)\citenamefont {Chen},
  \citenamefont {Ma},\ and\ \citenamefont {Cai}}]{Chen:2020uhe}%
  \BibitemOpen
  \bibfield  {author} {\bibinfo {author} {\bibfnamefont {C.}~\bibnamefont
  {Chen}}, \bibinfo {author} {\bibfnamefont {X.-H.}\ \bibnamefont {Ma}}, \ and\
  \bibinfo {author} {\bibfnamefont {Y.-F.}\ \bibnamefont {Cai}},\ }\href
  {\doibase 10.1103/PhysRevD.102.063526} {\bibfield  {journal} {\bibinfo
  {journal} {Phys. Rev. D}\ }\textbf {\bibinfo {volume} {102}},\ \bibinfo
  {pages} {063526} (\bibinfo {year} {2020})},\ \Eprint
  {http://arxiv.org/abs/2003.03821} {arXiv:2003.03821 [astro-ph.CO]}
  \BibitemShut {NoStop}%
\bibitem [{\citenamefont {Garcia-Bellido}\ \emph {et~al.}(1996)\citenamefont
  {Garcia-Bellido}, \citenamefont {Linde},\ and\ \citenamefont
  {Wands}}]{Garcia-Bellido:1996mdl}%
  \BibitemOpen
  \bibfield  {author} {\bibinfo {author} {\bibfnamefont {J.}~\bibnamefont
  {Garcia-Bellido}}, \bibinfo {author} {\bibfnamefont {A.~D.}\ \bibnamefont
  {Linde}}, \ and\ \bibinfo {author} {\bibfnamefont {D.}~\bibnamefont
  {Wands}},\ }\href {\doibase 10.1103/PhysRevD.54.6040} {\bibfield  {journal}
  {\bibinfo  {journal} {Phys. Rev. D}\ }\textbf {\bibinfo {volume} {54}},\
  \bibinfo {pages} {6040} (\bibinfo {year} {1996})},\ \Eprint
  {http://arxiv.org/abs/astro-ph/9605094} {arXiv:astro-ph/9605094} \BibitemShut
  {NoStop}%
\bibitem [{\citenamefont {Yokoyama}(1998)}]{Yokoyama:1998pt}%
  \BibitemOpen
  \bibfield  {author} {\bibinfo {author} {\bibfnamefont {J.}~\bibnamefont
  {Yokoyama}},\ }\href {\doibase 10.1103/PhysRevD.58.083510} {\bibfield
  {journal} {\bibinfo  {journal} {Phys. Rev. D}\ }\textbf {\bibinfo {volume}
  {58}},\ \bibinfo {pages} {083510} (\bibinfo {year} {1998})},\ \Eprint
  {http://arxiv.org/abs/astro-ph/9802357} {arXiv:astro-ph/9802357} \BibitemShut
  {NoStop}%
\bibitem [{\citenamefont {Kohri}\ \emph {et~al.}(2013)\citenamefont {Kohri},
  \citenamefont {Lin},\ and\ \citenamefont {Matsuda}}]{Kohri:2012yw}%
  \BibitemOpen
  \bibfield  {author} {\bibinfo {author} {\bibfnamefont {K.}~\bibnamefont
  {Kohri}}, \bibinfo {author} {\bibfnamefont {C.-M.}\ \bibnamefont {Lin}}, \
  and\ \bibinfo {author} {\bibfnamefont {T.}~\bibnamefont {Matsuda}},\ }\href
  {\doibase 10.1103/PhysRevD.87.103527} {\bibfield  {journal} {\bibinfo
  {journal} {Phys. Rev. D}\ }\textbf {\bibinfo {volume} {87}},\ \bibinfo
  {pages} {103527} (\bibinfo {year} {2013})},\ \Eprint
  {http://arxiv.org/abs/1211.2371} {arXiv:1211.2371 [hep-ph]} \BibitemShut
  {NoStop}%
\bibitem [{\citenamefont {Clesse}\ and\ \citenamefont
  {Garc\'\i{}a-Bellido}(2015)}]{Clesse:2015wea}%
  \BibitemOpen
  \bibfield  {author} {\bibinfo {author} {\bibfnamefont {S.}~\bibnamefont
  {Clesse}}\ and\ \bibinfo {author} {\bibfnamefont {J.}~\bibnamefont
  {Garc\'\i{}a-Bellido}},\ }\href {\doibase 10.1103/PhysRevD.92.023524}
  {\bibfield  {journal} {\bibinfo  {journal} {Phys. Rev. D}\ }\textbf {\bibinfo
  {volume} {92}},\ \bibinfo {pages} {023524} (\bibinfo {year} {2015})},\
  \Eprint {http://arxiv.org/abs/1501.07565} {arXiv:1501.07565 [astro-ph.CO]}
  \BibitemShut {NoStop}%
\bibitem [{\citenamefont {Cheng}\ \emph {et~al.}(2017)\citenamefont {Cheng},
  \citenamefont {Lee},\ and\ \citenamefont {Ng}}]{Cheng:2016qzb}%
  \BibitemOpen
  \bibfield  {author} {\bibinfo {author} {\bibfnamefont {S.-L.}\ \bibnamefont
  {Cheng}}, \bibinfo {author} {\bibfnamefont {W.}~\bibnamefont {Lee}}, \ and\
  \bibinfo {author} {\bibfnamefont {K.-W.}\ \bibnamefont {Ng}},\ }\href
  {\doibase 10.1007/JHEP02(2017)008} {\bibfield  {journal} {\bibinfo  {journal}
  {JHEP}\ }\textbf {\bibinfo {volume} {02}},\ \bibinfo {pages} {008} (\bibinfo
  {year} {2017})},\ \Eprint {http://arxiv.org/abs/1606.00206} {arXiv:1606.00206
  [astro-ph.CO]} \BibitemShut {NoStop}%
\bibitem [{\citenamefont {Espinosa}\ \emph
  {et~al.}(2018{\natexlab{a}})\citenamefont {Espinosa}, \citenamefont {Racco},\
  and\ \citenamefont {Riotto}}]{Espinosa:2017sgp}%
  \BibitemOpen
  \bibfield  {author} {\bibinfo {author} {\bibfnamefont {J.~R.}\ \bibnamefont
  {Espinosa}}, \bibinfo {author} {\bibfnamefont {D.}~\bibnamefont {Racco}}, \
  and\ \bibinfo {author} {\bibfnamefont {A.}~\bibnamefont {Riotto}},\ }\href
  {\doibase 10.1103/PhysRevLett.120.121301} {\bibfield  {journal} {\bibinfo
  {journal} {Phys. Rev. Lett.}\ }\textbf {\bibinfo {volume} {120}},\ \bibinfo
  {pages} {121301} (\bibinfo {year} {2018}{\natexlab{a}})},\ \Eprint
  {http://arxiv.org/abs/1710.11196} {arXiv:1710.11196 [hep-ph]} \BibitemShut
  {NoStop}%
\bibitem [{\citenamefont {Kannike}\ \emph {et~al.}(2017)\citenamefont
  {Kannike}, \citenamefont {Marzola}, \citenamefont {Raidal},\ and\
  \citenamefont {Veerm\"ae}}]{Kannike:2017bxn}%
  \BibitemOpen
  \bibfield  {author} {\bibinfo {author} {\bibfnamefont {K.}~\bibnamefont
  {Kannike}}, \bibinfo {author} {\bibfnamefont {L.}~\bibnamefont {Marzola}},
  \bibinfo {author} {\bibfnamefont {M.}~\bibnamefont {Raidal}}, \ and\ \bibinfo
  {author} {\bibfnamefont {H.}~\bibnamefont {Veerm\"ae}},\ }\href {\doibase
  10.1088/1475-7516/2017/09/020} {\bibfield  {journal} {\bibinfo  {journal}
  {JCAP}\ }\textbf {\bibinfo {volume} {09}},\ \bibinfo {pages} {020} (\bibinfo
  {year} {2017})},\ \Eprint {http://arxiv.org/abs/1705.06225} {arXiv:1705.06225
  [astro-ph.CO]} \BibitemShut {NoStop}%
\bibitem [{\citenamefont {Garcia-Bellido}\ and\ \citenamefont
  {Ruiz~Morales}(2017)}]{Garcia-Bellido:2017mdw}%
  \BibitemOpen
  \bibfield  {author} {\bibinfo {author} {\bibfnamefont {J.}~\bibnamefont
  {Garcia-Bellido}}\ and\ \bibinfo {author} {\bibfnamefont {E.}~\bibnamefont
  {Ruiz~Morales}},\ }\href {\doibase 10.1016/j.dark.2017.09.007} {\bibfield
  {journal} {\bibinfo  {journal} {Phys. Dark Univ.}\ }\textbf {\bibinfo
  {volume} {18}},\ \bibinfo {pages} {47} (\bibinfo {year} {2017})},\ \Eprint
  {http://arxiv.org/abs/1702.03901} {arXiv:1702.03901 [astro-ph.CO]}
  \BibitemShut {NoStop}%
\bibitem [{\citenamefont {Cheng}\ \emph {et~al.}(2018)\citenamefont {Cheng},
  \citenamefont {Lee},\ and\ \citenamefont {Ng}}]{Cheng:2018yyr}%
  \BibitemOpen
  \bibfield  {author} {\bibinfo {author} {\bibfnamefont {S.-L.}\ \bibnamefont
  {Cheng}}, \bibinfo {author} {\bibfnamefont {W.}~\bibnamefont {Lee}}, \ and\
  \bibinfo {author} {\bibfnamefont {K.-W.}\ \bibnamefont {Ng}},\ }\href
  {\doibase 10.1088/1475-7516/2018/07/001} {\bibfield  {journal} {\bibinfo
  {journal} {JCAP}\ }\textbf {\bibinfo {volume} {07}},\ \bibinfo {pages} {001}
  (\bibinfo {year} {2018})},\ \Eprint {http://arxiv.org/abs/1801.09050}
  {arXiv:1801.09050 [astro-ph.CO]} \BibitemShut {NoStop}%
\bibitem [{\citenamefont {Espinosa}\ \emph
  {et~al.}(2018{\natexlab{b}})\citenamefont {Espinosa}, \citenamefont {Racco},\
  and\ \citenamefont {Riotto}}]{Espinosa:2018eve}%
  \BibitemOpen
  \bibfield  {author} {\bibinfo {author} {\bibfnamefont {J.~R.}\ \bibnamefont
  {Espinosa}}, \bibinfo {author} {\bibfnamefont {D.}~\bibnamefont {Racco}}, \
  and\ \bibinfo {author} {\bibfnamefont {A.}~\bibnamefont {Riotto}},\ }\href
  {\doibase 10.1088/1475-7516/2018/09/012} {\bibfield  {journal} {\bibinfo
  {journal} {JCAP}\ }\textbf {\bibinfo {volume} {09}},\ \bibinfo {pages} {012}
  (\bibinfo {year} {2018}{\natexlab{b}})},\ \Eprint
  {http://arxiv.org/abs/1804.07732} {arXiv:1804.07732 [hep-ph]} \BibitemShut
  {NoStop}%
\bibitem [{\citenamefont {Inomata}\ \emph {et~al.}(2018)\citenamefont
  {Inomata}, \citenamefont {Kawasaki}, \citenamefont {Mukaida},\ and\
  \citenamefont {Yanagida}}]{Inomata:2018cht}%
  \BibitemOpen
  \bibfield  {author} {\bibinfo {author} {\bibfnamefont {K.}~\bibnamefont
  {Inomata}}, \bibinfo {author} {\bibfnamefont {M.}~\bibnamefont {Kawasaki}},
  \bibinfo {author} {\bibfnamefont {K.}~\bibnamefont {Mukaida}}, \ and\
  \bibinfo {author} {\bibfnamefont {T.~T.}\ \bibnamefont {Yanagida}},\ }\href
  {\doibase 10.1103/PhysRevD.97.043514} {\bibfield  {journal} {\bibinfo
  {journal} {Phys. Rev. D}\ }\textbf {\bibinfo {volume} {97}},\ \bibinfo
  {pages} {043514} (\bibinfo {year} {2018})},\ \Eprint
  {http://arxiv.org/abs/1711.06129} {arXiv:1711.06129 [astro-ph.CO]}
  \BibitemShut {NoStop}%
\bibitem [{\citenamefont {Atal}\ and\ \citenamefont
  {Germani}(2019)}]{Atal:2018neu}%
  \BibitemOpen
  \bibfield  {author} {\bibinfo {author} {\bibfnamefont {V.}~\bibnamefont
  {Atal}}\ and\ \bibinfo {author} {\bibfnamefont {C.}~\bibnamefont {Germani}},\
  }\href {\doibase 10.1016/j.dark.2019.100275} {\bibfield  {journal} {\bibinfo
  {journal} {Phys. Dark Univ.}\ }\textbf {\bibinfo {volume} {24}},\ \bibinfo
  {pages} {100275} (\bibinfo {year} {2019})},\ \Eprint
  {http://arxiv.org/abs/1811.07857} {arXiv:1811.07857 [astro-ph.CO]}
  \BibitemShut {NoStop}%
\bibitem [{\citenamefont {Ng}\ and\ \citenamefont {Wu}(2021)}]{Ng:2021hll}%
  \BibitemOpen
  \bibfield  {author} {\bibinfo {author} {\bibfnamefont {K.-W.}\ \bibnamefont
  {Ng}}\ and\ \bibinfo {author} {\bibfnamefont {Y.-P.}\ \bibnamefont {Wu}},\
  }\href {\doibase 10.1007/JHEP11(2021)076} {\bibfield  {journal} {\bibinfo
  {journal} {JHEP}\ }\textbf {\bibinfo {volume} {11}},\ \bibinfo {pages} {076}
  (\bibinfo {year} {2021})},\ \Eprint {http://arxiv.org/abs/2102.05620}
  {arXiv:2102.05620 [astro-ph.CO]} \BibitemShut {NoStop}%
\bibitem [{\citenamefont {Carrilho}\ \emph {et~al.}(2019)\citenamefont
  {Carrilho}, \citenamefont {Malik},\ and\ \citenamefont
  {Mulryne}}]{Carrilho:2019oqg}%
  \BibitemOpen
  \bibfield  {author} {\bibinfo {author} {\bibfnamefont {P.}~\bibnamefont
  {Carrilho}}, \bibinfo {author} {\bibfnamefont {K.~A.}\ \bibnamefont {Malik}},
  \ and\ \bibinfo {author} {\bibfnamefont {D.~J.}\ \bibnamefont {Mulryne}},\
  }\href {\doibase 10.1103/PhysRevD.100.103529} {\bibfield  {journal} {\bibinfo
   {journal} {Phys. Rev. D}\ }\textbf {\bibinfo {volume} {100}},\ \bibinfo
  {pages} {103529} (\bibinfo {year} {2019})},\ \Eprint
  {http://arxiv.org/abs/1907.05237} {arXiv:1907.05237 [astro-ph.CO]}
  \BibitemShut {NoStop}%
\bibitem [{\citenamefont {Palma}\ \emph {et~al.}(2020)\citenamefont {Palma},
  \citenamefont {Sypsas},\ and\ \citenamefont {Zenteno}}]{Palma:2020ejf}%
  \BibitemOpen
  \bibfield  {author} {\bibinfo {author} {\bibfnamefont {G.~A.}\ \bibnamefont
  {Palma}}, \bibinfo {author} {\bibfnamefont {S.}~\bibnamefont {Sypsas}}, \
  and\ \bibinfo {author} {\bibfnamefont {C.}~\bibnamefont {Zenteno}},\ }\href
  {\doibase 10.1103/PhysRevLett.125.121301} {\bibfield  {journal} {\bibinfo
  {journal} {Phys. Rev. Lett.}\ }\textbf {\bibinfo {volume} {125}},\ \bibinfo
  {pages} {121301} (\bibinfo {year} {2020})},\ \Eprint
  {http://arxiv.org/abs/2004.06106} {arXiv:2004.06106 [astro-ph.CO]}
  \BibitemShut {NoStop}%
\bibitem [{\citenamefont {Fumagalli}\ \emph
  {et~al.}(2020{\natexlab{b}})\citenamefont {Fumagalli}, \citenamefont
  {Renaux-Petel}, \citenamefont {Ronayne},\ and\ \citenamefont
  {Witkowski}}]{Fumagalli:2020adf}%
  \BibitemOpen
  \bibfield  {author} {\bibinfo {author} {\bibfnamefont {J.}~\bibnamefont
  {Fumagalli}}, \bibinfo {author} {\bibfnamefont {S.}~\bibnamefont
  {Renaux-Petel}}, \bibinfo {author} {\bibfnamefont {J.~W.}\ \bibnamefont
  {Ronayne}}, \ and\ \bibinfo {author} {\bibfnamefont {L.~T.}\ \bibnamefont
  {Witkowski}},\ }\href@noop {} {\  (\bibinfo {year} {2020}{\natexlab{b}})},\
  \Eprint {http://arxiv.org/abs/2004.08369} {arXiv:2004.08369 [hep-th]}
  \BibitemShut {NoStop}%
\bibitem [{\citenamefont {Inomata}\ \emph {et~al.}(2022)\citenamefont
  {Inomata}, \citenamefont {McDonough},\ and\ \citenamefont
  {Hu}}]{Inomata:2021tpx}%
  \BibitemOpen
  \bibfield  {author} {\bibinfo {author} {\bibfnamefont {K.}~\bibnamefont
  {Inomata}}, \bibinfo {author} {\bibfnamefont {E.}~\bibnamefont {McDonough}},
  \ and\ \bibinfo {author} {\bibfnamefont {W.}~\bibnamefont {Hu}},\ }\href
  {\doibase 10.1088/1475-7516/2022/02/031} {\bibfield  {journal} {\bibinfo
  {journal} {JCAP}\ }\textbf {\bibinfo {volume} {02}},\ \bibinfo {pages} {031}
  (\bibinfo {year} {2022})},\ \Eprint {http://arxiv.org/abs/2110.14641}
  {arXiv:2110.14641 [astro-ph.CO]} \BibitemShut {NoStop}%
\bibitem [{\citenamefont {Cole}\ \emph {et~al.}(2022)\citenamefont {Cole},
  \citenamefont {Gow}, \citenamefont {Byrnes},\ and\ \citenamefont
  {Patil}}]{Cole:2022xqc}%
  \BibitemOpen
  \bibfield  {author} {\bibinfo {author} {\bibfnamefont {P.~S.}\ \bibnamefont
  {Cole}}, \bibinfo {author} {\bibfnamefont {A.~D.}\ \bibnamefont {Gow}},
  \bibinfo {author} {\bibfnamefont {C.~T.}\ \bibnamefont {Byrnes}}, \ and\
  \bibinfo {author} {\bibfnamefont {S.~P.}\ \bibnamefont {Patil}},\ }\href@noop
  {} {\  (\bibinfo {year} {2022})},\ \Eprint {http://arxiv.org/abs/2204.07573}
  {arXiv:2204.07573 [astro-ph.CO]} \BibitemShut {NoStop}%
\bibitem [{\citenamefont {Zhou}\ \emph {et~al.}(2020)\citenamefont {Zhou},
  \citenamefont {Jiang}, \citenamefont {Cai}, \citenamefont {Sasaki},\ and\
  \citenamefont {Pi}}]{Zhou:2020kkf}%
  \BibitemOpen
  \bibfield  {author} {\bibinfo {author} {\bibfnamefont {Z.}~\bibnamefont
  {Zhou}}, \bibinfo {author} {\bibfnamefont {J.}~\bibnamefont {Jiang}},
  \bibinfo {author} {\bibfnamefont {Y.-F.}\ \bibnamefont {Cai}}, \bibinfo
  {author} {\bibfnamefont {M.}~\bibnamefont {Sasaki}}, \ and\ \bibinfo {author}
  {\bibfnamefont {S.}~\bibnamefont {Pi}},\ }\href {\doibase
  10.1103/PhysRevD.102.103527} {\bibfield  {journal} {\bibinfo  {journal}
  {Phys. Rev. D}\ }\textbf {\bibinfo {volume} {102}},\ \bibinfo {pages}
  {103527} (\bibinfo {year} {2020})},\ \Eprint
  {http://arxiv.org/abs/2010.03537} {arXiv:2010.03537 [astro-ph.CO]}
  \BibitemShut {NoStop}%
\bibitem [{\citenamefont {Leach}\ and\ \citenamefont
  {Liddle}(2001)}]{Leach:2000yw}%
  \BibitemOpen
  \bibfield  {author} {\bibinfo {author} {\bibfnamefont {S.~M.}\ \bibnamefont
  {Leach}}\ and\ \bibinfo {author} {\bibfnamefont {A.~R.}\ \bibnamefont
  {Liddle}},\ }\href {\doibase 10.1103/PhysRevD.63.043508} {\bibfield
  {journal} {\bibinfo  {journal} {Phys. Rev. D}\ }\textbf {\bibinfo {volume}
  {63}},\ \bibinfo {pages} {043508} (\bibinfo {year} {2001})},\ \Eprint
  {http://arxiv.org/abs/astro-ph/0010082} {arXiv:astro-ph/0010082} \BibitemShut
  {NoStop}%
\bibitem [{\citenamefont {Inomata}\ \emph {et~al.}(2021)\citenamefont
  {Inomata}, \citenamefont {McDonough},\ and\ \citenamefont
  {Hu}}]{Inomata:2021uqj}%
  \BibitemOpen
  \bibfield  {author} {\bibinfo {author} {\bibfnamefont {K.}~\bibnamefont
  {Inomata}}, \bibinfo {author} {\bibfnamefont {E.}~\bibnamefont {McDonough}},
  \ and\ \bibinfo {author} {\bibfnamefont {W.}~\bibnamefont {Hu}},\ }\href
  {\doibase 10.1103/PhysRevD.104.123553} {\bibfield  {journal} {\bibinfo
  {journal} {Phys. Rev. D}\ }\textbf {\bibinfo {volume} {104}},\ \bibinfo
  {pages} {123553} (\bibinfo {year} {2021})},\ \Eprint
  {http://arxiv.org/abs/2104.03972} {arXiv:2104.03972 [astro-ph.CO]}
  \BibitemShut {NoStop}%
\bibitem [{\citenamefont {\"Ozsoy}(2021)}]{Ozsoy:2020ccy}%
  \BibitemOpen
  \bibfield  {author} {\bibinfo {author} {\bibfnamefont {O.}~\bibnamefont
  {\"Ozsoy}},\ }\href {\doibase 10.1088/1475-7516/2021/04/040} {\bibfield
  {journal} {\bibinfo  {journal} {JCAP}\ }\textbf {\bibinfo {volume} {04}},\
  \bibinfo {pages} {040} (\bibinfo {year} {2021})},\ \Eprint
  {http://arxiv.org/abs/2005.10280} {arXiv:2005.10280 [astro-ph.CO]}
  \BibitemShut {NoStop}%
\bibitem [{\citenamefont {Aghanim}\ \emph {et~al.}(2020)\citenamefont {Aghanim}
  \emph {et~al.}}]{Aghanim:2018eyx}%
  \BibitemOpen
  \bibfield  {author} {\bibinfo {author} {\bibfnamefont {N.}~\bibnamefont
  {Aghanim}} \emph {et~al.} (\bibinfo {collaboration} {Planck}),\ }\href
  {\doibase 10.1051/0004-6361/201833910} {\bibfield  {journal} {\bibinfo
  {journal} {Astron. Astrophys.}\ }\textbf {\bibinfo {volume} {641}},\ \bibinfo
  {pages} {A6} (\bibinfo {year} {2020})},\ \Eprint
  {http://arxiv.org/abs/1807.06209} {arXiv:1807.06209 [astro-ph.CO]}
  \BibitemShut {NoStop}%
\bibitem [{\citenamefont {Kohri}\ and\ \citenamefont
  {Terada}(2018{\natexlab{b}})}]{Kohri:2018awv}%
  \BibitemOpen
  \bibfield  {author} {\bibinfo {author} {\bibfnamefont {K.}~\bibnamefont
  {Kohri}}\ and\ \bibinfo {author} {\bibfnamefont {T.}~\bibnamefont {Terada}},\
  }\href {\doibase 10.1103/PhysRevD.97.123532} {\bibfield  {journal} {\bibinfo
  {journal} {Phys. Rev. D}\ }\textbf {\bibinfo {volume} {97}},\ \bibinfo
  {pages} {123532} (\bibinfo {year} {2018}{\natexlab{b}})},\ \Eprint
  {http://arxiv.org/abs/1804.08577} {arXiv:1804.08577 [gr-qc]} \BibitemShut
  {NoStop}%
\bibitem [{\citenamefont {Lucchin}\ and\ \citenamefont
  {Matarrese}(1985)}]{Lucchin:1984yf}%
  \BibitemOpen
  \bibfield  {author} {\bibinfo {author} {\bibfnamefont {F.}~\bibnamefont
  {Lucchin}}\ and\ \bibinfo {author} {\bibfnamefont {S.}~\bibnamefont
  {Matarrese}},\ }\href {\doibase 10.1103/PhysRevD.32.1316} {\bibfield
  {journal} {\bibinfo  {journal} {Phys. Rev. D}\ }\textbf {\bibinfo {volume}
  {32}},\ \bibinfo {pages} {1316} (\bibinfo {year} {1985})}\BibitemShut
  {NoStop}%
\bibitem [{\citenamefont {Cai}\ \emph {et~al.}(2019{\natexlab{c}})\citenamefont
  {Cai}, \citenamefont {Pi},\ and\ \citenamefont {Sasaki}}]{Cai:2018dig}%
  \BibitemOpen
  \bibfield  {author} {\bibinfo {author} {\bibfnamefont {R.-g.}\ \bibnamefont
  {Cai}}, \bibinfo {author} {\bibfnamefont {S.}~\bibnamefont {Pi}}, \ and\
  \bibinfo {author} {\bibfnamefont {M.}~\bibnamefont {Sasaki}},\ }\href
  {\doibase 10.1103/PhysRevLett.122.201101} {\bibfield  {journal} {\bibinfo
  {journal} {Phys. Rev. Lett.}\ }\textbf {\bibinfo {volume} {122}},\ \bibinfo
  {pages} {201101} (\bibinfo {year} {2019}{\natexlab{c}})},\ \Eprint
  {http://arxiv.org/abs/1810.11000} {arXiv:1810.11000 [astro-ph.CO]}
  \BibitemShut {NoStop}%
\bibitem [{\citenamefont {Unal}(2019)}]{Unal:2018yaa}%
  \BibitemOpen
  \bibfield  {author} {\bibinfo {author} {\bibfnamefont {C.}~\bibnamefont
  {Unal}},\ }\href {\doibase 10.1103/PhysRevD.99.041301} {\bibfield  {journal}
  {\bibinfo  {journal} {Phys. Rev. D}\ }\textbf {\bibinfo {volume} {99}},\
  \bibinfo {pages} {041301} (\bibinfo {year} {2019})},\ \Eprint
  {http://arxiv.org/abs/1811.09151} {arXiv:1811.09151 [astro-ph.CO]}
  \BibitemShut {NoStop}%
\bibitem [{\citenamefont {Adshead}\ \emph {et~al.}(2021)\citenamefont
  {Adshead}, \citenamefont {Lozanov},\ and\ \citenamefont
  {Weiner}}]{Adshead:2021hnm}%
  \BibitemOpen
  \bibfield  {author} {\bibinfo {author} {\bibfnamefont {P.}~\bibnamefont
  {Adshead}}, \bibinfo {author} {\bibfnamefont {K.~D.}\ \bibnamefont
  {Lozanov}}, \ and\ \bibinfo {author} {\bibfnamefont {Z.~J.}\ \bibnamefont
  {Weiner}},\ }\href@noop {} {\  (\bibinfo {year} {2021})},\ \Eprint
  {http://arxiv.org/abs/2105.01659} {arXiv:2105.01659 [astro-ph.CO]}
  \BibitemShut {NoStop}%
\bibitem [{\citenamefont {Atal}\ \emph {et~al.}(2020)\citenamefont {Atal},
  \citenamefont {Cid}, \citenamefont {Escriv\`a},\ and\ \citenamefont
  {Garriga}}]{Atal:2019erb}%
  \BibitemOpen
  \bibfield  {author} {\bibinfo {author} {\bibfnamefont {V.}~\bibnamefont
  {Atal}}, \bibinfo {author} {\bibfnamefont {J.}~\bibnamefont {Cid}}, \bibinfo
  {author} {\bibfnamefont {A.}~\bibnamefont {Escriv\`a}}, \ and\ \bibinfo
  {author} {\bibfnamefont {J.}~\bibnamefont {Garriga}},\ }\href {\doibase
  10.1088/1475-7516/2020/05/022} {\bibfield  {journal} {\bibinfo  {journal}
  {JCAP}\ }\textbf {\bibinfo {volume} {05}},\ \bibinfo {pages} {022} (\bibinfo
  {year} {2020})},\ \Eprint {http://arxiv.org/abs/1908.11357} {arXiv:1908.11357
  [astro-ph.CO]} \BibitemShut {NoStop}%
\bibitem [{\citenamefont {Pi}\ and\ \citenamefont {Sasaki}(2020)}]{Pi:2020otn}%
  \BibitemOpen
  \bibfield  {author} {\bibinfo {author} {\bibfnamefont {S.}~\bibnamefont
  {Pi}}\ and\ \bibinfo {author} {\bibfnamefont {M.}~\bibnamefont {Sasaki}},\
  }\href {\doibase 10.1088/1475-7516/2020/09/037} {\bibfield  {journal}
  {\bibinfo  {journal} {JCAP}\ }\textbf {\bibinfo {volume} {09}},\ \bibinfo
  {pages} {037} (\bibinfo {year} {2020})},\ \Eprint
  {http://arxiv.org/abs/2005.12306} {arXiv:2005.12306 [gr-qc]} \BibitemShut
  {NoStop}%
\bibitem [{\citenamefont {Cai}\ \emph {et~al.}(2020)\citenamefont {Cai},
  \citenamefont {Pi},\ and\ \citenamefont {Sasaki}}]{Cai:2019cdl}%
  \BibitemOpen
  \bibfield  {author} {\bibinfo {author} {\bibfnamefont {R.-G.}\ \bibnamefont
  {Cai}}, \bibinfo {author} {\bibfnamefont {S.}~\bibnamefont {Pi}}, \ and\
  \bibinfo {author} {\bibfnamefont {M.}~\bibnamefont {Sasaki}},\ }\href
  {\doibase 10.1103/PhysRevD.102.083528} {\bibfield  {journal} {\bibinfo
  {journal} {Phys. Rev. D}\ }\textbf {\bibinfo {volume} {102}},\ \bibinfo
  {pages} {083528} (\bibinfo {year} {2020})},\ \Eprint
  {http://arxiv.org/abs/1909.13728} {arXiv:1909.13728 [astro-ph.CO]}
  \BibitemShut {NoStop}%
\bibitem [{\citenamefont {Yuan}\ \emph {et~al.}(2020)\citenamefont {Yuan},
  \citenamefont {Chen},\ and\ \citenamefont {Huang}}]{Yuan:2019wwo}%
  \BibitemOpen
  \bibfield  {author} {\bibinfo {author} {\bibfnamefont {C.}~\bibnamefont
  {Yuan}}, \bibinfo {author} {\bibfnamefont {Z.-C.}\ \bibnamefont {Chen}}, \
  and\ \bibinfo {author} {\bibfnamefont {Q.-G.}\ \bibnamefont {Huang}},\ }\href
  {\doibase 10.1103/PhysRevD.101.043019} {\bibfield  {journal} {\bibinfo
  {journal} {Phys. Rev. D}\ }\textbf {\bibinfo {volume} {101}},\ \bibinfo
  {pages} {043019} (\bibinfo {year} {2020})},\ \Eprint
  {http://arxiv.org/abs/1910.09099} {arXiv:1910.09099 [astro-ph.CO]}
  \BibitemShut {NoStop}%
\bibitem [{\citenamefont {{Detweiler}}(1979)}]{1979ApJ...234.1100D}%
  \BibitemOpen
  \bibfield  {author} {\bibinfo {author} {\bibfnamefont {S.}~\bibnamefont
  {{Detweiler}}},\ }\href {\doibase 10.1086/157593} {\bibfield  {journal}
  {\bibinfo  {journal} {ApJ}\ }\textbf {\bibinfo {volume} {234}},\ \bibinfo
  {pages} {1100} (\bibinfo {year} {1979})}\BibitemShut {NoStop}%
\bibitem [{\citenamefont {Desvignes}\ \emph {et~al.}(2016)\citenamefont
  {Desvignes} \emph {et~al.}}]{Desvignes:2016yex}%
  \BibitemOpen
  \bibfield  {author} {\bibinfo {author} {\bibfnamefont {G.}~\bibnamefont
  {Desvignes}} \emph {et~al.},\ }\href {\doibase 10.1093/mnras/stw483}
  {\bibfield  {journal} {\bibinfo  {journal} {Mon. Not. Roy. Astron. Soc.}\
  }\textbf {\bibinfo {volume} {458}},\ \bibinfo {pages} {3341} (\bibinfo {year}
  {2016})},\ \Eprint {http://arxiv.org/abs/1602.08511} {arXiv:1602.08511
  [astro-ph.HE]} \BibitemShut {NoStop}%
\bibitem [{\citenamefont {Hobbs}(2013)}]{Hobbs:2013aka}%
  \BibitemOpen
  \bibfield  {author} {\bibinfo {author} {\bibfnamefont {G.}~\bibnamefont
  {Hobbs}},\ }\href {\doibase 10.1088/0264-9381/30/22/224007} {\bibfield
  {journal} {\bibinfo  {journal} {Class. Quant. Grav.}\ }\textbf {\bibinfo
  {volume} {30}},\ \bibinfo {pages} {224007} (\bibinfo {year} {2013})},\
  \Eprint {http://arxiv.org/abs/1307.2629} {arXiv:1307.2629 [astro-ph.IM]}
  \BibitemShut {NoStop}%
\bibitem [{\citenamefont {McLaughlin}(2013)}]{McLaughlin:2013ira}%
  \BibitemOpen
  \bibfield  {author} {\bibinfo {author} {\bibfnamefont {M.~A.}\ \bibnamefont
  {McLaughlin}},\ }\href {\doibase 10.1088/0264-9381/30/22/224008} {\bibfield
  {journal} {\bibinfo  {journal} {Class. Quant. Grav.}\ }\textbf {\bibinfo
  {volume} {30}},\ \bibinfo {pages} {224008} (\bibinfo {year} {2013})},\
  \Eprint {http://arxiv.org/abs/1310.0758} {arXiv:1310.0758 [astro-ph.IM]}
  \BibitemShut {NoStop}%
\bibitem [{\citenamefont {Verbiest}\ \emph {et~al.}(2016)\citenamefont
  {Verbiest}, \citenamefont {Lentati}, \citenamefont {Hobbs}, \citenamefont
  {van Haasteren}, \citenamefont {Demorest}, \citenamefont {Janssen},
  \citenamefont {Wang}, \citenamefont {Desvignes}, \citenamefont {Caballero},
  \citenamefont {Keith},\ and\ \citenamefont {et~al.}}]{IPTA2016}%
  \BibitemOpen
  \bibfield  {author} {\bibinfo {author} {\bibfnamefont {J.~P.~W.}\
  \bibnamefont {Verbiest}}, \bibinfo {author} {\bibfnamefont {L.}~\bibnamefont
  {Lentati}}, \bibinfo {author} {\bibfnamefont {G.}~\bibnamefont {Hobbs}},
  \bibinfo {author} {\bibfnamefont {R.}~\bibnamefont {van Haasteren}}, \bibinfo
  {author} {\bibfnamefont {P.~B.}\ \bibnamefont {Demorest}}, \bibinfo {author}
  {\bibfnamefont {G.~H.}\ \bibnamefont {Janssen}}, \bibinfo {author}
  {\bibfnamefont {J.-B.}\ \bibnamefont {Wang}}, \bibinfo {author}
  {\bibfnamefont {G.}~\bibnamefont {Desvignes}}, \bibinfo {author}
  {\bibfnamefont {R.~N.}\ \bibnamefont {Caballero}}, \bibinfo {author}
  {\bibfnamefont {M.~J.}\ \bibnamefont {Keith}}, \ and\ \bibinfo {author}
  {\bibnamefont {et~al.}},\ }\href {\doibase 10.1093/mnras/stw347} {\bibfield
  {journal} {\bibinfo  {journal} {Monthly Notices of the Royal Astronomical
  Society}\ }\textbf {\bibinfo {volume} {458}},\ \bibinfo {pages} {1267–1288}
  (\bibinfo {year} {2016})}\BibitemShut {NoStop}%
\bibitem [{\citenamefont {Maggiore}\ \emph {et~al.}(2020)\citenamefont
  {Maggiore} \emph {et~al.}}]{Maggiore:2019uih}%
  \BibitemOpen
  \bibfield  {author} {\bibinfo {author} {\bibfnamefont {M.}~\bibnamefont
  {Maggiore}} \emph {et~al.},\ }\href {\doibase 10.1088/1475-7516/2020/03/050}
  {\bibfield  {journal} {\bibinfo  {journal} {JCAP}\ }\textbf {\bibinfo
  {volume} {03}},\ \bibinfo {pages} {050} (\bibinfo {year} {2020})},\ \Eprint
  {http://arxiv.org/abs/1912.02622} {arXiv:1912.02622 [astro-ph.CO]}
  \BibitemShut {NoStop}%
\bibitem [{\citenamefont {Barausse}\ \emph {et~al.}(2020)\citenamefont
  {Barausse} \emph {et~al.}}]{Barausse:2020rsu}%
  \BibitemOpen
  \bibfield  {author} {\bibinfo {author} {\bibfnamefont {E.}~\bibnamefont
  {Barausse}} \emph {et~al.},\ }\href {\doibase 10.1007/s10714-020-02691-1}
  {\bibfield  {journal} {\bibinfo  {journal} {Gen. Rel. Grav.}\ }\textbf
  {\bibinfo {volume} {52}},\ \bibinfo {pages} {81} (\bibinfo {year} {2020})},\
  \Eprint {http://arxiv.org/abs/2001.09793} {arXiv:2001.09793 [gr-qc]}
  \BibitemShut {NoStop}%
\bibitem [{\citenamefont {Yagi}\ and\ \citenamefont
  {Seto}(2011)}]{Yagi:2011wg}%
  \BibitemOpen
  \bibfield  {author} {\bibinfo {author} {\bibfnamefont {K.}~\bibnamefont
  {Yagi}}\ and\ \bibinfo {author} {\bibfnamefont {N.}~\bibnamefont {Seto}},\
  }\href {\doibase 10.1103/PhysRevD.95.109901, 10.1103/PhysRevD.83.044011}
  {\bibfield  {journal} {\bibinfo  {journal} {Phys. Rev.}\ }\textbf {\bibinfo
  {volume} {D83}},\ \bibinfo {pages} {044011} (\bibinfo {year} {2011})},\
  \bibinfo {note} {[Erratum: Phys. Rev.D95,no.10,109901(2017)]},\ \Eprint
  {http://arxiv.org/abs/1101.3940} {arXiv:1101.3940 [astro-ph.CO]} \BibitemShut
  {NoStop}%
%%CITATION = ARXIV:1101.3940;%%
\bibitem [{\citenamefont {Kawamura}\ \emph {et~al.}(2020)\citenamefont
  {Kawamura} \emph {et~al.}}]{Kawamura:2020pcg}%
  \BibitemOpen
  \bibfield  {author} {\bibinfo {author} {\bibfnamefont {S.}~\bibnamefont
  {Kawamura}} \emph {et~al.},\ }\href@noop {} {\  (\bibinfo {year} {2020})},\
  \Eprint {http://arxiv.org/abs/2006.13545} {arXiv:2006.13545 [gr-qc]}
  \BibitemShut {NoStop}%
\bibitem [{\citenamefont {Schmitz}(2021)}]{Schmitz:2020syl}%
  \BibitemOpen
  \bibfield  {author} {\bibinfo {author} {\bibfnamefont {K.}~\bibnamefont
  {Schmitz}},\ }\href {\doibase 10.1007/JHEP01(2021)097} {\bibfield  {journal}
  {\bibinfo  {journal} {JHEP}\ }\textbf {\bibinfo {volume} {01}},\ \bibinfo
  {pages} {097} (\bibinfo {year} {2021})},\ \Eprint
  {http://arxiv.org/abs/2002.04615} {arXiv:2002.04615 [hep-ph]} \BibitemShut
  {NoStop}%
\bibitem [{\citenamefont {Schmitz}(2020)}]{schmitz_kai_2020_3689582}%
  \BibitemOpen
  \bibfield  {author} {\bibinfo {author} {\bibfnamefont {K.}~\bibnamefont
  {Schmitz}},\ }\href {\doibase 10.5281/zenodo.3689582} {\enquote {\bibinfo
  {title} {{New Sensitivity Curves for Gravitational-Wave Experiments}},}\ }
  (\bibinfo {year} {2020})\BibitemShut {NoStop}%
\bibitem [{\citenamefont {Thrane}\ and\ \citenamefont
  {Romano}(2013)}]{Thrane:2013oya}%
  \BibitemOpen
  \bibfield  {author} {\bibinfo {author} {\bibfnamefont {E.}~\bibnamefont
  {Thrane}}\ and\ \bibinfo {author} {\bibfnamefont {J.~D.}\ \bibnamefont
  {Romano}},\ }\href {\doibase 10.1103/PhysRevD.88.124032} {\bibfield
  {journal} {\bibinfo  {journal} {Phys. Rev.}\ }\textbf {\bibinfo {volume}
  {D88}},\ \bibinfo {pages} {124032} (\bibinfo {year} {2013})},\ \Eprint
  {http://arxiv.org/abs/1310.5300} {arXiv:1310.5300 [astro-ph.IM]} \BibitemShut
  {NoStop}%
%%CITATION = ARXIV:1310.5300;%%
\bibitem [{\citenamefont {Gow}\ \emph {et~al.}(2021)\citenamefont {Gow},
  \citenamefont {Byrnes}, \citenamefont {Cole},\ and\ \citenamefont
  {Young}}]{Gow:2020bzo}%
  \BibitemOpen
  \bibfield  {author} {\bibinfo {author} {\bibfnamefont {A.~D.}\ \bibnamefont
  {Gow}}, \bibinfo {author} {\bibfnamefont {C.~T.}\ \bibnamefont {Byrnes}},
  \bibinfo {author} {\bibfnamefont {P.~S.}\ \bibnamefont {Cole}}, \ and\
  \bibinfo {author} {\bibfnamefont {S.}~\bibnamefont {Young}},\ }\href
  {\doibase 10.1088/1475-7516/2021/02/002} {\bibfield  {journal} {\bibinfo
  {journal} {JCAP}\ }\textbf {\bibinfo {volume} {02}},\ \bibinfo {pages} {002}
  (\bibinfo {year} {2021})},\ \Eprint {http://arxiv.org/abs/2008.03289}
  {arXiv:2008.03289 [astro-ph.CO]} \BibitemShut {NoStop}%
\bibitem [{\citenamefont {Robson}\ \emph {et~al.}(2019)\citenamefont {Robson},
  \citenamefont {Cornish},\ and\ \citenamefont {Liu}}]{Robson:2018ifk}%
  \BibitemOpen
  \bibfield  {author} {\bibinfo {author} {\bibfnamefont {T.}~\bibnamefont
  {Robson}}, \bibinfo {author} {\bibfnamefont {N.~J.}\ \bibnamefont {Cornish}},
  \ and\ \bibinfo {author} {\bibfnamefont {C.}~\bibnamefont {Liu}},\ }\href
  {\doibase 10.1088/1361-6382/ab1101} {\bibfield  {journal} {\bibinfo
  {journal} {Class. Quant. Grav.}\ }\textbf {\bibinfo {volume} {36}},\ \bibinfo
  {pages} {105011} (\bibinfo {year} {2019})},\ \Eprint
  {http://arxiv.org/abs/1803.01944} {arXiv:1803.01944 [astro-ph.HE]}
  \BibitemShut {NoStop}%
\bibitem [{\citenamefont {Moore}\ \emph {et~al.}(2015)\citenamefont {Moore},
  \citenamefont {Cole},\ and\ \citenamefont {Berry}}]{Moore:2014lga}%
  \BibitemOpen
  \bibfield  {author} {\bibinfo {author} {\bibfnamefont {C.~J.}\ \bibnamefont
  {Moore}}, \bibinfo {author} {\bibfnamefont {R.~H.}\ \bibnamefont {Cole}}, \
  and\ \bibinfo {author} {\bibfnamefont {C.~P.~L.}\ \bibnamefont {Berry}},\
  }\href {\doibase 10.1088/0264-9381/32/1/015014} {\bibfield  {journal}
  {\bibinfo  {journal} {Class. Quant. Grav.}\ }\textbf {\bibinfo {volume}
  {32}},\ \bibinfo {pages} {015014} (\bibinfo {year} {2015})},\ \Eprint
  {http://arxiv.org/abs/1408.0740} {arXiv:1408.0740 [gr-qc]} \BibitemShut
  {NoStop}%
\bibitem [{\citenamefont {Caprini}\ \emph {et~al.}(2019)\citenamefont
  {Caprini}, \citenamefont {Figueroa}, \citenamefont {Flauger}, \citenamefont
  {Nardini}, \citenamefont {Peloso}, \citenamefont {Pieroni}, \citenamefont
  {Ricciardone},\ and\ \citenamefont {Tasinato}}]{Caprini:2019pxz}%
  \BibitemOpen
  \bibfield  {author} {\bibinfo {author} {\bibfnamefont {C.}~\bibnamefont
  {Caprini}}, \bibinfo {author} {\bibfnamefont {D.~G.}\ \bibnamefont
  {Figueroa}}, \bibinfo {author} {\bibfnamefont {R.}~\bibnamefont {Flauger}},
  \bibinfo {author} {\bibfnamefont {G.}~\bibnamefont {Nardini}}, \bibinfo
  {author} {\bibfnamefont {M.}~\bibnamefont {Peloso}}, \bibinfo {author}
  {\bibfnamefont {M.}~\bibnamefont {Pieroni}}, \bibinfo {author} {\bibfnamefont
  {A.}~\bibnamefont {Ricciardone}}, \ and\ \bibinfo {author} {\bibfnamefont
  {G.}~\bibnamefont {Tasinato}},\ }\href {\doibase
  10.1088/1475-7516/2019/11/017} {\bibfield  {journal} {\bibinfo  {journal}
  {JCAP}\ }\textbf {\bibinfo {volume} {11}},\ \bibinfo {pages} {017} (\bibinfo
  {year} {2019})},\ \Eprint {http://arxiv.org/abs/1906.09244} {arXiv:1906.09244
  [astro-ph.CO]} \BibitemShut {NoStop}%
\end{thebibliography}%

\end{document}